\documentclass[12pt,twoside]{article}

\newlength{\dinwidth}
\newlength{\dinmargin}
\usepackage{amsmath, amsthm, amsfonts, amssymb, bm, mathrsfs, indentfirst,
sectsty,
graphicx, fancyhdr, slashed, fullpage, color, authblk}

\theoremstyle{definition}

\theoremstyle{remark}

\def\a{\alpha}
\def\n{\nu}
\def\b{\beta}
\def\c{\chi}

\def\l{\lambda}
\def\m{\mu}

\def\r{\rho}
\def\s{\sigma}

\def\w{\omega}
\def\y{\psi}

\def\Th{\Theta}

\def\F{\Phi}
\def\Y{\Psi}

\def\beq{\begin{equation}}
\def\enq{\end{equation}}

\def\erf#1{(\ref{#1})} 
\newcommand{\cA}{{\cal A}}  \newcommand{\cB}{{\cal B}}
\newcommand{\cC}{{\cal C}}  
  \newcommand{\cF}{{\cal F}}

\newcommand{\cM}{{\cal M}}  
\newcommand{\cO}{{\cal O}}  \newcommand{\cP}{{\cal P}}
  \newcommand{\cR}{{\cal R}}

  \newcommand{\bq}{{\mathbf q}}
\newcommand{\bx}{{\mathbf x}}

\newcommand{\be}{\begin{equation}} \newcommand{\ee}{\end{equation}}
\newcommand{\bea}{\begin{eqnarray}} \newcommand{\eea}{\end{eqnarray}}
\newcommand{\beann}{\begin{eqnarray*}}  \newcommand{\eeann}{\end{eqnarray*}}
\newcommand{\bfig}{\begin{figure}} \newcommand{\efig}{\end{figure}}
\newcommand{\ba}{\begin{array}} \newcommand{\ea}{\end{array}}
\newcommand{\bcen}{\begin{center}} \newcommand{\ecen}{\end{center}}
\newcommand{\btab}{\begin{tabular}} \newcommand{\etab}{\end{tabular}}
\newcommand{\nn}{\nonumber}
\newcommand{\matt}{\left ( \begin{array}{ccc}}
\newcommand{\ematt}{\end{array} \right )} \newcommand{\matf}{\left
(\begin{array}{cccc}}
    \newcommand{\ematf}{\end{array} \right )} \newcommand{\vect}{\left (
\begin{array}{c}}
    \newcommand{\evect}{\end{array} \right )}    \def\beqn{\begin{eqnarray}}
 \def\eeqn{\end{eqnarray}}  
     
\renewcommand{\Re}{\mathop{\rm Re}}   \renewcommand{\Im}{\mathop{\rm Im}}
\newcommand{\vev}[1]{\left\langle{#1}\right\rangle}

\newtheorem{Proposition}{Proposition}[section]

\newtheorem{Theorem}{Theorem}[section]
\newtheorem{Lemma}{Lemma}[section]
\newtheorem{Corrolary}{Corrolary}[section]

\newcommand{\bp}{\begin{Proposition}}   \newcommand{\ep}{\end{Proposition}}
\newcommand{\bt}{\begin{Theorem}}   \newcommand{\et}{\end{Theorem}}
\newcommand{\bl}{\begin{Lemma}}     \newcommand{\el}{\end{Lemma}}
\newcommand{\bc}{\begin{Corrolary}} \newcommand{\ec}{\end{Corrolary}}

\begin{document}
\begin{titlepage}

\flushright{IFT-UAM/CSIC-13-01}\\[2cm]
\begin{center}
 {\Large \bf \sc Holographic Type II Goldstone bosons}
\\[1.52cm]
{\large Irene Amado$^{a,}$\footnote{irene.r.amado@gmail.com}, Daniel
Are\'an$^{b,e,}$\footnote{arean@sissa.it},
Amadeo Jimenez-Alba$^{c,}$\footnote{amadeo.j@gmail.com},\\
Karl Landsteiner$^{c,}$\footnote{karl.landsteiner@csic.es}, Luis
Melgar$^{c,}$\footnote{luis.melgar@csic.es} and Ignacio Salazar
Landea$^{d,b,}$\footnote{peznacho@gmail.com} }

\bigskip

{}$^{a}${\small \it Department of Physics, Technion, Haifa 32000, Israel}\\
{}$^{b}${\small \it International Centre for Theoretical Physics
(ICTP), Strada Costiera 11; I 34014 Trieste, Italy}\\
{}$^{c}${\small \it Instituto de F\'\i sica Te\'orica IFT-UAM/CSIC, Universidad
Aut\'onoma de Madrid,
28049 Cantoblanco, Spain}\\
{}$^{d}${\small \it Instituto de F\'\i sica La Plata (IFLP) and
Departamento de F\'\i sica Universidad Nacional de La Plata, CC 67,
1900 La Plata, Argentina}\\
 {}$^{e}${\small \it INFN - Sezione di
Trieste, Strada Costiera 11; I 34014 Trieste, Italy}
\end{center}

\bigskip

\begin{abstract}
The Goldstone theorem implies the appearance of an ungapped mode whenever a
continuous global symmetry is spontaneously broken. In general it does not
say anything about the precise form of the dispersion relation nor does it
imply that there is one massless mode for each broken symmetry generator.
It is a well-established fact that even for relativistic field theories
in the presence of a chemical potential Goldstone modes with quadratic
dispersion relation, the type II Goldstone bosons,  appear in the spectrum.
We develop two holographic models that feature type II Goldstone modes as
part of the quasinormal mode spectrum. The models are based on simple
generalizations
with $U(2)$ symmetry of the well-studied holographic s-wave superfluid.
Our results include Goldstone modes without broken generators but with
unusual realization of symmetries and a frequency dependent conductivity of
striking resemblance to the one of Graphene.

\end{abstract}

\end{titlepage}

\setcounter{footnote}{0}
\section{Introduction}

The AdS/CFT correspondence has proved to be a powerful tool to study strongly
coupled quantum systems, with applications to QCD and condensed matter systems.
Among the applications to condensed matter physics one of the most
important examples is the construction of the so called holographic
superfluids \cite{Gubser:2008px, Hartnoll:2008vx, Hartnoll:2008kx}.

The ingredients of such a model are a charged black
hole describing a CFT at finite temperature and charge density.
At sufficiently low temperature (or high chemical potential) a charged scalar
field
develops an expectation value
and triggers a symmetry breaking phase transition towards a
superfluid phase. In the simplest model the order parameter is a scalar and
therefore
we speak of an s-wave superfluid. Furthermore it is possible to go to a {\it
decoupling limit}
in which the fluctuations of the bulk metric are suppressed. The dynamics in
this
limit is completely specified by the fluctuations of the gauge field and the
scalar field \cite{Hartnoll:2008kx}.
Generalizations to p-wave \cite{Gubser:2008wv} and d-wave \cite{Benini:2010pr,
Chen:2010mk}
superconductors,
for which the
order parameter has angular momentum $l=1$ and $l=2$ respectively are also
known.
Much of this development is reviewed in
\cite{Horowitz:2010gk,Kaminski:2010zu}.

The strength of the gauge/gravity correspondence is that it allows to study the
real time dynamics of strongly coupled field theories rather easily.
Linear response theory captures the behavior of a quantum system after an
initial, small perturbation. It also applies to the late time behavior when
an initially large perturbation has already sufficiently died out and enters
the linear regime. The basic ingredient of linear response theory is the
retarded Green's function. In the context of the AdS/CFT correspondence it
was shown in \cite{Son:2002sd, Herzog:2002pc} how to calculate retarded Green's
functions by imposing
infalling boundary conditions at the black hole horizon. For black holes with
non-degenerate horizons the retarded Green's functions are analytic in the upper
half of the complexified frequency plane and have (an infinite series of)
isolated
poles in the lower half. These poles are the holographic quasinormal modes (QNM)
of the black hole \cite{Horowitz:1999jd, Birmingham:2001pj, Berti:2009kk,
Landsteiner:2012gn}.
Within the QNM
spectrum, the ungapped modes play a special role, since they give the dominant
contribution to the retarded Green's functions at low frequency and small
momentum.
Therefore they determine the
hydrodynamic description of the system.
In this paper we will
study the quasinormal mode spectrum of a multi-component generalization
of the simple, holographic, non-backreacted s-wave superfluid with particular
focus on the low lying and hydrodynamic modes.

The study of the QNMs for the s-wave $U(1)$ superfluid was first carried out
in \cite{Amado:2009ts}. Since the basic physics of superfluids is the one
of spontaneous symmetry breaking it can be expected that known results such as
the existence of a Goldstone boson\footnote{It is also often called {\em
Nambu-Goldstone} boson.
For simplicity we will refer to it as Goldstone boson or Goldstone mode
throughout the text.} carry over to the holographic models.
Indeed, one of the main results of \cite{Amado:2009ts} was that the QNM spectrum
in the superfluid phase contains such an ungapped Goldstone mode with
dispersion relation $\omega = \pm v_s k + O(k^2)$. This mode can also be
understood
as the sound mode of the superfluid and $v_s$ is the sound
velocity\footnote{
In \cite{Yarom:2009uq} it was pointed out that this mode corresponds to the {\it
fourth} sound. Whereas second sound is defined through temperature oscillations
fourth sound
is the phenomenon of sound propagation in the superfluid component only
\cite{Khalatnikov}.
This is the mode that survives the probe limit in which 
propagation in the normal component of the dual fluid is prohibited. In this
model
there is no other sound mode so we will simplfy refer to it as the sound mode.
Due to the nature of fourth sound it interpolates between second sound at
$T=T_c$
and normal sound at $T=0$.}.
In the non-backreacted model these are the only hydrodynamic modes in the broken
phase.
In the unbroken phase in contrast there exists a single hydrodynamic mode
signaling
the usual diffusive behavior of a normal fluid. Its dispersion relation is
$\omega = -i D k^2$,
where $D$ is the diffusion constant. The question what happens to this diffusive
mode
in the broken phase was also answered in \cite{Amado:2009ts}: it develops a
purely
imaginary gap $\omega = -i \gamma - i\tilde D k^2$. This is quite natural
because
the single
purely imaginary mode can not move off the imaginary axis\footnote{Quasinormal
modes
are bound to come either in pairs $\omega_n$ and $\tilde \omega_n = -
\omega_n^*$ or
are fixed on the imaginary axis. This follows from rather generic symmetry
considerations for retarded Green's functions, see appendix \ref{app.a}.}.
The hydrodynamics of the broken phase is fully captured by the
Goldstone
mode and the diffusion mode does the simplest thing it can to drop out of the
hydrodynamic
regime by growing the gap $\gamma$. Since this purely imaginary gapped mode has
its origin in
the universal diffusive mode of the unbroken phase we expect that it is a
universal feature
of a large class of superfluids, not only holographic ones.
This mode will necessarily dominate the
late time response in the order parameter to homogeneous perturbations and in
regimes
close but below the critical temperature where the gap $\gamma$ is rather small.
Therefore the order parameter is bound to show a purely
exponential decay towards its equilibrium value without any oscillation.
In contrast for lower temperatures where $\gamma$ becomes large there are other
low lying QNMs with real and imaginary parts in their frequency. In this low
temperature
regime the response in the order parameter is then an exponentially damped
oscillation rather than a
purely imaginary decay.
This universal aspect of the late time response of superfluids was also
emphasized
in recent numerical studies of quenches of holographic superfluids in
\cite{Bhaseen:2012gg}.

In this paper we generalize the results on the QNM spectrum to models
with $U(2)$ symmetry.
In the first model we simply add a second scalar
field of the same mass, we will call this the {\em ungauged model}.
A second model also includes gauge fields for the whole $U(2)$ symmetry.
The difference between the two models is as follows. In the ungauged model
only the $U(1)$ symmetry is local in the bulk. It has however a {\em
global} $SU(2)$ symmetry under which the scalar fields transform as a
doublet. According to the holographic dictionary this model contains
only one conserved current, corresponding to the single gauge field in
the bulk. The dual field theory inherits of course the global $SU(2)$ symmetry
of the bulk but this symmetry is not generated by operators in the dual
conformal field theory. This is similar to the decoupling limit in which we
are working and in which the fluctuations of the metric are suppressed. The
dual field theory has then strictly speaking no energy momentum tensor.
In usual four dimensional Lagrangian field theories Noether's theorem guarantees
that we can always construct a conserved charge generating a given symmetry
of the Lagrangian. In holographically defined field theories the existence of a
four dimensional Lagrangian is a priori not guaranteed and therefore Noether's
theorem
does not straightforwardly apply. This is the case here. Although the
dual field theory has the $SU(2)$ symmetry (and Poincar{\'e} covariance)
it does not contain operators generating these symmetries. We can speak
of these symmetries as an outer automorphism of the operator algebra
of the dual field theory\footnote{A string theory example
for such a situation is the theory based on the small ${\cal N}=4$
superconformal algebra on the world sheet. This algebra possesses a
large $SO(4)=SU(2)\times SU(2)$ symmetry acting on the four supercharges
of which only one $SU(2)$ is represented through chiral currents on the
worldsheet.}. Physically the difference between the two models is that the
ungauged
one is a one-component fluid (there is only one notion of charge)
whereas the gauged one is a two component fluid. In the latter case the charges
are
the expectation values of the zero-component of the currents in the Cartan
subalgebra
of the $U(2)$ symmetry.

Although this ungauged model does not contain conserved currents for the $SU(2)$
symmetry
and therefore many of the standard proofs about existence of Goldstone bosons do
not
strictly apply we find a new ungapped mode in the QNM spectrum of the scalars.
This mode is however not a standard Goldstone boson with linear dispersion
relation
but a so-called type II Goldstone mode whose energy depends quadratically on
momentum.

The second model we consider has a scalar field doublet coupled to
the full set of $U(2)$ gauge fields.  We switch on a chemical potential only
for the overall $U(1)$ symmetry. Therefore the high temperature phase
has the full $U(2)$ symmetry. At low temperatures this symmetry is broken
to $U(1)$. In this model the dual field theory contains
currents for all the $U(2)$ symmetries. We can therefore also study the
conductivities.

In the context
of condensed matter physics it has been pointed out long ago in \cite{Halperin}
that such multicomponent superfluids have unusual Goldstone modes with
quadratic dispersion relation. In the high energy context such models
have been considered as models for Kaon condensation in the color-flavor
locked phase of QCD in \cite{Schafer:2001bq, Miransky:2001tw} again emphasizing
the existence of the quadratic Goldstone mode.
Our gauged holographic model is a straightforward holographic analogue of
the model in \cite{Schafer:2001bq, Miransky:2001tw} and indeed we also find the
presence of a Goldstone mode
with quadratic dispersion relation. Let us also note that in the holographic
context a type II Goldstone boson was found before in magnetized $D3/D5$ defect
theory \cite{Filev:2009xp}.

It seems useful to collect now some of the known theorems on Goldstone bosons
(a very useful review on symmetry breaking and Goldstone modes is
\cite{Brauner:2010wm}).
First we have of course the actual {\bf Goldstone theorem}. Its proof assumes
the existence of a conserved current $j^\mu$ such that the broken charge is
$Q=\int d^dx j^0$ (with $d$ spatial dimensions). The theorem then states that
spontaneous breaking of a continuous global symmetry implies the existence of
a mode whose energy fulfills
\begin{equation}\label{eq:gapless}
 \lim_{k\rightarrow 0}\omega(k) =0\,.
\end{equation}

The theorem by itself does not make any statement about the number of these
modes, nor
does it fix the $k$-dependence of the frequency. In the presence of Poincar\'e
symmetry
one can make however a stronger statement, namely that the dispersion relation
of
the Goldstone mode has to be linear and that the number of Goldstone bosons
equals the
number of broken generators.

Lorentz symmetry might be absent however, either in principle such as in
non-relativistic
field theories or the system under consideration might be in a Lorentz symmetry
breaking
state, such as being at finite temperature or density.  In these cases another
theorem
classifies {\bf Goldstone bosons} as {\bf type I} if their energy vanishes as an
{\bf odd power}
of the momentum or as {\bf type II} if their energy vanishes as an {\bf even
power} of the
momentum  in the zero momentum limit. The number of type I and type II Goldstone
bosons has to fulfill then
\begin{equation}\label{eq:chaniel}
 n_I + 2 n_{II} \geq N_{BG}\,,
\end{equation}
where $N_{BG}$ is the number of broken generators \cite{Nielsen:1975hm}.
The number of type I and type II Goldstone bosons can be further constrained.
Upon
assuming that the broken symmetry generators obey $\langle [ Q_a, Q_b ] \rangle
= B_{ab}$
the number of Goldstone bosons has to fulfill \cite{Watanabe:2011ec,
Watanabe:2012hr, Hidaka:2012ym} (see also
\cite{Brauner:2010wm, Kapustin:2012cr, Watanabe:2013iia} for more on counting
rules of Goldstone bosons).
\begin{equation}\label{eq:watbrau}
 n_I + n_{II} = N_{BG} - \frac 1 2 \mathrm{rank}(B)\,.
\end{equation}

We note that the gauged holographic model fulfills all these theorems. We have
in total
four symmetry generators. The symmetry is broken from $U(2)$ to $U(1)$ and so
there
are three broken generators. In the broken phase the charges corresponding to
the overall
$U(1)$ and the Cartan $U(1)$ generator inside $SU(2)$ receive vacuum expectation
values.
Therefore the rank of the matrix $B$ is two and so the number of type I and type
II Goldstone
bosons should add up to two. This is precisely what we find in the QNM spectrum,
one ungapped
mode with linear dispersion relation and one ungapped mode with quadratic
dispersion relation.

We also note that the ungauged model satisfies Goldstone's theorem and the
counting rule of
Chadha and Nielsen \erf{eq:chaniel}. It violates however the more refined
counting rule
(\ref{eq:watbrau}).
In a strict sense this model only has one symmetry generator since it has only
one $U(1)$ gauge
field in the bulk. Therefore the counting rule (\ref{eq:watbrau}) would suggest
the existence
of only one Goldstone boson, the number of broken generators is one and the
matrix $B$ vanishes
trivially.

This paper is organized as follows.
In section \ref{sec:fieldth} we briefly review a simple field theoretical model
featuring
type II Goldstone modes. This model has been introduced in the context
of Kaon condensation in color-flavor locked QCD. It serves us as inspiration
for constructing the holographic models.

Section \ref{sec:ungauged} is devoted to the analysis of the ungauged model.
Since the well-known s-wave superconductor is a subsector of both the
ungauged and the gauged model we also briefly review first the
findings of \cite{Amado:2009ts}.
Then we show that even with this drastic simplification, i.e. not gauging
the global $SU(2)$ symmetry in the bulk, the model presents
Goldstone modes with quadratic dispersion relation. Hence, within this model a
type II NG boson is found as a consequence of having broken \emph{just one}
charge generator (the one associated to the $U(1)$ symmetry).

In section \ref{sec:gauged} we study the fully gauged $U(2)$ model.
Then we analyze the fluctuation equations
to linear order. They decompose into three decoupled sectors. One
being the already encountered $U(1)$ s-wave superfluid, the other
describing the non-Abelian sector in which the type II Goldstone mode
resides and a third one with the unbroken $U(1)$ symmetry.
We proceed to study the conductivities
which now arrange naturally into a two by two matrix.
We show that the diagonal conductivities have delta-functions at zero
frequency and are in this sense superconducting. Furthermore upon a change
of basis we find a frequency dependence that is strikingly similar to the
one of Graphene \cite{Graphene}.
Furthermore we find indications that for temperatures below $T=0.4T_c$ another
instability arises in the gauge field sector leading to an additional p-wave
condensate.
Then we study the low lying quasinormal modes and analyze the results. We
find the type II Goldstone mode and also study the fate of the
diffusion
modes in the broken phase. Since now two symmetries participate there are two
diffusion
modes that in the broken phase pair up and can move away from the imaginary
axis. We find
that this is precisely what happens. Therefore the response  in this sector
does not show the purely exponential decay induced by the gapped pseudo
diffusion
mode of
the $U(1)$ sector.

We conclude in section \ref{sec:conclusions} with a discussion of our results
and an outlook to
further possible studies of holographic type II Goldstone modes.

Finally, in appendix \ref{app.a} we present some general properties of
matrix-valued Green's
functions from which constraints on the quasinormal mode spectrum follow . In
appendix \ref{appB}
we collect technical details on how to actually compute the QNMs for coupled
systems.

\section{A field theoretical model with type II Goldstone boson}
\label{sec:fieldth}

Motivated by the physics of Kaon condensation in the color-flavor locked phase
of QCD
the authors of \cite{Schafer:2001bq, Miransky:2001tw} studied QCD at a nonzero
chemical potential for strangeness.
It was shown that at a critical value of the chemical potential equal to the
Kaon mass, Kaon
condensation occurs through a continuous phase transition. Moreover, a Goldstone
boson with
the non-relativistic dispersion relation $\omega\sim p^2$ appears in the Kaon
condensed phase.
To illustrate this fact, they considered the following (Euclidean) toy model:
\begin{equation}
  {\cal L} = (\partial_0 + \mu)\phi^\dagger
  (\partial_0 - \mu)\phi + \partial_i\phi^\dagger\partial_i\phi
  + M^2 \phi^\dagger \phi + \lambda
  (\phi^\dagger \phi)^2\,,
  \label{Lagr}
\end{equation}
where $\phi$ is a complex scalar doublet,
\begin{equation}
   \phi = \vect \phi_1 \\ \phi_2 \evect.
\end{equation}

As long as $\mu<M$ the masses of the four excitations in the model are
the positive roots in $\omega$ of
\begin{align}
 (\omega \pm \mu)^2=M^2\,.
\end{align}
All are doubly degenerate.
It is straightforward to check that at $\mu=M$ the global $U(2)$ symmetry gets
broken and the new vacuum can be chosen to be:
\begin{equation}
 \phi = \frac{1}{\sqrt2} \vect 0
  \\ v \evect, \qquad {\rm with} \quad v^2 = \frac{\mu^2 -M^2}\lambda\,.
  \label{choice}
\end{equation}

\begin{figure}[htp!]
\centering
\includegraphics[width=250pt]{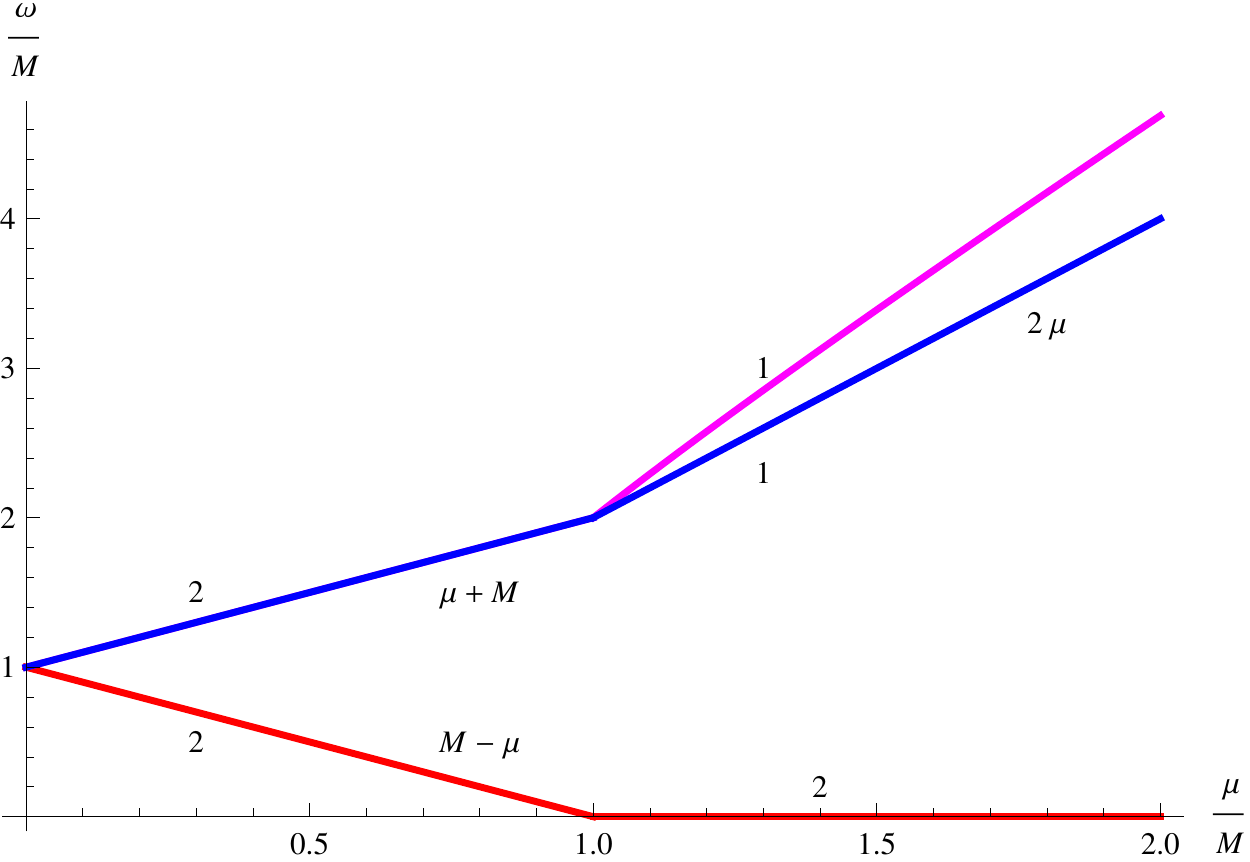}
\caption{\label{fig:FTspec} The spectrum of the field theoretical model. Below
the critical value $\mu=M$ there are four massive modes. The masses
are $M-\mu$ and $M+\mu$, the numbers indicate that they are doubly degenerate.
In the broken phase $\mu>M$ there are two Goldstone
modes with exactly zero mass and two gapped modes. The special gapped mode has
mass $2\mu$.}
\end{figure}

Studying the fluctuations of the doublet $\phi$ around this background one finds
two massless and two massive modes with the following dispersion relations:
\begin{align}
\omega_1^2&=\frac{\mu^2-M^2}{3\mu^2-M^2}\,p^2+O(p^4)\,,  \label{eq:typeI}\\
\omega_2^2&=6\mu^2-2M^2+O(p^2)\,,\label{eq:gapped}\\
\omega_3^2&= p^2 - 2 \mu \omega_3\,,\label{eq:E3}\\
\omega_4^2&= p^2 + 2\mu \omega_4\,.\label{eq:E4}
\end{align}
If we concentrate on the positive roots we see that $\omega_1$ is a normal,
linear Goldstone mode.
The positive root of equation (\ref{eq:E3}) is
\begin{equation}
 \omega_3 = \frac{p^2}{2\mu}+O(p^4)\,.
\end{equation}
This is the type II Goldstone mode. It has formally a non-relativistic
dispersion relation. Since the
underlying theory has however Lorentz invariance there is of course also a
negative energy mode with
quadratic dispersion. This arises as the negative root of $\omega_4$.
Finally $\omega_2$ and $\omega_4$ are gapped modes with
\begin{equation}
 \omega_4 = 2\mu + O(p^2)\,.\label{eq:spmode}
\end{equation}
Since the symmetry breaking pattern is $U(2) \rightarrow U(1)$ there are three
broken generators but
only two massless Goldstone modes in the spectrum. This model fulfills all the
counting theorems
noted in the introduction. In particular the Chadha-Nielsen rule
(\ref{eq:chaniel})
is exactly saturated.
The role of $\omega_4$ is special. It is the mode that pairs up with the type II
Goldstone mode in
the dispersion relations (\ref{eq:E3}) and  (\ref{eq:E4}). It has been argued
that this mode is a universal feature
and that its energy at zero momentum is exact and protected against quantum
corrections \cite{ Kapustin:2012cr, Brauner:2006xm, Nicolis:2012vf}.
The spectrum obtained from this model is summed up in Figure \ref{fig:FTspec}.
In our holographic models we will look for this
special gapped partner mode of the type II Goldstone mode. It will turn out that
the gauged and ungauged models
differ significantly here: only the mode in the gauged model shows the
characteristic linear dependence on the
chemical potential.

This simple Lagrangian model serves as our motivation and guideline to construct
a holographic model featuring type II Goldstone modes.
In fact we can use the same kind of matter Lagrangian in a holographic setup.
According to the usual holographic dictionary a local bulk symmetry corresponds
to a global symmetry in the boundary conformal field theory. We would therefore
most naturally be led to a model in which we gauge the global $U(2)$ symmetry
of
(\ref{Lagr}) and put it into an AdS Schwarzschild background.
In order to trigger spontaneous symmetry breaking we introduce a chemical
potential
via a boundary value for the temporal component of the overall, Abelian $U(1)$
gauge
field. This is then our {\em gauged model}.

Alternatively we might ask what are the minimal ingredients necessary to trigger
spontaneous
symmetry breaking. The chemical potential resides entirely in the overall $U(1)$
factor. The other three $SU(2)$ gauge fields are not needed to achieve
symmetry breaking.
Therefore we can choose as a sort of minimal setup a model in which the $SU(2)$
symmetry stays
global in the bulk of AdS. As already mentioned in the introduction this is a
somewhat unusual
realization of the symmetry from the boundary conformal field theory point of
view. There are
no conserved currents associated to this $SU(2)$ symmetry, nevertheless all
states and operators
fall naturally into representations of this symmetry group since it is a global
symmetry of the bulk
and it is also not broken by any of the boundary conditions. This setup
constitutes our {\em ungauged
model} and we will study it in detail in the next section.

Let us note here one more technical detail: the field theoretic model of this
section is
most naturally viewed as living in four space time dimensions. In the following
our holographic models will be dual to field theories living in three space time
dimensions
in order to stay as close as possible to the well-studied holographic $U(1)$
s-wave
superfluid of \cite{Hartnoll:2008vx, Amado:2009ts}. This is however of no
relevance to the
essential features of the models, i.e. the existence and the nature of the
hydrodynamic
and Goldstone modes.

\section{The ungauged model}\label{sec:ungauged}

We will now study the holographic model where the condensation of a charged
scalar breaks a global
$SU(2)$ symmetry in the bulk. We shall look at the spectrum of quasinormal
modes on both sides
of the phase transition and study their dispersion relations.
Since the simple $U(1)$ s-wave holographic superfluid constitutes a subsector
of this as well as of the gauged model we will also use the opportunity to
briefly
review the most salient features of its QNM spectrum.

The minimal holographic model containing a type II Goldstone boson consists of a
scalar doublet of $SU(2)$ charged under a $U(1)$ gauge field. The Lagrangian is
given by
\be
\mathcal{L}=\left(-\frac{1}{4}F^{\mu\nu}F_{\mu\nu}
-m^2\Psi^\dagger\Psi-(D^\mu\Psi)^\dagger D_\mu\Psi\right)\,,
\ee
where
\be
\Psi=
\begin{pmatrix}
\lambda\\
\psi
\end{pmatrix} \,,
\hspace{2cm}
D_\mu=\partial_\mu-iA_\mu\,,
\ee
and $A_\mu$ is the Abelian gauge field. The mass of the scalar field is chosen
to be $m^2 = -2/L^2$.
This is basically the same as the model in \cite{Hartnoll:2008vx} except that we
have added a second
scalar field $\lambda$ with the same mass. Because of the degeneracy in the mass
the model possesses
in addition to the bulk-local $U(1)$ symmetry a bulk-global $SU(2)$ symmetry.
Note that the global $SU(2)$ symmetry is a priori not enough to set the field
$\lambda(r)=0$. But we are interested in un-sourced static solutions for the
scalar fields,
i.e. we assume that the leading non-normalizable mode is not switched on. The
solution
space is then a two dimensional complex vector space spanned by the vevs of the
operators
dual to the scalar fields. On this parameter space we can act with the global
$SU(2)$ symmetry to set the operator corresponding to the field $\lambda$ equal
to zero. Since now the non-normalizable and the normalizable mode of $\lambda$
are set to zero it follows that $\lambda(r)=0$.

We will be working in the probe limit, in which the coupling of the gauge field
is very large
and the backreaction of the matter fields onto the metric can be neglected. The
background metric
is then taken to be the Schwarzschild-AdS black brane
\begin{eqnarray}
\nonumber ds^2=-f(r)dt^2 +\frac{dr^2}{f(r)} + \frac{r^2}{L^2} (dx^2+dy^2)\,, \\
\label{eq:metric}f(r)= \frac{r^2}{L^2} - \frac{M}{r}\,.
\end{eqnarray}
The horizon of the black hole is located at $r_H=M^{1/3} L^{2/3}$ and its
Hawking temperature is $T = \frac{3 r_H}{4 \pi L^2}$.
In the following we use dimensionless coordinates
\be
\label{eq:rescale}\left(r,\,t,\,x,\,y\right)\,
\to\,\left(r_H\,\rho,\,\frac{L^2}{r_H}\bar t,\,\frac{L^2}{r_H}\bar
x,\,\frac{L^2}{r_H}\bar y\right)\,.
\ee
These rescalings allow us to set $M=r_H=1$ in the dimensionless system.
In order to switch on a finite chemical potential in the boundary theory, the
bulk Maxwell field
\be
A = \chi (\rho) d\bar t \,,
\ee
must take a non-zero value at the boundary.

The equations of motion for the background fields are
\begin{eqnarray}
\label{eq:gauge} &&\chi''+\frac{2}{\rho}\c'-\frac{2 \psi^2}{f}\c=0 \,,\\
\label{eq:scalar}&&\psi''+\left(\frac{f'}{f}+\frac{2}{\rho}\right)\psi'+\frac{
\chi^2}{f^2}\psi-\frac{m^2}{f}\psi=0\,.
\end{eqnarray} Notice that the
system above is precisely
the original $U(1)$ holographic superconductor first studied in
\cite{Hartnoll:2008vx}. To ensure
finiteness of the norm of the current at the horizon, we have to demand the
scalar field to be regular
whereas the gauge field has to vanish $\c(\rho=1)=0$. With these boundary
conditions, the asymptotic
behavior of the fields at the conformal boundary is
\begin{eqnarray}
\label{eq:expchi} \chi= \bar \mu-\frac{\bar
n}{\rho}+O\left(\frac{1}{\rho^2}\right)\,,\\
\label{eq:exppsi} \psi=
\frac{\psi_1}{\rho}+\frac{\psi_2}{\rho^2}+O\left(\frac{1}{\rho^3}
\right)\,.
\end{eqnarray}
For the chosen value of the scalar mass, both terms in the scalar asymptotics
correspond to normalizable modes  \cite{Klebanov:1999tb}.
Considering one or the other as the vacuum expectation value of a dual boundary
operator leads to two different theories. In what follows we will consider only
the case in which 
$\psi_1$ is interpreted as the coupling and
$\psi_2$ as the vev of a mass dimension two operator.

The dimensionless parameters are related with the physical quantities by
\begin{eqnarray}
\label{eq:barmu}\bar\mu&=&\frac{3 }{4\pi T}\, \mu\,,\\
\bar n&=&\frac{9}{16\pi^2 T^2 L^2}\,  n\,,\\
\psi_1&=&\frac{3}{4\pi T L^2}\, J_{\mathcal O}\,,\\
\label{eq:psi2}\psi_2&=&\frac{9}{16\pi^2 T^2 L^4}\,
\langle{\mathcal O}\rangle\,,
\end{eqnarray}
where $\mu$, $n$ and $J_{\mathcal O}$, $\langle{\mathcal O}\rangle$ are the
chemical potential, charge density and source and
expectation value
of an operator ${\mathcal O}$ of dimension $2$, respectively. From now on we set
$L=1$.
In the following we will work in the grand canonical ensemble. In practice we
vary the
dimensionless parameter $\bar \mu$. Because of the underlying conformal symmetry
this
can then be thought of as either fixing the chemical potential $\mu$ and varying
the temperature $T$ or fixing the temperature
and varying the chemical potential. We define the temperature by $T/T_c =
\bar\mu_c/\bar{\mu}$ and
fix $\mu=1$.

Spontaneous symmetry breaking is driven by low temperature or high chemical
potential.
It triggers a non trivial expectation value for the scalar field without
switching on any
source $J_\mathcal{O}$. For small $\bar \mu$
the scalar field is trivial and the gauge equation is solved by
$\chi = \bar \mu (1-1/\rho)$ and $\psi=0$. The system is then in the symmetric
phase.
However, by decreasing the temperature the system becomes unstable
towards condensation of the scalar \cite{Gubser:2008px, Hartnoll:2008vx}.
In \cite{Amado:2009ts} it was shown that at the critical temperature indeed
the lowest quasinormal mode of the scalar field becomes unstable, i. e.
it crosses over to the upper half plane.

\begin{figure}[htp!]
 \centering
 \includegraphics[width=230pt]{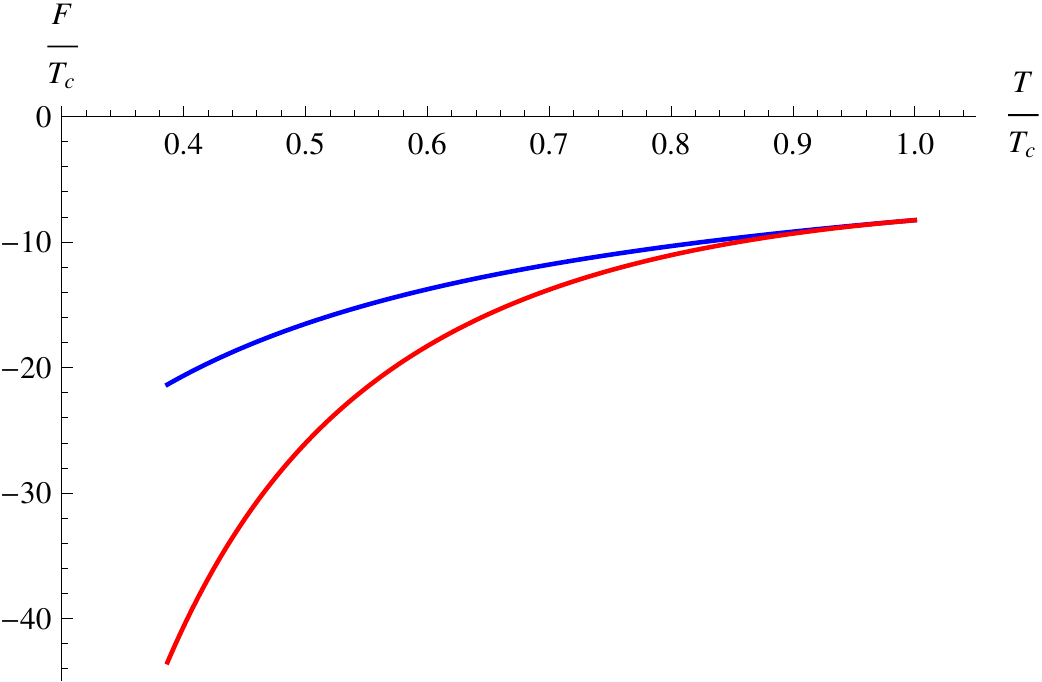}\hfill
 \includegraphics[width=230pt]{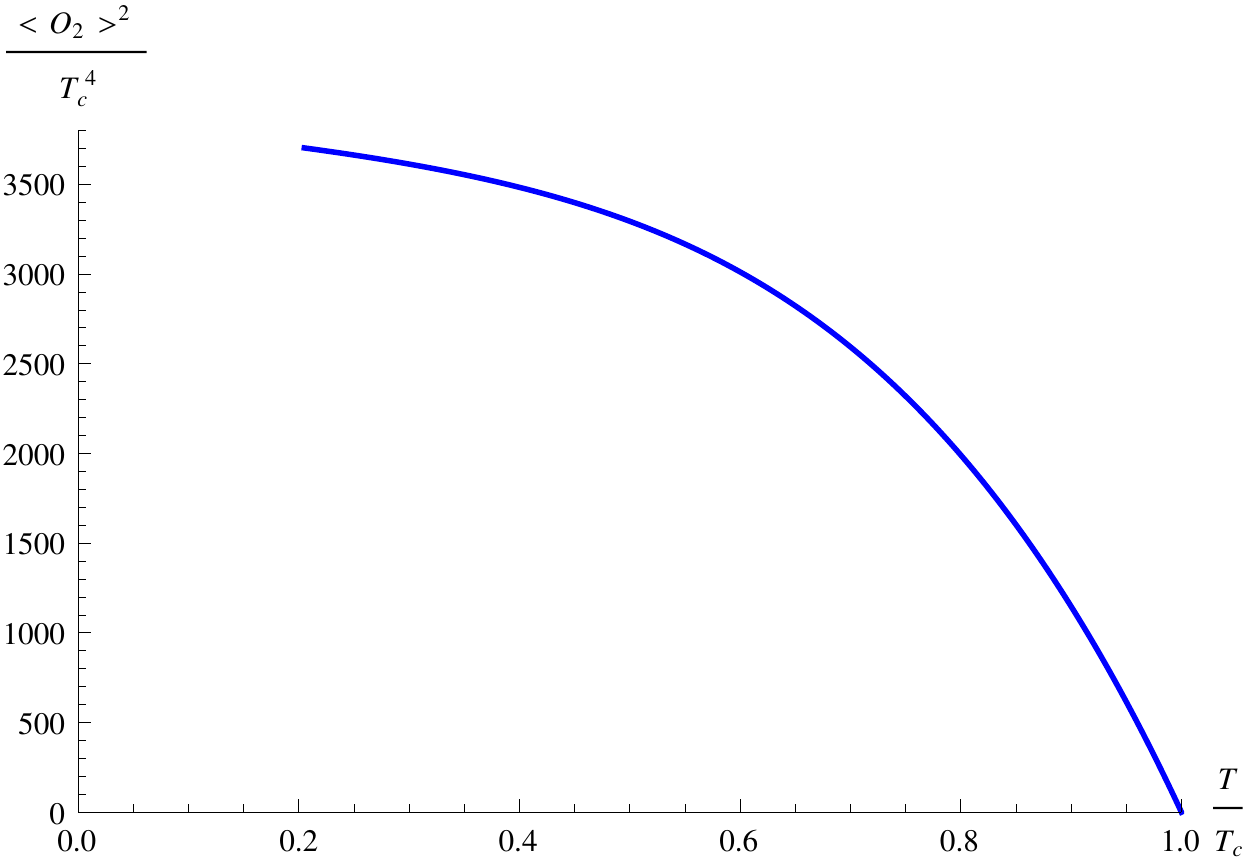}
 \caption{\label{fig:FE} (Left) The free energy of
the trivial (blue) and condensate (red) background solutions at low
temperatures, $T<T_c$.
 (Right) Value of the condensate in the grand canonical ensemble as a function
of
$T/T_c$.}
\end{figure}

The free energy density of the system is given by the on-shell renormalized
action,
\be
F=-T S_{ren} = - T \left(\frac{1}{2} \mu \, n-\int_{r_H}^\infty dr\, \frac{r^2
\psi^2\chi^2}{f}\right)\,.
\ee
The second term vanishes in the absence of a condensate and it
works against the phase transition if
it is present. In Figure \ref{fig:FE}
the free energies for the symmetric and broken phase are compared. It is clear
that for $T<T_c$ the condensate solution is
always preferred and therefore the system undergoes a second order phase
transition
to the superconducting phase.
Note that the presence of the second scalar plays no role for the phase
structure. It simply vanishes in the broken
and unbroken phase $\lambda=0$.

In order to extract the quasinormal mode spectrum, we switch on fluctuations of
the background fields
\begin{eqnarray}
\Psi^{\rm T} &=& \left(\eta(\rho,t,x) ,\, \psi(\rho) + \sigma(\rho,t,x)
\right)\,,\\
A &=&\left(\chi(\rho)+a_t(\rho,t,x)\right)dt + a_x(\rho,t,x)dx\,.
\end{eqnarray}
We do not include transverse fluctuations because they decouple from the
interesting physical features of the model at hand.

In the normal phase, i.e. expanding around $\psi(\rho)=0$, the system reduces to
the $U(1)$
holographic superconductor studied in \cite{Amado:2009ts} with two copies of the
scalar fluctuations,
\begin{align}
f s'' +s'\left(f'+\frac{2
f}{\rho}\right)+\left(\frac{(\chi+\omega)^2}{f} -\frac{k^2}{\rho^2}
-m^2\right)s=&0\,,\label{eq:flucunbs}\\
f a_t '' +\frac{2 f}{\rho} a_t ' -\frac{k^2}{\rho^2} a_t -\frac{\omega
k}{\rho^2} a_x=&0\,,\label{eq:flucubx}\\
f a_x '' + f' a_x ' +\frac{\omega^2}{f}a_x+\frac{\omega k}{f}
a_t=&0\,,\label{eq:flucunbt}\\
\frac{i \omega}{f} a_t ' +\frac{i k}{\rho^2} a_x ' =&0 \label{eq:flucunbcon}\,,
\end{align}
where $s$ stands for both $\eta$ and $\sigma$ fluctuations. The equation
for the complex conjugate scalar $\bar s$ can be obtained by changing the sign
of the potential $\chi$ in
\erf{eq:flucunbs}.
The frequency and momentum are related to the physical ones by
\begin{align}\label{eq:physical}
\omega  &= \frac{3}{4\pi T} \omega_{ph} \,,\\
k &= \frac{3}{4\pi T}  k_{ph} \,.
\end{align}

The scalar and gauge fluctuations completely decouple in the symmetric phase.
This is a consequence of working in the probe limit. The quasinormal mode
spectrum of the $U(1)$ field in the normal phase is just that of an
electromagnetic field on an AdS-Sch background.
The longitudinal fluctuations contain one hydrodynamic mode, $\omega=-i D k^2$,
reflecting the diffusive behavior of normal fluids. In physical units $D=3/(4\pi
T)$.
Due to the lack of an energy-momentum tensor for the dual field theory in the
probe limit, the diffusion pole is the only hydrodynamic mode in the unbroken
phase.

There are two copies of the scalar fluctuations. The quasinormal modes of $\eta$
and $\sigma$ move towards the origin when
decreasing the temperature, whereas the modes of $\bar \eta$ and $\bar \sigma$
have larger masses and widths the smaller the temperature.
As we approach the critical temperature $T=T_c$, the lowest quasinormal modes of
$\eta$
and of $\sigma$ become massless, triggering the phase transition:
the scalar field acquires a non trivial vev in order to avoid its fluctuations
to become tachyonic.
By symmetry we can choose the condensate to reside completely in the $\psi$
field.
The fluctuations $\sigma$ couple then to the gauge field fluctuations
just as in \cite{Amado:2009ts}.
Therefore the QNM spectrum in this sector contains a Goldstone mode with
linear dispersion relation $\omega = \pm v_s k +O(k^2) $.
This is the usual type I Goldstone boson
associated with the breaking of the gauge $U(1)$ symmetry. As shown in
\cite{Amado:2009ts} it can be interpreted as the sound mode of the dual
superfluid in the broken
phase. What happens then to the QNMs in the fluctuations of the second scalar
$\eta$?
At the critical temperature there is also an ungapped mode present since its QNM
spectrum is
simply another copy of the scalar sector.
Since there are no operators generating the $SU(2)$ symmetry in the dual field
theory
standard arguments about the appearance of Goldstone modes do a priori not
apply.
Three logical possibilities arise then: the mode could become unstable for
$T<T_c$, it
could become gapped again or it stays ungapped, playing the role of an
unexpected Goldstone
boson for the broken bulk-global $SU(2)$ symmetry.
Shortly we will see that the last possibility is realized and
that the massless mode of $\eta$ will indeed
correspond to a type II Goldstone boson with quadratic
dispersion relation, $\omega\propto k^2$.

In the broken phase, the equations of motion read
\begin{eqnarray}
\label{eq:eta}&&0= f \eta'' +\eta'\left(f'+\frac{2
f}{\rho}\right)+\left(\frac{(\chi+\omega)^2}{f} -\frac{k^2}{\rho^2}
-m^2\right)\eta\,,\\
&&0= f \delta'' +\delta' \left(f'+\frac{2 f}{\rho}\right)+\left(\frac{\chi^2}{f}
+\frac{\omega^2}{f}-\frac{k^2}{\rho^2} -m^2\right)\delta
-\frac{2 i \omega \chi}{f} \zeta- i\psi \left(\frac{\omega}{f}
a_t+\frac{k}{\rho^2} a_x\right)\,,\nonumber\\
\label{eq:delta}&&\\
&&0=f \zeta'' +\zeta' \left(f'+\frac{2 f}{r}\right)+\left(\frac{\chi^2}{f}
+\frac{\omega^2}{f}-\frac{k^2 L^2}{r^2} -m^2\right)\zeta+\frac{2 i \omega
\chi}{f} \delta+\frac{2 \chi\psi}{f} a_t\,,\\
 &&0=f a_t '' +\frac{2 f}{\rho} a_t ' -\left(\frac{k^2}{\rho^2}+2 \psi^2 \right)
a_t -\frac{\omega k}{\rho^2} a_x -2 i\omega \psi\delta-4\chi\psi \zeta\,,\\
 &&0=f a_x '' + f' a_x ' +\left(\frac{\omega^2}{f}-2 \psi\right)a_x+\frac{\omega
k}{f} a_t +2i k \psi \delta\,,\\
\label{eq:cons} &&0=\frac{i \omega}{f} a_t ' +\frac{i k}{\rho^2} a_x ' + 2 \psi'
\delta - 2 \psi\delta' \,,
\end{eqnarray}
where we have divided $\sigma=\zeta+i\delta$ into real and imaginary part. The
system
\erf{eq:delta}-\erf{eq:cons}
is again the one studied in \cite{Amado:2009ts}. This sector, that also appears
in the
gauged model that will be presented afterwards, decouples from the additional
scalar
fluctuation $\eta$. Notice that even if \erf{eq:eta} is formally the same as in
the normal
phase, the background $\chi$ is different leading to non trivial features in the
$\eta$
sector such
as the presence of a massless excitation.

\begin{figure}[htp!]
\centering
\includegraphics[width=230pt]{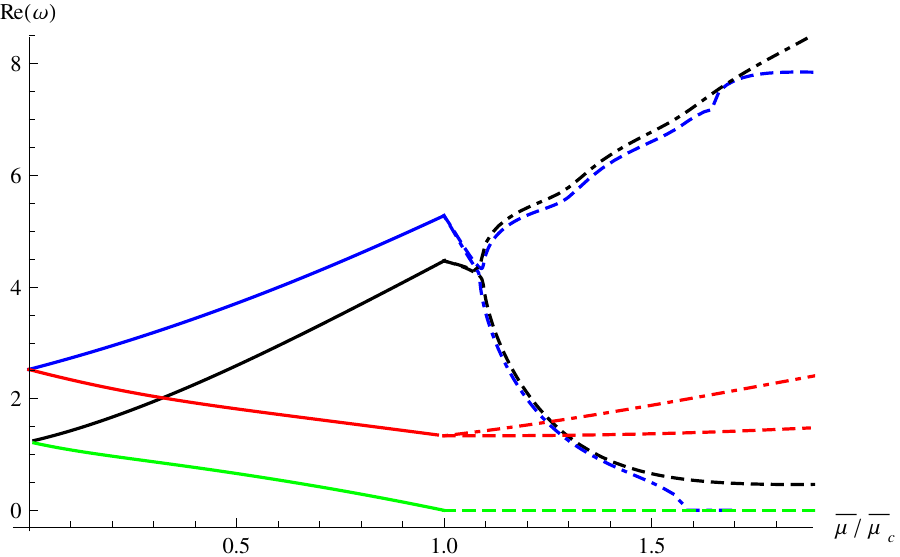}\hfill
\includegraphics[width=230pt]{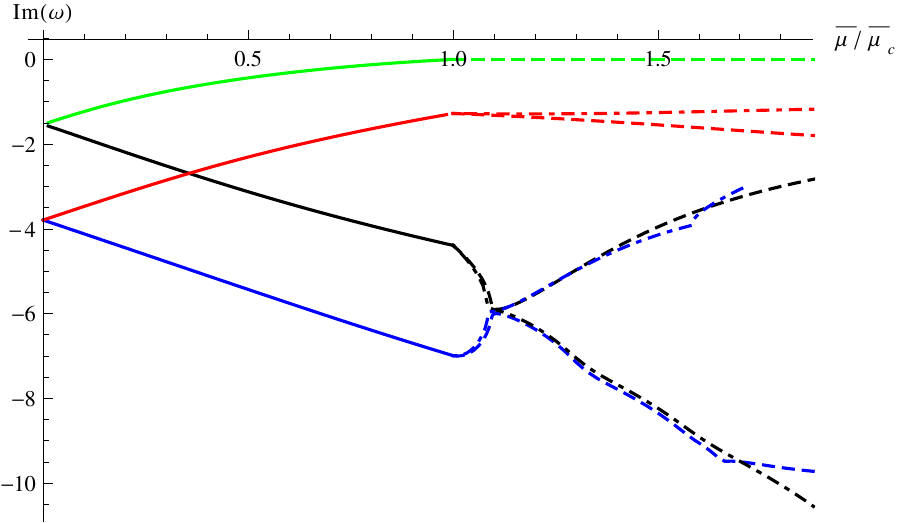}
\caption{\label{fig:bivio} Real (left) and imaginary (right) parts of the lowest
scalar
QNMs as a function of the chemical potential. Solid lines correspond to the
unbroken phase.
For the broken phase dashed lines stand for modes of the additional scalar while
dotdashed lines represent the modes common
to the $U(1)$ holographic superconductor.}
\end{figure}

Figure \ref{fig:bivio} shows the spectrum of quasinormal excitations of
the scalar
doublet. In the normal phase we have two degenerate copies of the spectrum that
partially split after the phase transition.
It is clear that the two lowest excitations become massless at the critical
chemical potential
and then remain massless in the superconducting phase. They can be identified
with the two Goldstone
bosons at the phase transition.
The rest of the excitations remain gapped in the broken phase.
Notice that the first $\bar \eta$ excitation (dashed black line in figure
\ref{fig:bivio})
does not follow the expected universal behavior in the broken
phase, i.e. it is not linear in $\mu$. This mode is the equivalent of the
special gapped mode
$\omega_4$ in the field theoretical model of section \ref{sec:fieldth}. However,
it has already been
mentioned that the {\it ungauged model} does not satisfy all the theorems about
symmetry breaking and
therefore deviations from the universal behavior should not be surprising. The
behavior of the gapped modes is actually similar to that of the $U(1)$ model
modes. In the unbroken phase we can distinguish the modes that come from the
$s$-type of fluctuations from the ones that come from the complex conjugate
$\bar s$
fluctuations. The former become lighter whereas the latter become
heavier\footnote{This
behavior is reversed if we had taken the chemical potential to be negative.}.
In the broken phase it is more useful to use real and imaginary parts, at least
for the scalar that mixes with the gauge fields fluctuations, i.e. the lower
component
of the scalar in our conventions. So we can not a priori talk of $s$ and $\bar
s$ type
fluctuations. We still can study to which modes the $s$ and $\bar s$ type modes
connect
to in the broken phase. Here we see an interesting pattern: the $s$ type modes
split in the broken phase whereas the $\bar s$ type modes stay almost degenerate
close to the phase transition (at least at
zero momentum). This is surprising given the fact that the fluctuations
correspond
to two completely different systems, one coming from a single differential
equation whereas
the others come from a complicated system of coupled equations. However, for
small temperatures they split and actually the real part
of the lowest one for the $U(1)$ sector goes to zero at a finite temperature.
For temperatures below $T\approx 0.63\, T_c$ it becomes a purely imaginary mode.

\paragraph{Sound mode:}

There are two massless modes in the broken phase. The first one is the type I
Goldstone boson appearing because of the spontaneous breaking of
the $U(1)$ gauge symmetry. In \cite{Amado:2009ts}, it was shown that this mode
corresponds to the sound mode of the
dual superfluid and that in the hydrodynamic limit it has a linear dispersion
relation
\be\label{eq:soundmode}
\omega_I = \pm \left(v_s\, k\, +\,\bar b\,k^2\right) - i \Gamma_s \, k^2 \,,
\ee
where $v_s$ is the speed of sound and $\Gamma_s$ is its attenuation.
It turns out that the quadratic part of the dispersion relation also has a real
component. This component is very small and subleading compared to the linear
term
that determines the speed of sound. In \cite{Amado:2009ts} this real quadratic
part
has not been studied.

\begin{figure}[htp!]
\centering
\includegraphics[width=230pt]{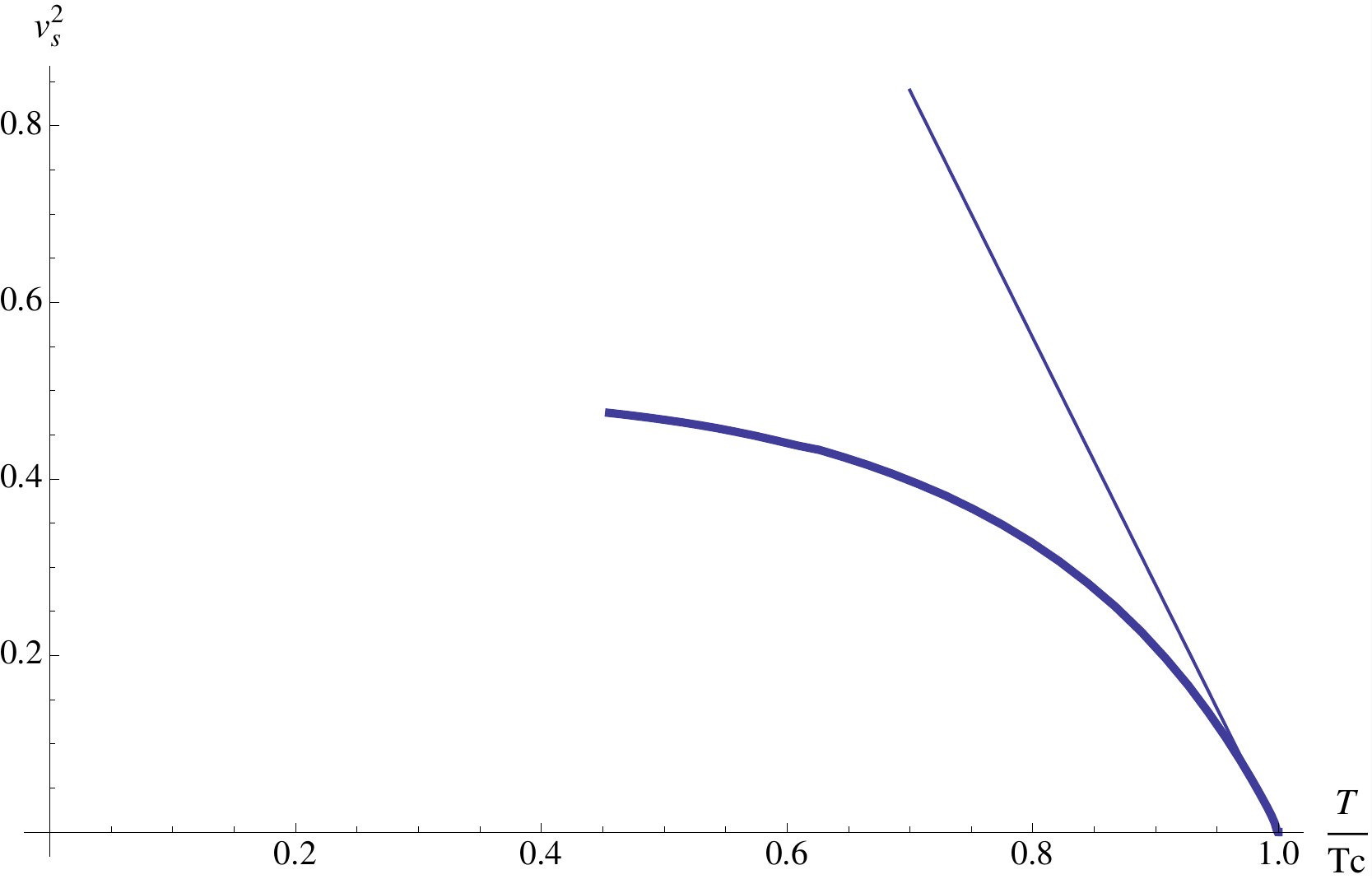}\hfill
\includegraphics[width=230pt]{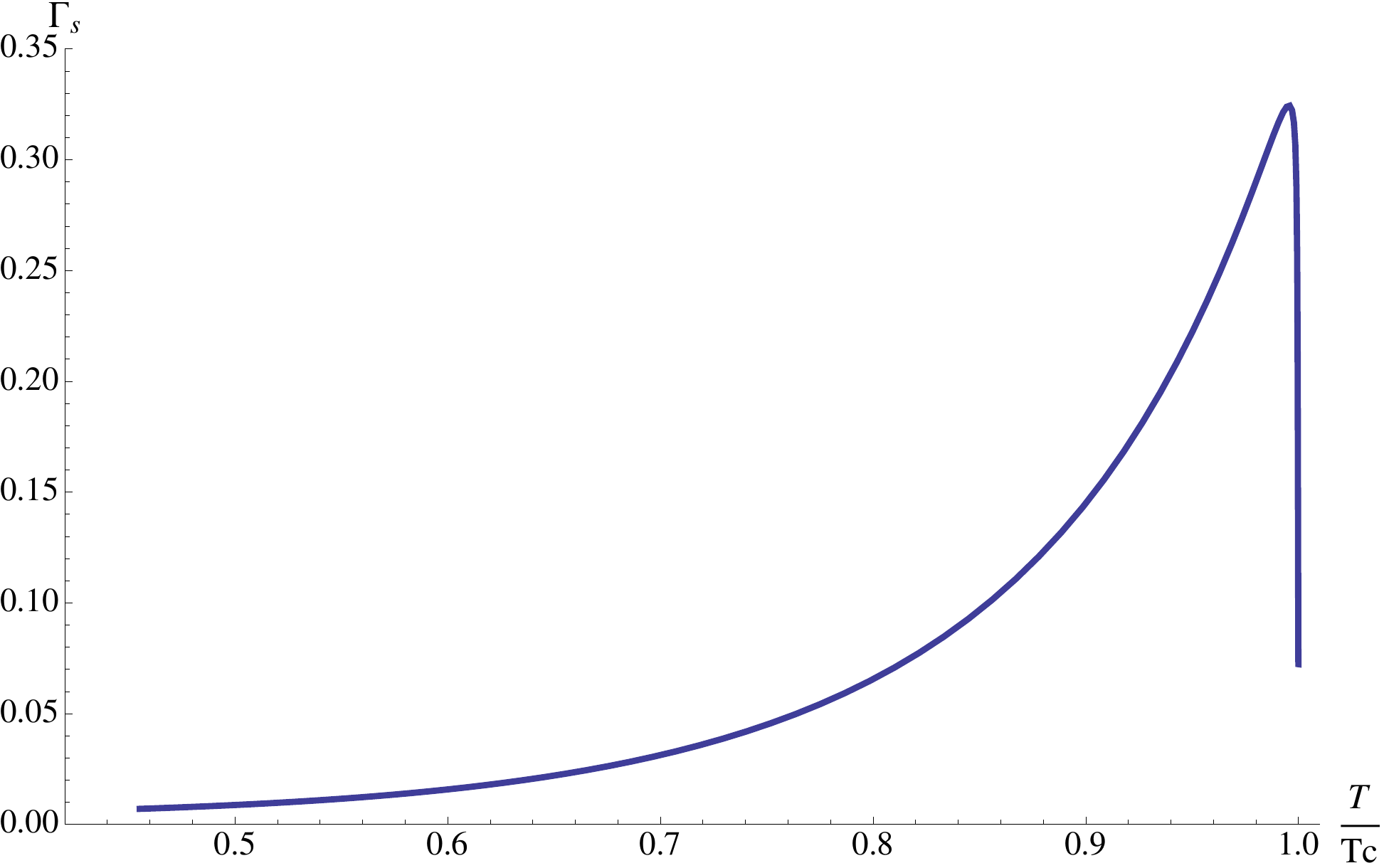}
\caption{\label{fig:vsandGammas}
Speed of sound and damping for the sound mode. The speed of sound goes to zero
at the critical temperature.
The damping constant first rises quickly and then falls off again. Precisely at
the critical temperature its
value is such that the sound modes connect continuously to the scalar modes that
become massless there. The peak
in the damping constant sits close to the critical temperature and was not
resolved in \cite{Amado:2009ts}.}
\end{figure}

For very small
temperatures the velocity approaches its conformal value $v_s^2=1/2$ while the
width goes to zero, see figure \ref{fig:vsandGammas}.
Close to the phase transition, the speed of sound has a mean field behavior as a
function of temperature
\be
v_s^2 \approx 2.8\,\left(1-\frac{T}{T_c}\right)\,.
\ee

As expected, at $T=T_c$ the speed of sound vanishes. This can be traced
back to the fact that at the phase transition the total mass $m_{*}^{2}= M^2
-\mu^2$ fulfills $m_{*}^{2} = v^2= 0$, as expected, and hence the complex scalar
field, charged under a $U(1)$ symmetry, becomes massless.

\begin{figure}[htp!]
\centering
\includegraphics[width=230pt]{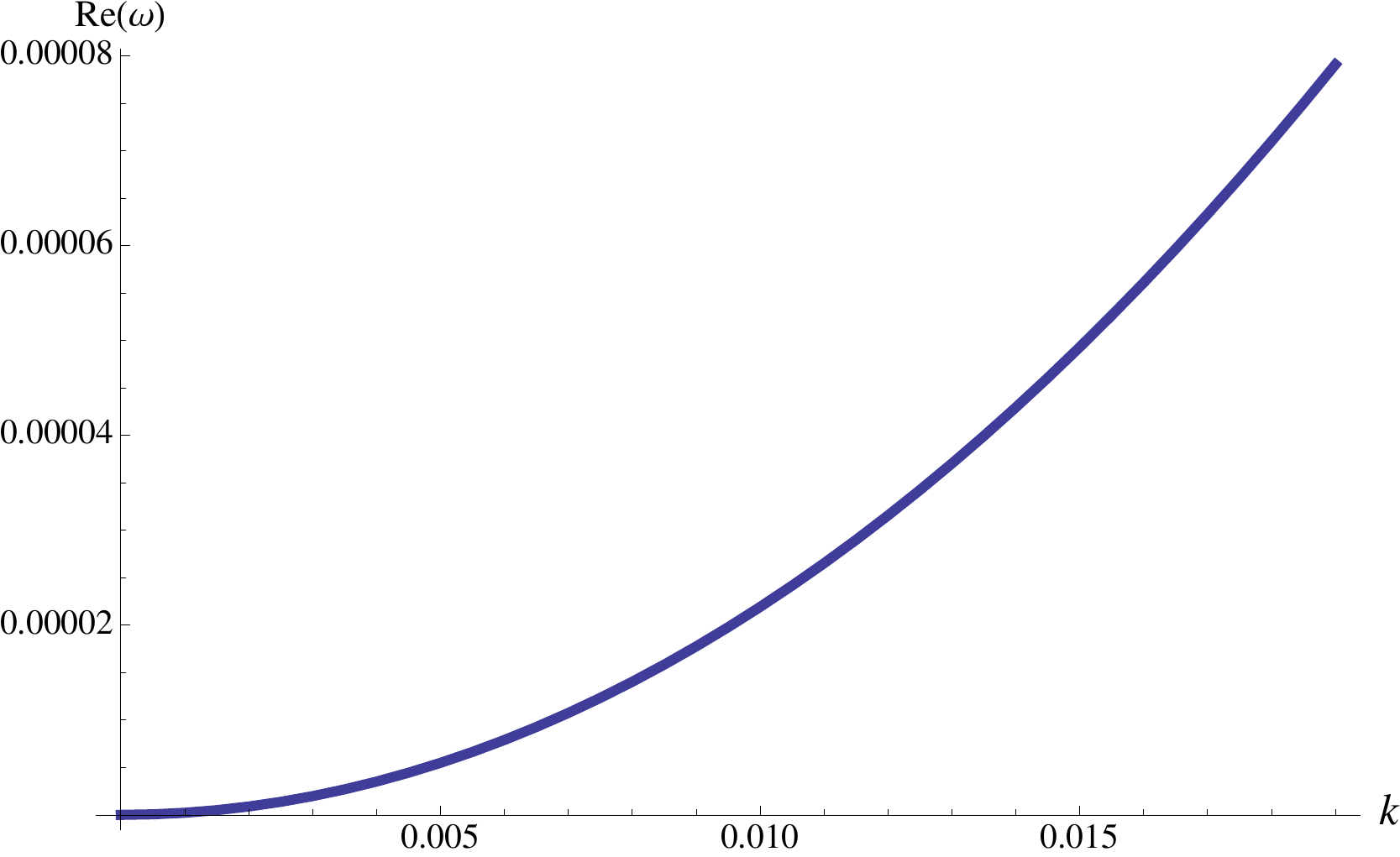}\hfill
\includegraphics[width=230pt]{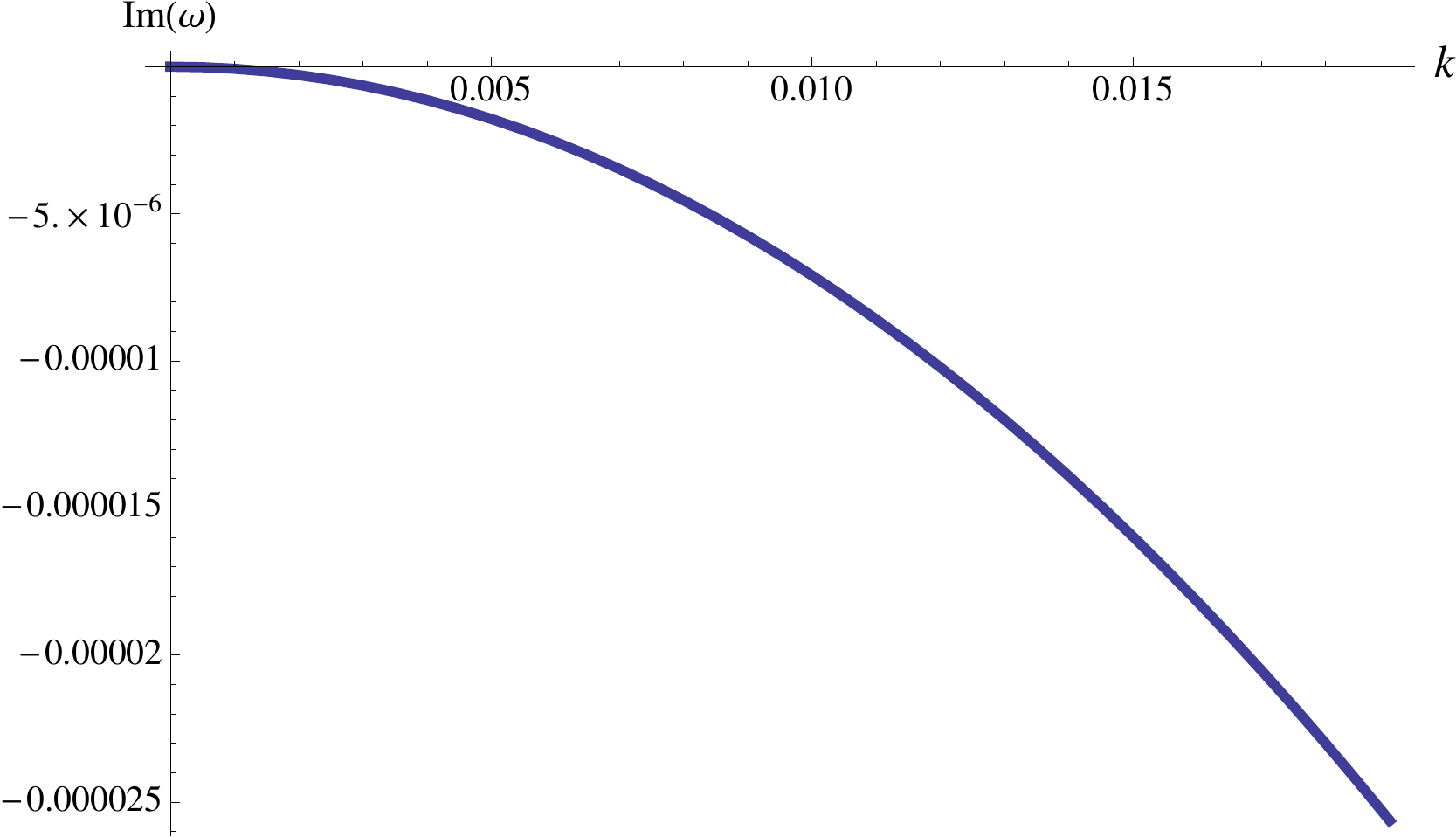}
\caption{\label{fig:Goldstone03quadratic} Dispersion relations of $\Re{\w}$
(left) and $\Im{\w}$ (right) at $T=T_c$ for the type I Goldstone boson in the
system studied by \cite{Amado:2009ts}. The behavior $\Re{\w} \sim k$ becomes
quadratic right at this temperature: $\Re{\w}=\bar b k^2$. The coefficient is
$\bar b =0.22$, which in turn matches the value that one finds if
approaches $T_c$ from above (i.e. from the unbroken phase).}
\end{figure}

Indeed, one can write down the effective action, analogous to (\ref{Lagr}), for
a complex scalar field with mass $M$, in the presence of a chemical potential
for a $U(1)$ symmetry that is spontaneously broken. The excitations on top of
the
$U(1)$-breaking background have a dispersion relation equal to
(\ref{eq:typeI})-(\ref{eq:gapped}), being (\ref{eq:typeI}) the type I Goldstone
boson.
It is a general feature of these linear sigma models that the coefficient in
front of the linear term in the momentum depends on $m_{*}^{2}$, as can be
explicitly checked for the case at hand (see (\ref{eq:typeI})). Therefore, at
the phase transition the leading term in the dispersion relation is of
$O(k^2)$; this effect can be seen very clearly with numerical methods,
as shown in Figure \ref{fig:Goldstone03quadratic}.
Since the quasinormal mode spectrum has to vary continuously through the second
order phase transition the real and complex coefficients of the $k^2$ term
have to coincide
at $T=T_c$ with the ones obtained from the massless scalars in the unbroken
phase.
Numerically we find $\bar b(T_c)= 0.22$ and $\Gamma_s(T_c) = 0.071$.

\paragraph{Pseudo diffusion mode:}

\begin{figure}[htp!]
\centering
\includegraphics[width=230pt]{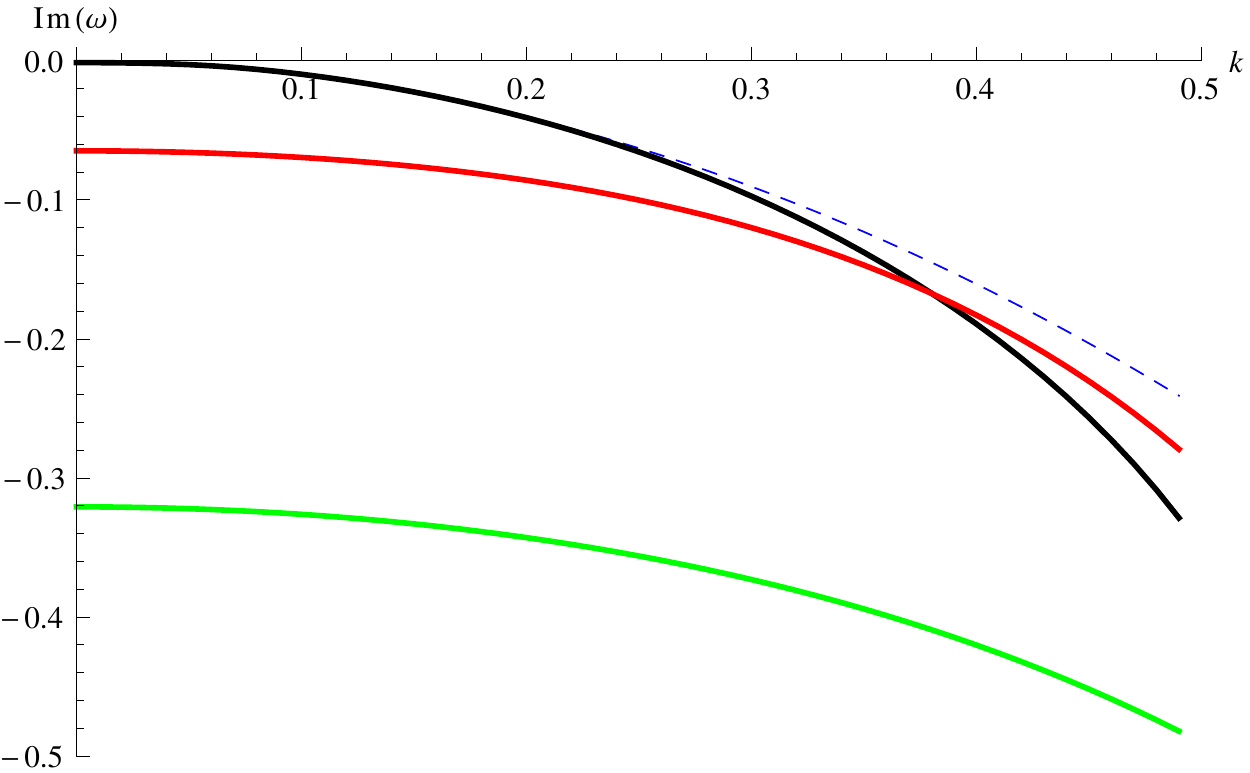}
\hfill
\includegraphics[width=230pt]{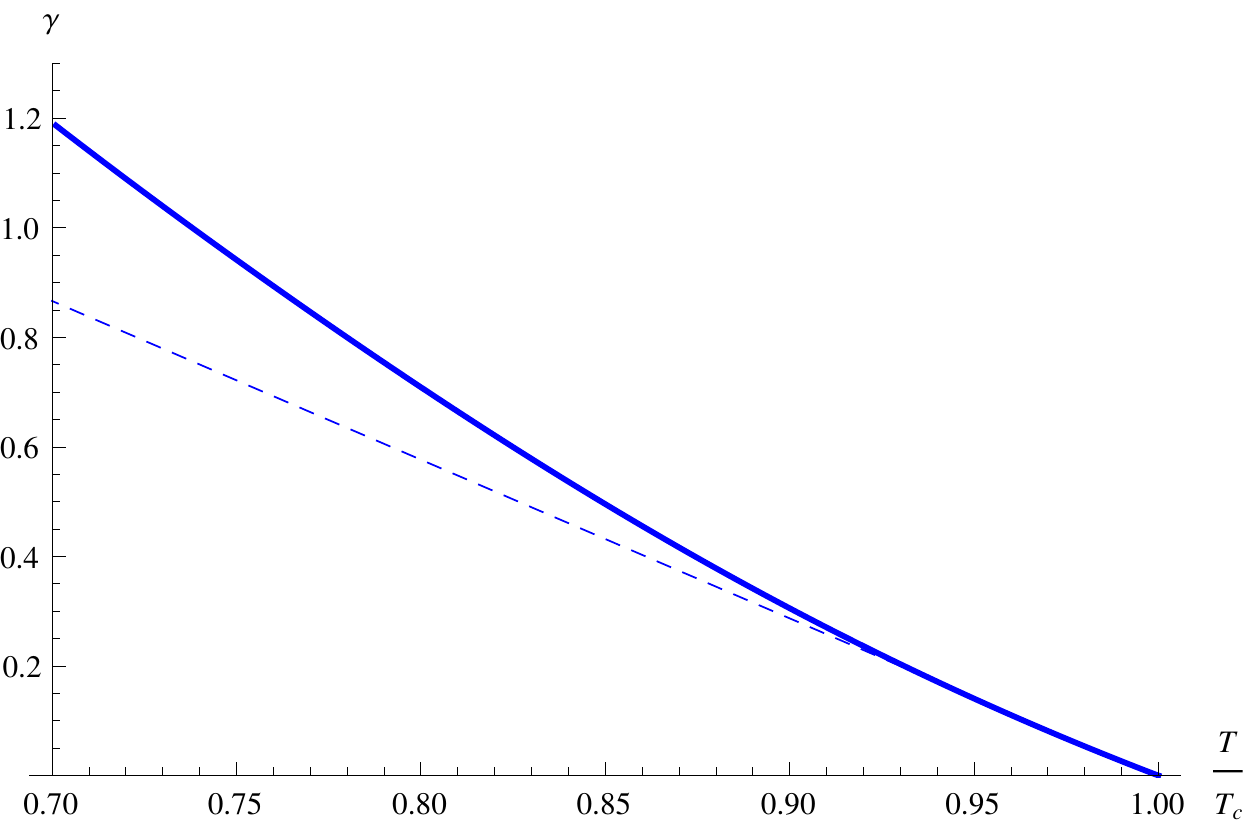}
\caption{\label{fig:PseudoDiff} (Left) Dispersion relation of the gapped
pseudo diffusion mode in the broken phase for three different temperatures. The
gap widens as the temperature
is lowered. (Right) Gap $\gamma$ as a function of the reduced temperature
$T/T_c$.
As one approaches the critical temperature from below the gap vanishes
linearly.}
\end{figure}

In the unbroken phase our model has only one hydrodynamic mode, the diffusion
mode $\omega = -i D k^2 + O(k^4)$
with $D=3/(4\pi T)$ in physical units. The shear and normal sound modes have
their origin in the metric fluctuations and
therefore are absent in the decoupling limit we are studying. The phase
transition to the broken phase is
second order. For the spectrum of quasinormal modes this implies that the modes
of the broken and unbroken
phase must connect continuously through the phase transition. In the case of the
diffusion mode there must
therefore exist a quasinormal mode with purely imaginary frequency.
Hydrodynamics implies however that the
only ungapped modes are the sound modes corresponding to the type I Goldstone
mode. Not too far from the phase
transition, i.e. for $T\lesssim T_c$ the diffusion mode of the broken phase must
develop into a mode with
dispersion relation
\begin{equation}
 \omega = -i \gamma(T) - i D(T) k^2 + O(k^4)\,,
\end{equation}
as shown in Figure \ref{fig:PseudoDiff}.

We might say that the diffusion mode develops a gap in the broken phase and
becomes what has been called
a {\em pseudo diffusion mode} in \cite{Amado:2009ts}.
Precisely at zero momentum $k=0$ this gapped pseudo diffusion mode explains the
observation made in \cite{Bhaseen:2012gg}
on the late time response of holographic superconductors. For temperatures
$T\lesssim T_c$ the pseudo diffusion mode
is the mode that lies closest to the real axis and therefore it dominates the
long time response to any
perturbation, such as the quenches studied in \cite{Bhaseen:2012gg}. It follows
that the order parameter shows a purely
exponential decay since this mode does not have a real frequency. The existence
of that mode can ultimately
be traced back to the universality of the diffusion mode in the unbroken phase.
We expect therefore the pseudo diffusion
mode to be a universal feature of a wide class of superfluids (not necessarily
holographic ones).

\begin{figure}[htp!]
\centering
\includegraphics[width=230pt]{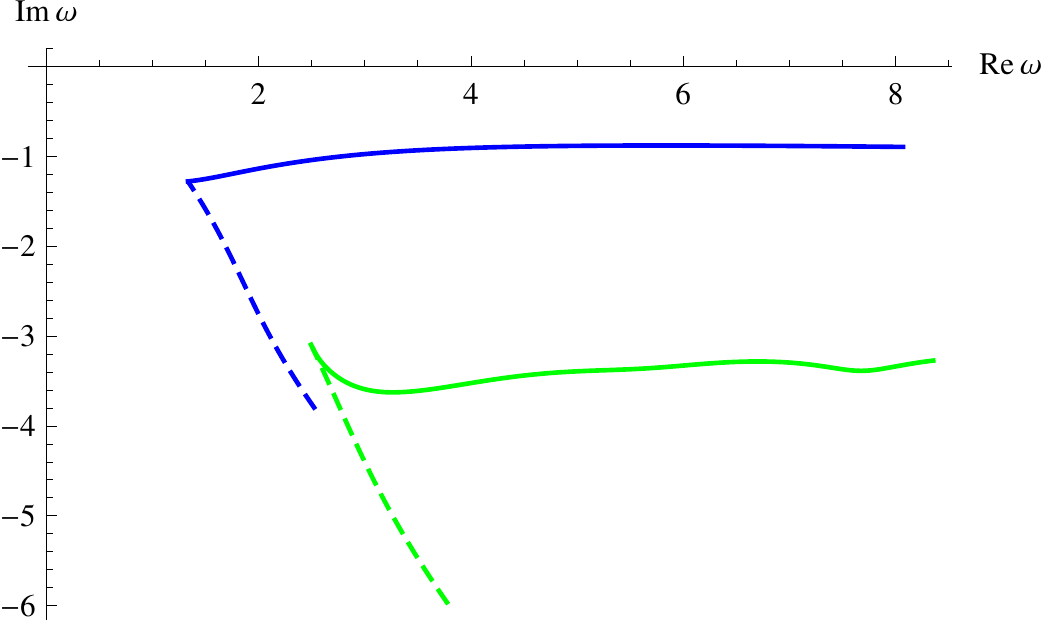}\hfill
\includegraphics[width=230pt]{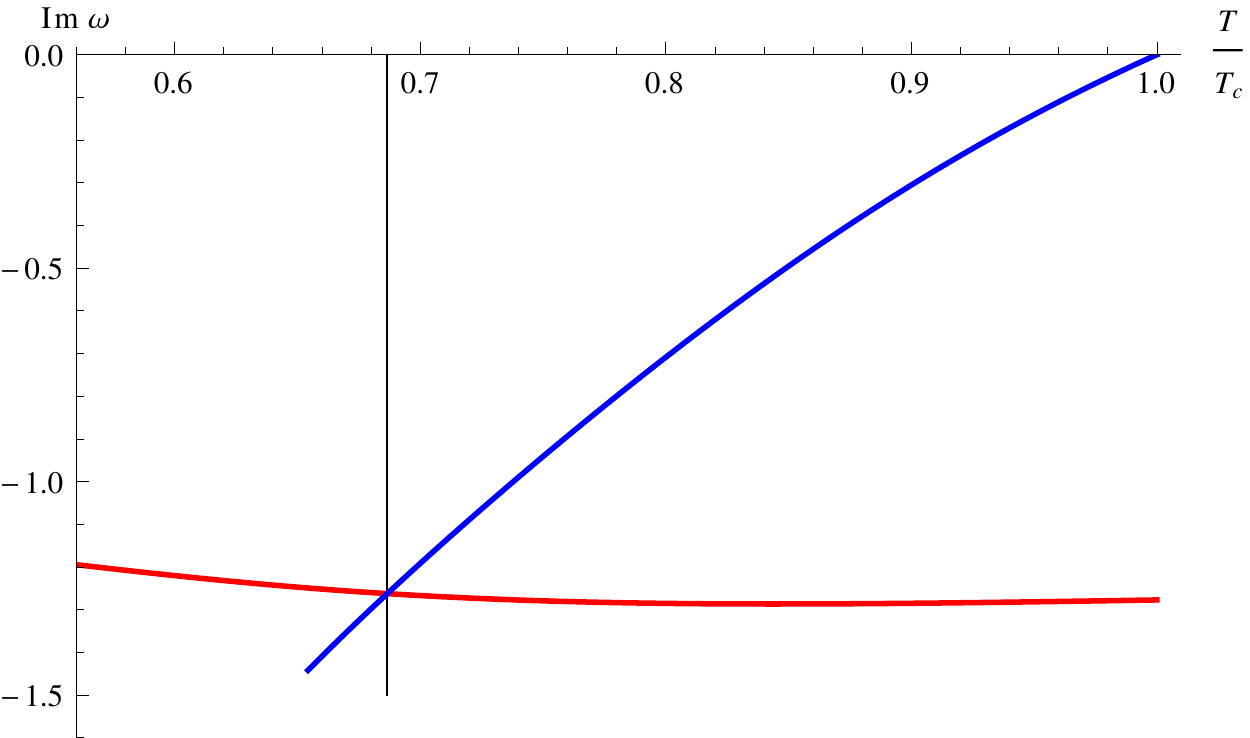}
\caption{\label{fig:Tstar} (Left) Continuation of the second and third scalar
QNM into the
broken phase. The real part grows as the temperature is lowered whereas the
imaginary part shows
very little dependence on $T$. (Right) The gap $\gamma$ (blue line) and the
imaginary part
of the lowest (scalar) mode fluctuation (red line) in the broken
phase are shown as function of $T/T_c$. At $T_*\approx 0.69 \,T_c$ the imaginary
parts cross. For lower temperatures the late time response is not dominated
anymore by the pseudo diffusion mode and consequently is in form of a
exponentially decaying oscillation.
}
\end{figure}

The gap $\gamma$ grows as the temperature decreases.
On the other hand there are quasinormal modes (connecting to the QNMs in the
scalar sector of the unbroken phase) whose
imaginary part is only weakly dependent on the temperature. At a certain
crossover temperature $T_*$ the gap of
the pseudo diffusion mode is bigger than the imaginary part of these modes, as
shown in Figure \ref{fig:Tstar}. Then the response pattern changes from
a purely exponential decay to an exponentially damped oscillation. Numerically
we find that the crossover temperature is
$T_*=0.69\,T_c$.\footnote{This is lower than in the model of
\cite{Bhaseen:2012gg}. The difference is
presumably due to the fact that we work in the decoupling limit.}
Such crossover changes in the long term response
appear frequently in the details of the quasinormal mode spectrum of holographic
field theories, \cite{Amado:2008ji,Kaminski:2009dh, Davison:2011ek}.
In fact this purely exponential decay applies not only to the order parameter
but to all operators that correspond
to the fields participating in the fluctuation system
\erf{eq:delta}-\erf{eq:cons}, e.g. charge density or $x$-component
of the current.

For finite momentum the response pattern is expected to be different however.
Now one also has to take into account
the sound mode. While precisely at zero momentum the sound mode, i.e. the
Goldstone mode, degenerates to
a constant phase change of the condensate at small but non-zero momentum the
long time response should be dominated
by the complex frequencies \erf{eq:soundmode}. If one looks however only to the
response in the gauge invariant
order parameter $|\cO|$ the Goldstone modes, being local phase rotations of the
order parameter, are projected out.

\begin{figure}[htp!]
\centering
\includegraphics[width=230pt]{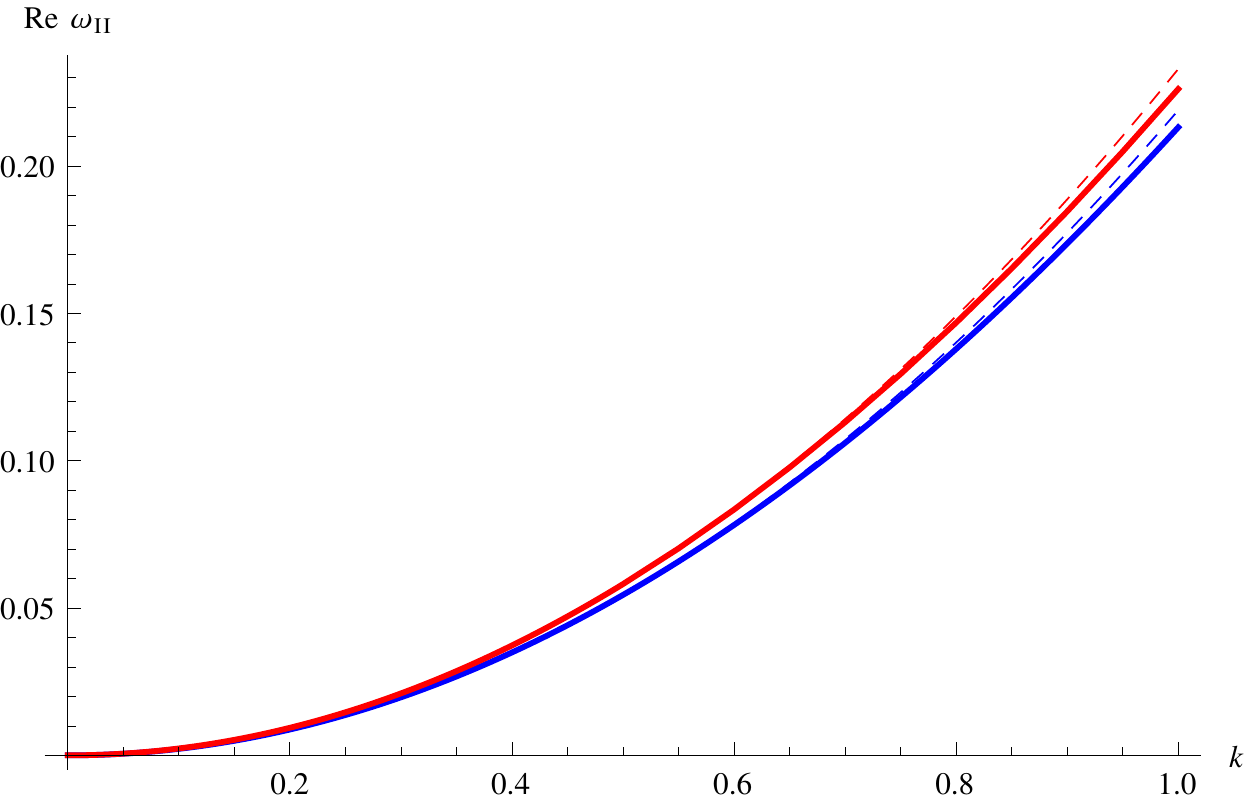}\hfill
\includegraphics[width=230pt]{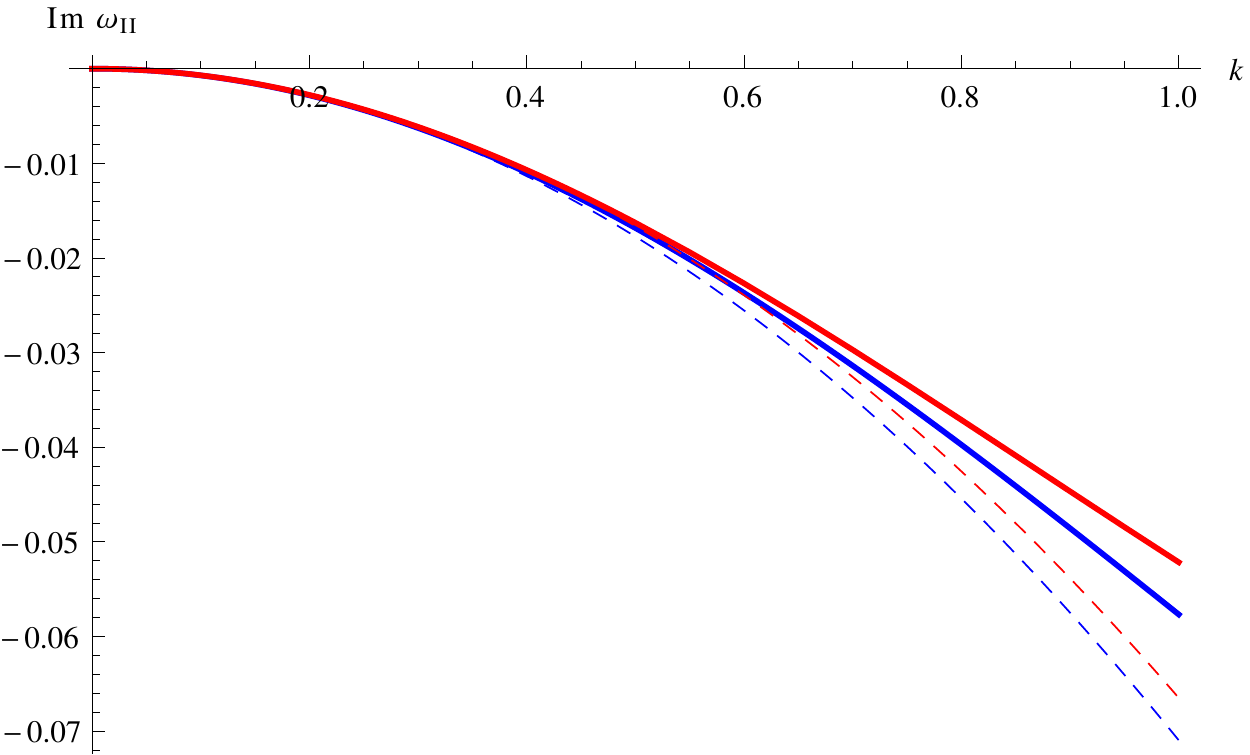}
\caption{\label{fig:IIGoldbivio} Real (left) and imaginary (right) parts of the
type II Goldstone mode as
a function of the momentum for $T/T_c = 0.9998$ (blue) and  $T/T_c = 0.704$
(red) . The solid lines correspond to
the numerical result while
the dashed lines are the quadratic fit to a dispersion relation $\omega_{II} =
b\,k^2 - i c\,k^2$.}
\end{figure}

\paragraph{Type II Goldstone mode:}
The second massless mode is the Goldstone boson associated with the breaking of
the bulk-global
$SU(2)$ symmetry.
It can
be fit to a quadratic dispersion relation of the form
\be
\label{eq:gII}\omega_{II} = \pm b\,k^2 - i c\,k^2\,+ O(k^4)\,,
\ee
in the long wavelength limit. Therefore it has the characteristic of a type II
Goldstone mode.
In Figure
\ref{fig:IIGoldbivio} the dispersion relation for the $\eta$
massless mode is shown for various temperatures as well as its fit to the
hydrodynamic form. It is
clear that there is a good agreement in the regime of validity of the low energy
limit.

\begin{figure}[htp!]
\centering
\includegraphics[width=230pt]{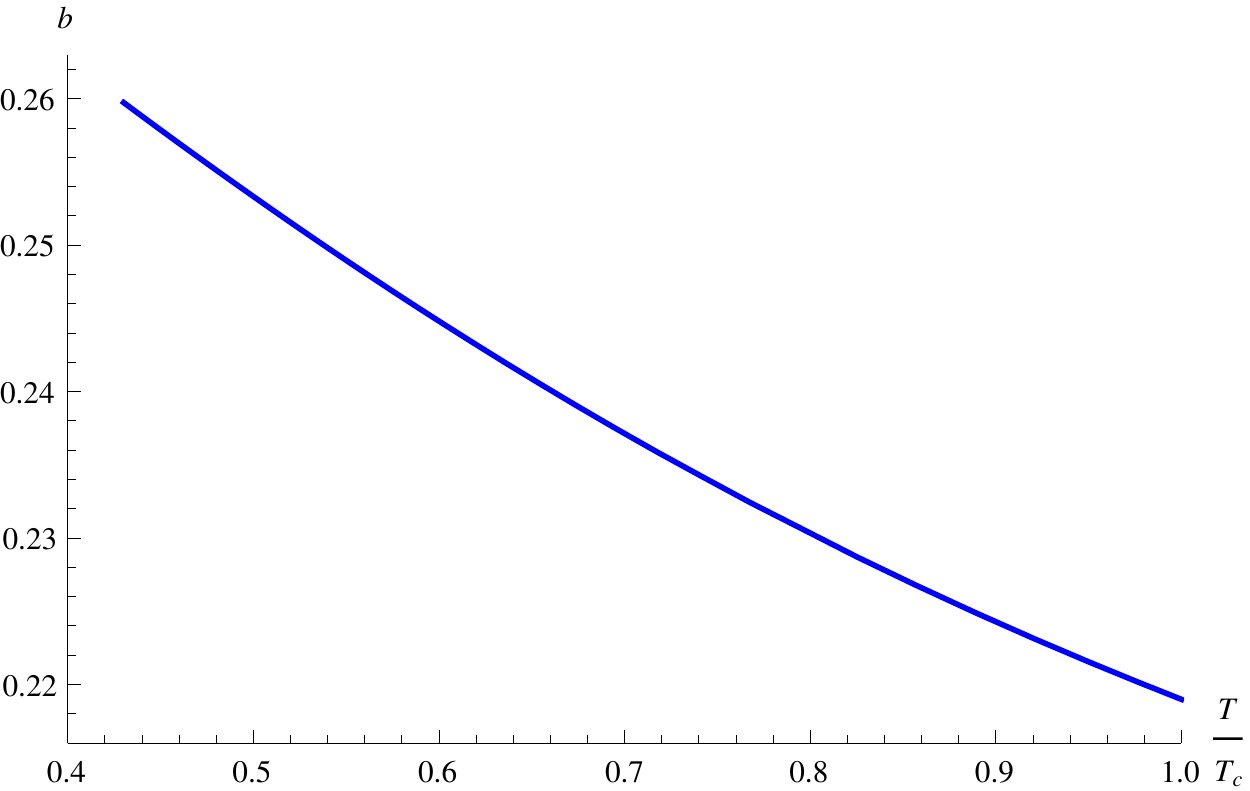}\hfill
\hfill
\includegraphics[width=230pt]{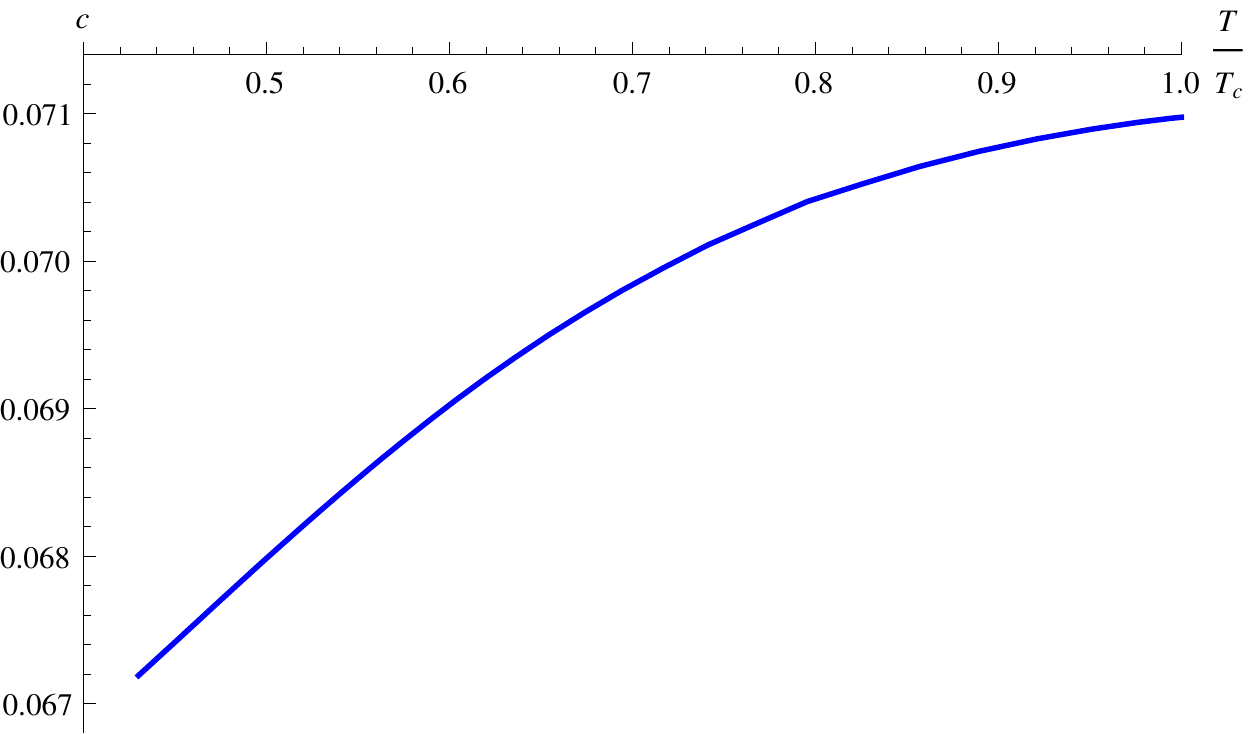}
\caption{\label{fig:t2gbbivio}Coefficients of the type II Goldstone mode
dispersion relation $\omega_{II} =
b\,k^2 - i c\,k^2$, as a function of the temperature. Dependence with
temperature is very mild.}
\end{figure}

The coefficients in the hydrodynamic dispersion relation \erf{eq:gII} as a
function of the temperature are shown in Figure \ref{fig:t2gbbivio}.
Close to the phase transition they have a linear dependence in the reduced
temperature
\bea
\label{bcIIung}b(T)&=& 0.22 + 0.049  \left(1-\frac{T}{T_c}\right)\,,\\
c(T)&=& 0.071 - 0.0014 \left(1-\frac{T}{T_c}\right)\quad{\rm
near}\,\,T_c\,.\nonumber
\eea
Notice that at the phase transition the sound mode and the type II
Goldstone must behave
in the same way due to continuity of the modes through the phase transition and
the fact that they are degenerate in the normal phase.
In fact, at the transition $b=\bar b = 0.22$ and $c=\Gamma_s=0.071$, values that
of course coincide with those of the lowest scalar mode in
the normal phase. On the other hand, it is interesting to notice that in the
broken phase the behavior of the coefficients of the type II
Goldstone is completely different from that of the
coefficients of the sound of the superfluid. Unlike the sound
velocity, that vanishes at the phase transition,
the coefficient $b$ of the type II Goldstone mode takes a finite value at the
critical temperature. This result of course persists for the {\it gauged model}.
The attenuation on the other hand, as it happens for the $U(1)$ sector, has a
finite value at the phase transition and then decreases with temperature,
reflecting the fact that the fluid is more ideal the lower the temperature.

\section{The gauged model}\label{sec:gauged}

Let us now discuss the fully gauged model. Consider the following Lagrangian for
a complex scalar field living in the
fundamental representation of a $U(2)$ gauge symmetry of the bulk,
\beq
S=\int \sqrt{-g}\mathcal{L}=\int d^4 x \sqrt{-g}\left(-\frac{1}{4}F^{\m\n
c}F_{\m\n}^c-m^2\Y^\dagger\Y-(D^\m\Y)^\dagger D_\m\Y\right)\,,
\enq
where
\beq
\Y=
\begin{pmatrix}
\l\\
\y
\end{pmatrix}\,,  \hspace{2cm}A_\m=A_{\m}^{c}T_c\,,
\hspace{2cm}
D_\m=\partial_\m-iA_\m\,,
\enq
and $c=0,1,2,3$ is the color index.
The field $\Y$ plays the role of the condensate. The expectation value of its
dual
operator thus triggers the spontaneous breaking of the $U(2)$
global symmetry of the boundary theory. For simplicity, we set $\l=0$ in the
background. $T_c$ are the generators of $U(2)$:
\begin{align}
T_0=\frac{1}{2}\mathbb{I}\,,\hspace{1cm}T_i=\frac{1}{2}\s_i\,,\nonumber\\
\lbrace T_i,T_j \rbrace =\frac{1}{2}\delta_{ij}\mathbb{I}\,,\hspace{1cm} \lbrace
T_0,T_i\rbrace=\frac{1}{2}\s_i\,.
\end{align}

Notice that we are again working in the probe limit, so the background metric is
taken to be
the Schwarzschild-AdS black brane of \erf{eq:metric}. On the other hand, the
gauge field is now
\be
A^{(0)}_0\equiv \Phi(r)\,,\quad   A^{(3)}_0 \equiv \Theta(r) \,.
\ee
The rest of the components of the gauge field being zero.
As in the previous section, we will
use dimensionless coordinates defined by the rescaling given in
\erf{eq:rescale}.

The equations of motion for our ansatz are
\bea
\label{scalar}
\y''+\left(\frac{f'}{f}+\frac{2}{\r}\right)\y'+\frac{(\F-\Th)^2}{4f^2}\y-\frac{
m^2}{f}\y=0\,,\\
\label{gauge1} \F''+\frac{2}{\r}\F'-\frac{\y^2}{2f}(\F-\Th)=0\,,\\
\label{gauge2} \Th''+\frac{2}{\r}\Th'+\frac{\y^2}{2f}(\F-\Th)=0\,.
\eea
Notice that from \erf{gauge2} it follows that we can not simply switch on
 $\Phi$ without also allowing for a non-trivial $\Theta$.
We are of course only interested in switching on a chemical potential in the
overall
$U(1)$, and therefore we will impose $\Th(\r\rightarrow \infty)=0$ and allow for
a finite boundary value of $\Phi$.

The coupled system of equations above can be simplified by defining $\c \equiv
\frac{1}{2}\left(\F-\Th\right)$ and $\xi\equiv \frac{1}{2}\left( \F+\Th\right)$.
Using (\ref{gauge1}) and (\ref{gauge2}), we see that the resulting equations for
these fields are\footnote{These equations of motion correspond
to the probe limit of the system studied in \cite{francescos} as a dual of
superconductors with chemical potential imbalance.
Notice however that in \cite{francescos} the gauge symmetry was $U(1)\times
U(1)$ instead of $U(2)$ as in the present setup.}
\bea
\label{scalar2}
\Y''+\left(\frac{f'}{f}+\frac{2}{\r}\right)\Y'+\frac{\c^2}{f^2}\Y-\frac{m^2}{f}
\Y=0\,,\\
\label{gauge3} \c''+\frac{2}{\r}\c'-\frac{2 \Y^2}{f}\c=0\,,\\
\label{gauge4} \xi''+\frac{2}{\r}\xi'=0\,,
\eea
where we have redefined $\y \rightarrow \sqrt{2} \Y$. As usual we choose the
boundary conditions
$\c(\r= 1)= 0$, $\xi(\r=1)=0$ along with regularity of $\Y$. Having a dual field
theory with only
one finite chemical potential switched on, implies
that $\c$ and $\xi$ must take the same non trivial value at the boundary in
order to ensure that $\Theta$ vanishes asymptotically. Notice that $\xi$
decouples completely. The remaining system (\ref{scalar2})-(\ref{gauge3}) is
again
the background found for the widely studied s-wave $U(1)$ holographic
superconductor. Therefore, the background of
the $U(2)$ gauge model contains the Abelian superconductor plus a decoupled
conserved $U(1)$ sector.

The field $\c$ lies in the direction of one of the broken generators, which is
the
linear combination $\frac{1}{2}(T_3-T_0)$, whereas $\xi$ lies
in the direction of the preserved $U(1)$ given by $\frac{1}{2}(T_3+T_0)$.

The asymptotic expansion of the fields near the conformal boundary reads
\begin{eqnarray}
\label{nearbexpchi} \chi= \bar \mu_{\chi}-\frac{\bar n_{\chi}}{\r}+
O\left(\frac{1}{\r^2}\right)\,,\\
\label{nearbexpsig} \xi= \bar \mu_{\xi}-\frac{\bar n_{\xi}}{\r}+
O\left(\frac{1}{\r^2}\right)\,,\\
\label{nearbexpY} \Y = \frac{\psi_1}{\r}+\frac{\psi_2}{\r^2}+
O\left(\frac{1}{\r^3}\right)\,.
\end{eqnarray}
The map of the various coefficients in the previous equations to the boundary
conditions is $\bar \mu_{\chi}=\bar \mu_{\xi}=\bar \mu$. We will again focus in
the
$\mathcal{O}_2$ theory exclusively, henceforth we will demand $\psi_1 = 0$.

Equations (\ref{scalar2})-(\ref{gauge3}) allow for solutions
with a non-vanishing condensate, and therefore $\frac{1}{2}(T_3-T_0)$ will be
spontaneously broken.
This solution must be found numerically, since the system is non-linear.
However, (\ref{gauge4}) \emph{does} have an analytic solution
\beq\label{eq:xisolution}
\xi = \bar\m\left(1-\frac{1}{\r}\right)
\enq
and thus $\bar n_{\xi} = \bar\mu$.

When the symmetry is not broken, $\Y=0$, the equation for
$\c$ has of course
\be\label{eq:chiunbroken}
\c=\bar\m\left(1-\frac{1}{\r}\right)
\ee
as a solution as well. Therefore, in the unbroken phase
\begin{align}
\label{eq:Thetaunbroken}
\Th&=0\,,\\
\label{eq:Phiunbroken}
\F&= 2\bar\m\left(1-\frac{1}{\r}\right) \,.
\end{align}
This behavior reflects the fact that $T_3$ is completely independent from $T_0$
in the unbroken phase. However, once we switch on the condensate, the
interplay between $T_3$ and $T_0$
(recall that the remaining symmetry is a combination of the two) makes it
impossible to set only one of the fields to zero.

Finally, let us mention that in order to relate the dimensionless parameters
with
the physical ones, we need to apply the same dictionary
\erf{eq:barmu}-\erf{eq:psi2} used for the ungauged model.

\subsection{Charge Density in the broken phase}

According to \cite{Schafer:2001bq,Watanabe:2011ec}
we can expect the presence of type II Goldstone modes if the broken symmetry
generators fulfill
\begin{equation}
\label{eq:Qcommutator} \langle [ Q_a, Q_b ] \rangle = B_{ab}
\end{equation}
with at least one $B_{ab}\neq0$.
In our case we have $[Q_1, Q_2] = i Q_3$. Therefore in the broken phase
we are interested in a non-vanishing  expectation value for the charge density
operator  $\langle Q_3 \rangle = n_\Th$. As we argued
previously, in the unbroken phase we necessarily have $\Th(r)=0$. This happened
since both $\c$
and $\xi$ obey the same differential equation and the integration constants had
to be set equal in order to do not switch on a source for $\Th$. Now we
would like to find out whether or not an expectation value for
$\Th$ will be spontaneously generated in the broken phase.

\begin{figure}[htp!]
\centering
\includegraphics[width=250pt]{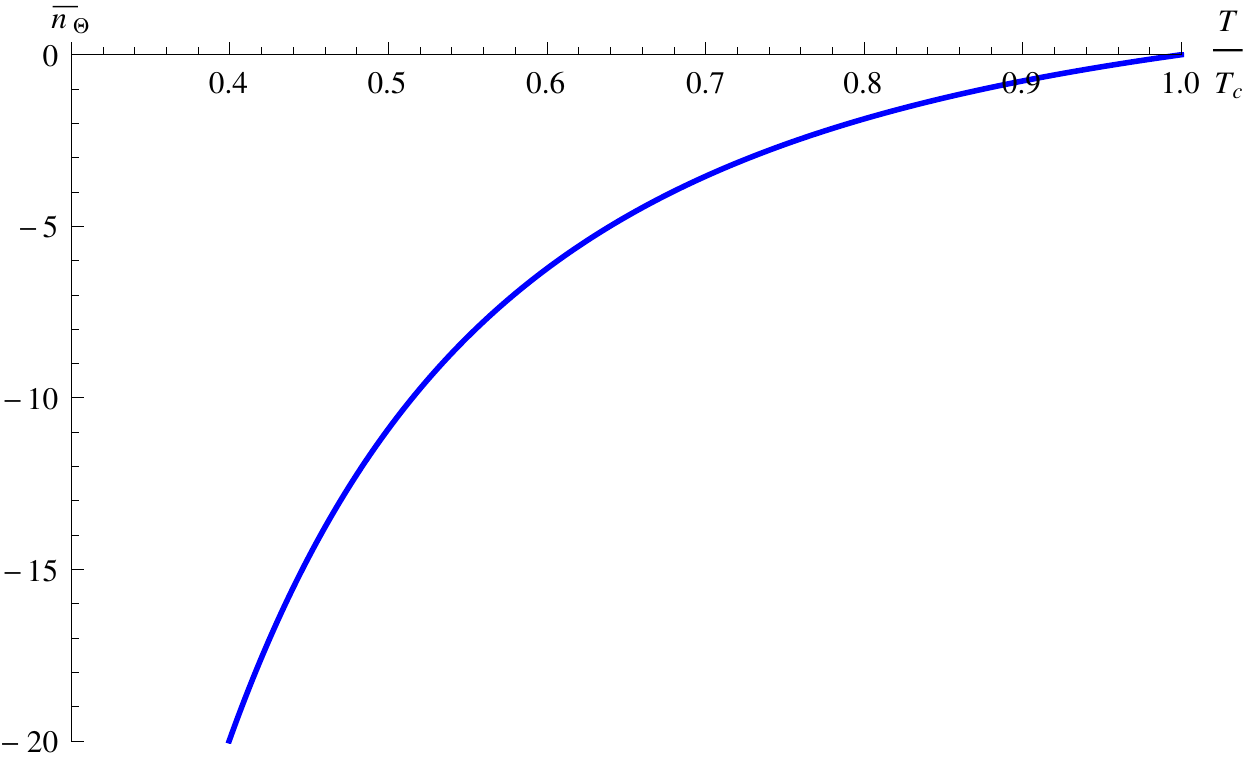}
\caption{\label{fig:subleading} Charge density of $\Th$, $\bar n_{\Th}$, as a
function of the temperature $T/T_c$.}
\end{figure}

Independently of the phase the field associated to the unbroken combination of
generators is given by (\ref{eq:xisolution}).
Since $\Th=\xi-\c$, then
\beq
\bar n_\Th= \bar \mu - \bar n_\chi \,.
\enq
Hence, what we want to check is the difference between the leading and the
subleading coefficients of $\c$ as a function of the temperature. The numerical
result is shown in Figure \ref{fig:subleading}.

So we conclude that precisely at $T \leq T_c$ this difference is switched on and
an expectation value for $\langle Q_3 \rangle$ appears. This can be taken as a
clear
indication for the appearance
of type II Goldstone bosons in the spectrum.

\subsection{Fluctuations of the gauged model}

In order to study the quasinormal spectrum and the conductivities of the system,
we switch on longitudinal
perturbations on top of the background, so that
\begin{eqnarray}
\hat \Y^T &=& (\eta(t,\r,x), \Y(\r)+ \s(t,\r,x)) \,,\\
A^{(0)} &=& (\Phi(\r) + a^{(0)}_t(t,\r,x))dt +  a^{(0)}_x(t,\r,x)dx\,,\\
A^{(1)} &=& a^{(1)}_t(t,\r,x)dt +  a^{(1)}_x(t,\r,x)dx\,,\\
A^{(2)} &=& a^{(2)}_t(t,\r,x)dt +  a^{(2)}_x(t,\r,x)dx\,,\\
A^{(3)} &=& (\Th(\r) + a^{(3)}_t(t,\r,x))dt +  a^{(3)}_x(t,\r,x)dx\,.
\end{eqnarray}

\subsubsection{Perturbations in the Unbroken Phase}

In the normal phase, the background value of the condensate vanishes.
Moreover, we have $\Theta(\r)=0$. The equations of motion for the perturbations
read
\begin{eqnarray}
\label{eqn1unb} s'' +s'\left(\frac{f'}{f}
+\frac{2}{\r}\right)+\left(\frac{(\frac{\Phi}{2}+\omega)^2}{f^2}
-\frac{k^2}{f\r^2} -\frac{m^2}{f}\right)s&=&0\,,\\
 a''^{(c)}_t+\frac{2}{\r}a'^{(c)}_t  - \frac{\omega k}{f\r^2}a^{(c)}_x -
\frac{k^2}{f\r^2}a^{(c)}_t&=&0\,,\\
\label{eqn5unb} a''^{(c)}_x+\frac{f'}{f}a'^{(c)}_x  +
\frac{\omega^2}{f^2}a^{(c)}_x +\frac{\omega k}{f^2}a^{(c)}_t&=&0\,,\\
\frac{\omega}{f}a'^{(c)}_t+ \frac{k}{\r^2}a'^{(c)}_x&=&0\,,
\end{eqnarray}
where $s\in\{ \eta, \sigma \}$. Since the color indices do not see each
other the system is the same one as \erf{eq:flucunbs}-\erf{eq:flucunbcon} except
that there
are four copies of the gauge field fluctuations. Due to the chosen normalization
of the $U(2)$ generators
the gauge field background $\Phi$ enters with an additional factor $\frac 1 2$
compared to \erf{eq:flucunbs}.
The quasinormal mode spectrum is the same as the one of the holographic s-wave
superconductor \cite{Amado:2009ts}
except that the scalar modes are doubly degenerate and the  gauge field modes
are fourfold degenerate. In particular
there are four copies of the hydrodynamic diffusion mode $\omega = -iD k ^2$.

\subsubsection{Perturbations in the Broken Phase}

The equations of motion in the broken phase decouple in two sets: one mixing the
$(0)-(3)$ colors of the gauge field and $\sigma$
fluctuations and the other mixing the $(1)-(2)$ colors and the $\eta$
fluctuations.

Writing $\sigma =\zeta + i \delta$, the equations of the $(0)-(3)$ sector are

\begin{eqnarray}
0&=&f\zeta'' +\left(f' +\frac{2f}{\r}\right) \zeta'
+\left(\frac{\omega^2}{f}+\frac{\chi^2}{f}-
\frac{k^2}{\r^2} -m^2\right)\zeta + \frac{2 i\omega\chi}{f}\delta
+(a^{(0)}_t -a^{(3)}_t)\Psi\frac{\chi}{f} \,,   \\
0&=&f\delta'' +\left(f' +\frac{2f}{\r}\right) \delta'
+\left(\frac{\omega^2}{f}+\frac{\chi^2}{f}-
\frac{k^2}{\r^2} -m^2\right)\delta - \frac{2i\omega\chi}{f}\zeta + i
\Psi\omega\frac{a^{(3)}_t-a^{(0)}_t}{2f} +\nn\\
&\,&+i \Psi k\frac{a^{(3)}_x-a^{(0)}_x}{2\r^2}\,,\\
0&=&fa''^{(0)}_t +\frac{2f}{\r}a'^{(0)}_t -\left(\Psi^2+
\frac{k^2}{\r^2}\right)a^{(0)}_t-\frac{\omega k}{\r^2}a^{(0)}_x +
\Psi^2a^{(3)}_t- 4
\zeta \Psi \c-2i\omega \Psi \delta\,,\\
0&=&fa''^{(0)}_x +f'a'^{(0)}_x +\left(\frac{\omega^2 }{f} -
\Psi^2\right)a^{(0)}_x+\frac{\omega k
}{f}a^{(0)}_t + \Psi^2 a^{(3)}_x+2ik \delta \Psi \,,\\
0&=&fa''^{(3)}_t +\frac{2f}{\r}a'^{(3)}_t  -\left(\Psi^2 +
\frac{k^2}{\r^2}\right)a^{(3)}_t-\frac{\omega k}{\r^2}a^{(3)}_x +
\Psi^2a^{(0)}_t+ 4 \zeta \Psi \chi + 2i\omega \Psi \delta\,,\\
0&=&fa''^{(3)}_x +f'a'^{(3)}_x +\left(\frac{\omega^2 }{f} -
\Psi^2\right)a^{(3)}_x+\frac{\omega k }{f}a^{(3)}_t + \Psi^2 a^{(0)}_x-2ik
\delta
\Psi\,,\\
0&=&\frac{ik}{\r^2}a'^{(0)}_x+\frac{i\omega}{f}a'^{(0)}_t+2 \Psi' \delta- 2\Psi
\delta'\,,\label{cons031}\\
0&=&\frac{ik}{\r^2}a'^{(3)}_x+\frac{i\omega}{f}a'^{(3)}_t-2 \Psi' \delta+ 2\Psi
\delta'\,.\label{cons032}
\end{eqnarray}
It is trivial to show that by defining new fields $a_t^{(\pm)}\equiv\frac{1}{2}
(a_t^{(0)}\pm a_t^{(3)})$ and
$a_x^{(\pm)}\equiv\frac{1}{2} (a_x^{(0)}\pm a_x^{(3)})$ the system further
decouples
into a coupled system for
the scalar fluctuations and $a_\mu^{(-)}$ and a background independent set of
equations for the $U(1)$ gauge
field $a_\mu^{(+)}$. The first subsystem reproduces the eoms
\erf{eq:delta}-\erf{eq:cons} and therefore
corresponds to the s-wave $U(1)$ superconductor contained in the $U(2)$ model.
On the other hand, the
field $a_\mu^{(+)}$ corresponds to the preserved gauge symmetry
surviving the $U(2)\to U(1)$ spontaneous
symmetry breaking.
The quasinormal mode spectrum in this sector is therefore the same one as
in \cite{Amado:2009ts} plus the QNMs that are stem from a $U(1)$ gauge field
in $AdS_4$. In particular the hydrodynamic modes in this sector are the sound
mode
and the diffusion mode of the unbroken $U(1)$.

 From now on we will
concentrate on the remaining fields. We will call this remaining, inherently
non-Abelian sector the $(1)-(2)$ sector and will show that the expected
type II Goldstone boson resides there.
Writing $\eta  =\alpha + i\beta $, we find the following equations in the
$(1)-(2)$ sector:

\begin{eqnarray}
\nonumber \label{qnmalpha}0&=&f\a'' +\left(f' +\frac{2f}{\r}\right) \a'
+\left(\frac{\omega^2}{f}+\frac{\left(\Phi+\Theta\right)^2}{4f}-
\frac{k^2}{\r^2} -m^2\right)\a + \frac{i\omega(\Phi+\Theta)}{f}\b -\\
&\,&-i \Psi
\left(\frac{k}{2\r^2}a^{(2)}_x + \frac{\omega}{2f} a^{(2)}_t
\right)+\frac{\Psi \Phi}{2f}a^{(1)}_t\,,\\
\nonumber \label{qnmbeta}0&=&f\b'' +\left(f' +\frac{2f}{\r}\right) \b'
+\left(\frac{\omega^2}{f}+\frac{\left(\Phi+\Theta\right)^2}{4f}-
\frac{k^2}{\r^2} -m^2\right)\b - \frac{i\omega(\Phi+\Theta)}{f}\a -\\
&\,&-  i \Psi
\left(\frac{k}{2\r^2}a^{(1)}_x + \frac{\omega}{2f} a^{(1)}_t \right)
-\frac{\Phi \Psi}{2f}a^{(2)}_t\,,\\
\label{qnmat1}0&=&fa''^{(1)}_t + \frac{2f}{\r}a'^{(1)}_t- \left(\Psi^2 +
\frac{k^2}{\r^2}\right)a^{(1)}_t - \frac{\omega k}{\r^2}a^{(1)}_x+ i\Theta
\frac{k}{\r^2}a^{(2)}_x -2 \Phi \Psi \alpha -2 i\omega \Psi \beta\,,\\
\label{qnmax1}0&=&fa''^{(1)}_x +f'a'^{(1)}_x+ \left(\frac{\omega^2}{f} - \Psi^2
+
\frac{\Theta^2}{f}\right)a^{(1)}_x - 2i\frac{\Theta \omega }{f}a^{(2)}_x-
i\Theta \frac{k}{f}a^{(2)}_t +\frac{\omega k}{f}a^{(1)}_t
\beta +\nn\\
&\,&+ 2ik \Psi \beta\,,\\
\label{qnmat2}0&=&fa''^{(2)}_t + \frac{2f}{\r}a'^{(2)}_t- \left(\Psi^2 +
\frac{k^2}{\r^2}\right)a^{(2)}_t -
\frac{\omega k}{\r^2}a^{(2)}_x- i\Theta\frac{k}{\r^2}a^{(1)}_x + 2\Phi \Psi \b -
2i\omega \Psi \a\,,\\
\label{qnmax2}0&=&fa''^{(2)}_x +f'a'^{(2)}_x+ \left(\frac{\omega^2}{f} - \Psi^2
+
\frac{\Theta^2}{f}\right)a^{(2)}_x + 2i\frac{\Theta \omega }{f}a^{(1)}_x+
i\Theta \frac{k}{f}a^{(1)}_t +\frac{\omega k}{f}a^{(2)}_t +\nn\\
&\,&+ 2ik \Psi \a\,,\\
\label{cons1}0&=&\frac{ik}{\r^2}a'^{(1)}_x +
\frac{i\omega}{f}a'^{(1)}_t+\frac{1}{f}\left(a'^{(2)}_t\Theta -
a^{(2)}_t\Theta'\right) +2 \Psi' \b-2 \b' \Psi\,,\\
\label{cons2}0&=&\frac{ik}{\r^2}a'^{(2)}_x +
\frac{i\omega}{f}a'^{(2)}_t-\frac{1}{f}\left(a'^{(1)}_t\Theta -
a^{(1)}_t\Theta'\right) +2 \Psi' \a-2\a' \Psi\,.
\end{eqnarray}

A comment is in order here. This system of equations could be written in
a more compact form by using complex field variables $\eta$ and $a^{(1)}_{t,x}
\pm i a^{(2)}_{t,x}$.
One has to keep in mind then that the field equations one needs to solve for
the QNM spectrum for the complex conjugate
fields are not the complex conjugate equations since one has to demand infalling
boundary conditions on the fields and on the complex conjugate fields
simultaneously.
This aspect is somewhat clearer if one works with the (formally) real field
variables
on paying the price of writing a somewhat lengthy system of equations.

Up to linear order in perturbations, there are three decoupled
sectors in the system. Two of them belong to the `$(0)-(3)$ sector'
and they are a copy of the $U(1)$ holographic superconductor, already
extensively
studied, and the preserved $U(1)$ gauge symmetry. The main features of the
spectrum of
this sector have already been presented in section \ref{sec:ungauged} since it
is also
a subsector of the ungauged model. On the other hand, the so called `$(1)-(2)$
sector' has not been
studied before. The physics in this sector is quite different from the
holographic
superconductors studied up to now and we will concentrate on it in the rest of
this paper.

Before studying the quasinormal modes we will focus on a simpler
problem, namely the conductivities.

\subsection{Conductivities}

In order to study the conductivities via Kubo formulae, it is enough to
solve the linearized equations in the limit $k=0$. The retarded correlators that
we are interested in have the form $\mathcal{G}_{\mathcal{R}} \sim
\left<J^x_{(c)},J^x_{(c')}\right>_{\cR}$, with $c,c'$ color indices.

We will be applying the prescription of \cite{Kaminski:2009dh}
for computing Green functions in the presence of holographic operator mixing. If
one has a set of fields $\Phi_I$, the two-point correlation functions will be
\be
\mathcal{G}_{IJ} = \lim_{\Lambda\to\infty} \left(\cA_{I M} \cF_k ^M \,_J
(\Lambda)' + \cB_{IJ}\right),
\ee
where the matrix $\cF_k (r)$ is nothing but the bulk-to-boundary propagator for
the fields, normalized to be the unit matrix at the boundary. The matrices $\cA$
and $\cB$ can be read off from the on-shell renormalized action. The
corresponding DC
conductivities are given by the following Kubo formula
\be
\sigma_{IJ} = \lim_{\omega\to0} \left(\frac{i}{\omega} \mathcal{G}_{IJ}
(\omega,0)\right)\,.
\ee

At
vanishing momentum the longitudinal components of the gauge field perturbations
decouple from the scalar perturbations, as well as from the temporal components
of the gauge fields. Moreover, the constraints (eqs.
(\ref{cons031}-\ref{cons032}) and(\ref{cons1}-\ref{cons2}))
become trivial. Since we know
that the system splits into the $(0)-(3)$ and the $(1)-(2)$ sectors we can
rearrange
the $a_x^{(c)}$ fields in two vectors
\begin{eqnarray}
\Phi^T_{k\ (0-3)}(\r)= (a^{(0)}_x(\r),a^{(3)}_x(\r))\quad{\rm and}\quad
\Phi^T_{k\ (1-2)}(\r)= (a^{(1)}_x(\r),a^{(2)}_x(\r))\,.
\end{eqnarray}
One can check that in our case the $\mathcal{A, B}$ matrices take the simple
form
\be
\label{amatrix} \mathcal{A} =-\frac{f(r)}{2}\,
\mathbb{I}\,,\qquad\mathcal{B}=0\,,
\ee
for both sectors. A priori we would have a $4\times 4$ matrix of conductivities.
We know
however that the fluctuations in the  $(0)-(3)$ and the $(1)-(2)$ sector
decouple from
each other. Therefore we can restrict ourselves to study two independent
$2\times 2$ matrices
of conductivities.

\subsection{Conductivities in the $(0)-(3)$ sector}

\begin{figure}[htp!]
\centering
\includegraphics[width=230pt]{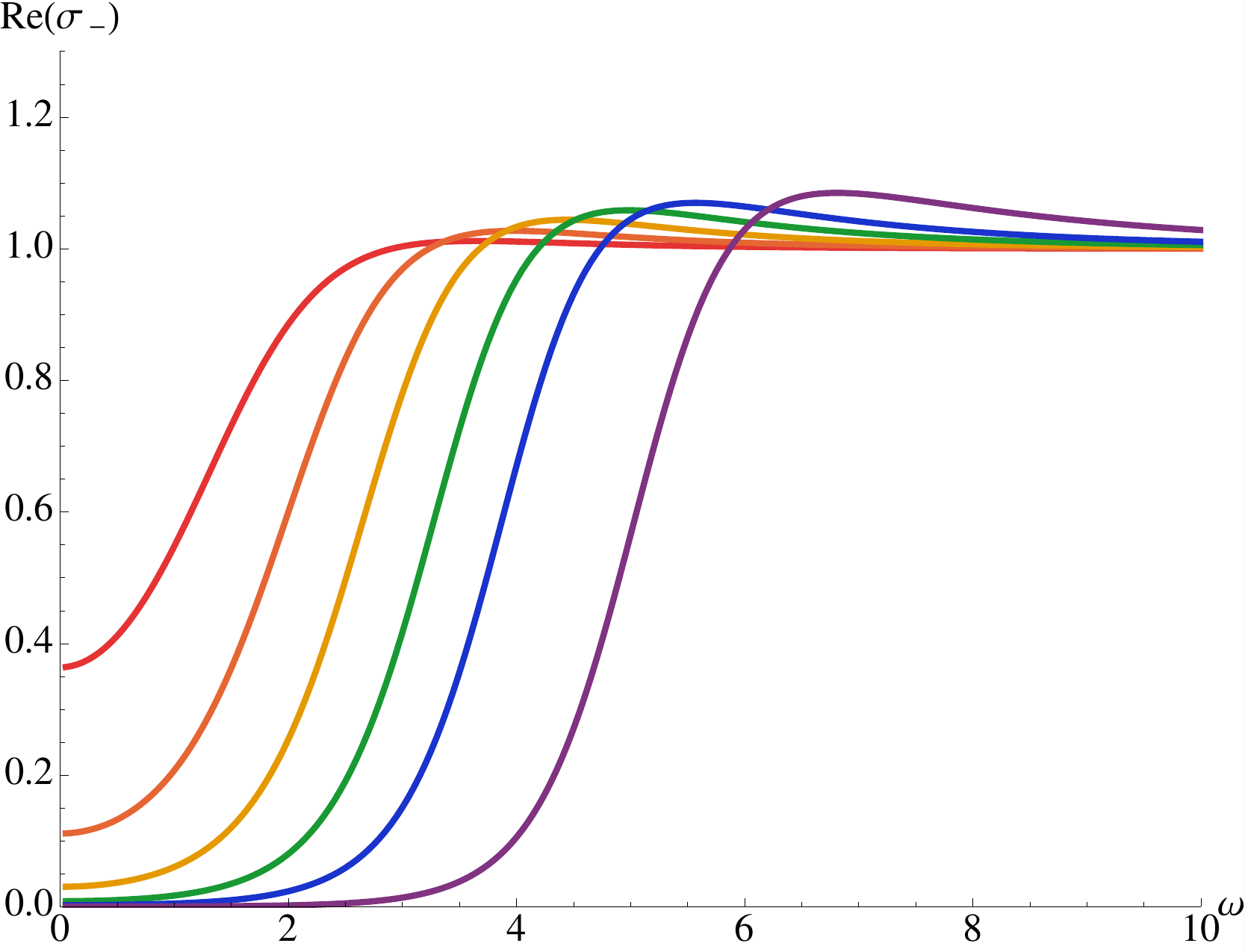}\hfill
\includegraphics[width=230pt]{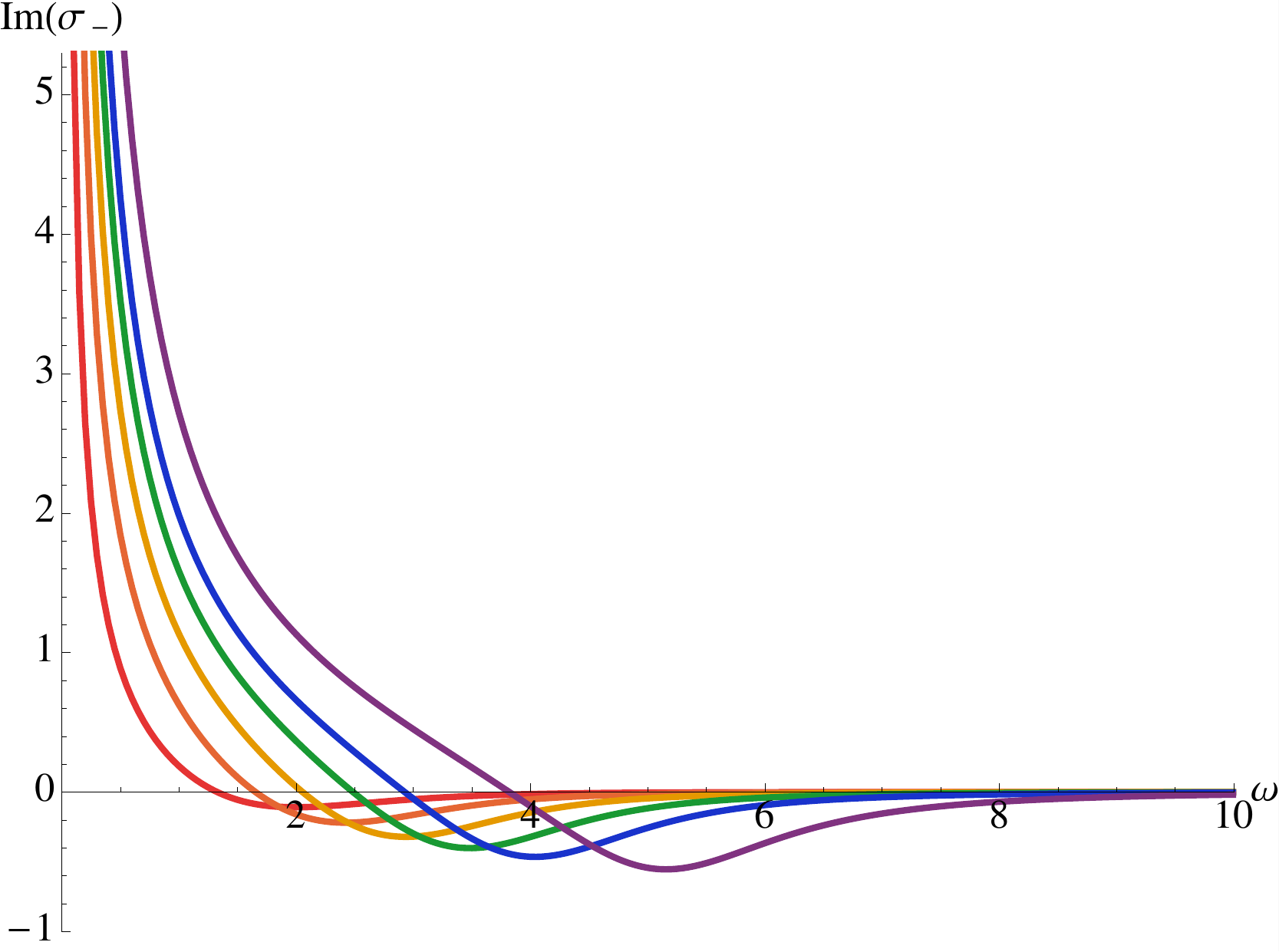}
\caption{\label{fig:supercond}Real part (left) and imaginary part (right) of the
conductivity as a function of frequency. The plots correspond to temperatures in
the range $T/T_c \approx 0.91-0.41$, from red to purple. As expected,
the plots reproduce the ones of \cite{Hartnoll:2008kx}.}
\end{figure}

The $k=0$ equations of motion for $a^{(0)}_x$ and $a^{(3)}_x$ can be simplified
by using the already defined $a^{(-)}_x$
and $a^{(+)}_x$ fields. This results in
\begin{eqnarray}
\label{cond03b}0&=&fa''^{(+)}_x +f'a'^{(+)}_x +\frac{\omega^2 }{f}
a^{(+)}_x\,,\\
 \label{cond03a}0&=&f a''^{(-)}_x +f' a'^{(-)}_x
+\left(\frac{\omega^2 }{f} -2 \Psi^2\right)a^{(-)}_x\,.
\end{eqnarray}
We see that the resulting system of equations is now completely decoupled.
We only have two diagonal conductivities $\sigma_{++}$ and $\sigma_{--}$,
corresponding to the unbroken $U(1)$ diffusive sector and a mode which is
associated to the broken $U(1)$ coupling to the condensate. The former
is the same as in the unbroken phase and of no further interest for us.
The latter is again the well-studied $U(1)$ s-wave superconductor.
Its conductivity has been already calculated in
\cite{Hartnoll:2008kx}. To check our numerics we have re-calculated it
and in Figure \ref{fig:supercond} we show its behavior. It coincides
completely with \cite{Hartnoll:2008kx}.
The real part shows the
$\omega=0$ delta function characteristic of superconductivity\footnote{In
general,
this behavior is also typical of translation invariant
charged media, in which accelerated charges cannot relax.
However, working in the probe limit we effectively break translation invariance
and therefore the infinite DC conductivity is a genuine sign of
superconductivity.}. Numerically
this can be seen through the  $1/\omega$ behavior in the imaginary part. The
Kramers-Kronig relation (see (\ref{KKrel}) in appendix \ref{app.a}) implies then
infinite DC conductivity.
The real part of the AC conductivity also exhibits a temperature dependent gap.

\subsection{Conductivities in the $(1)-(2)$ sector}

The relevant equations for the $(1)-(2)$ sector read
\begin{eqnarray}
\label{eqna1x}0=fa''^{(1)}_x +f'a'^{(1)}_x+ \left(\frac{\omega^2}{f} - \Psi^2 +
\frac{\Theta^2}{f}\right)a^{(1)}_x - 2i\frac{\Theta \omega }{f}a^{(2)}_x \,,\\
\label{eqna2x}0=fa''^{(2)}_x +f'a'^{(2)}_x+ \left(\frac{\omega^2}{f} -\Psi^2 +
\frac{\Theta^2}{f}\right)a^{(2)}_x + 2i\frac{\Theta \omega }{f}a^{(1)}_x\,.
\end{eqnarray}
These equations obey the symmetry
\begin{equation}\label{eq:condsym}
 (a_x^{(1)}\rightarrow a_x^{(2)}~, ~a_x^{(2)} \rightarrow -a_x^{(1)} ) \,.
\end{equation}

One can see that the fact that
$\Th(1)=0$ implies that $a^{(1)}_x(1)$ is independent of $a^{(2)}_x(1)$, so,
after imposing infalling boundary conditions at the horizon, the parameter space
of boundary conditions is two-dimensional, as expected.

In the unbroken phase the system completely decouples
\be
\label{eqnax1unb}0=fa''^{(c)}_x +f'a'^{(c)}_x+ \frac{\omega^2}{f}a^{(c)}_x  \,.
\ee

\subsubsection{Diagonal Conductivities $\s^{11}\  \&\  \s^{22}$}

The diagonal components of the conductivity, $\s^{11}$ and $\s^{22}$ have the
same behavior, as could be anticipated
from the equations (\ref{eqna1x}),(\ref{eqna2x}). Henceforth, we will only refer
to $\s^{11}$, but all the conclusions also apply to $\s^{22}$.

\begin{figure}[htp!]
\centering
\includegraphics[width=230pt]{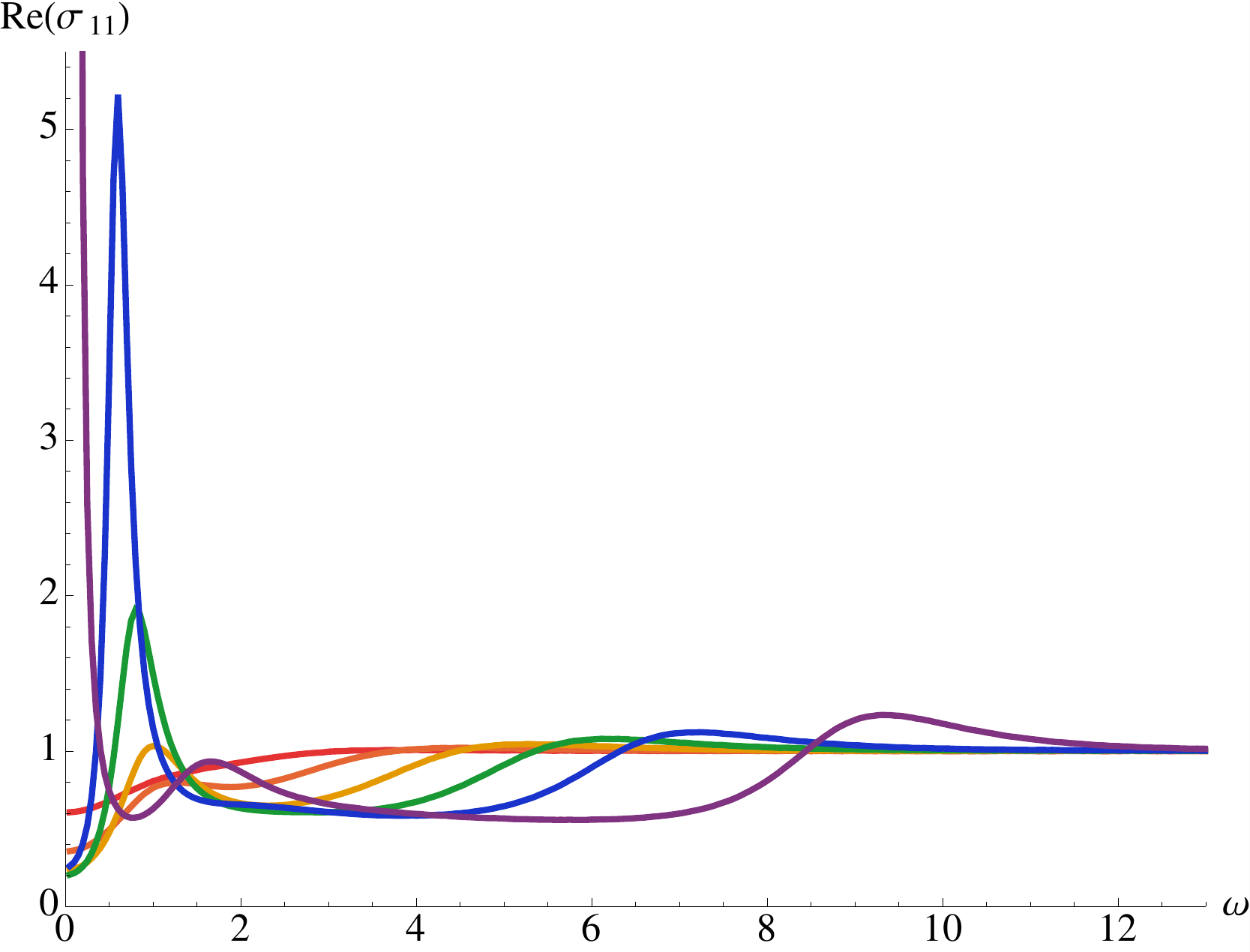}\hfill
\includegraphics[width=230pt]{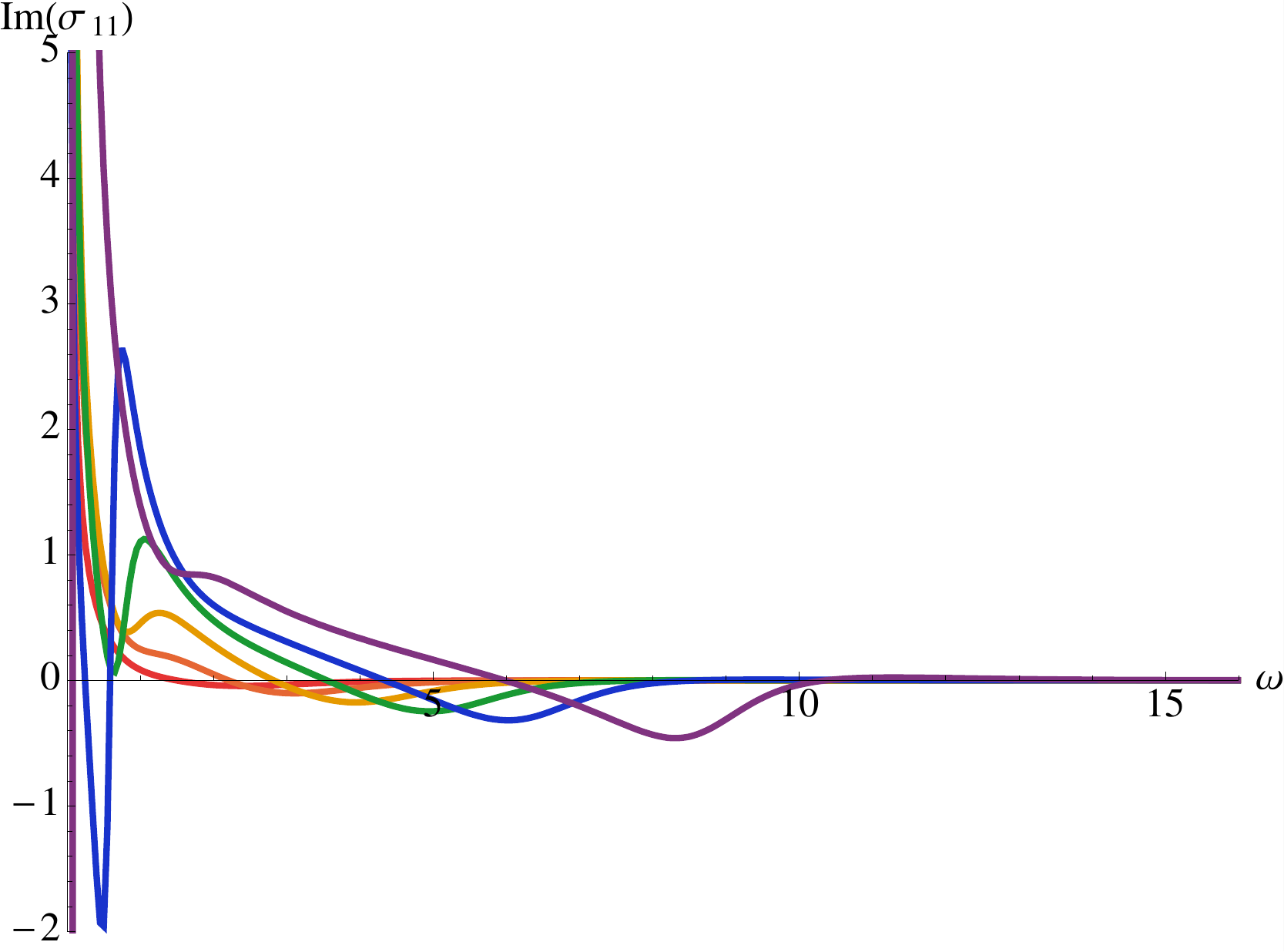}
\caption{\label{fig:sig11} Real (left) and imaginary (right) parts of $\s^{11}$
versus $\w$
for five different temperatures chosen in a range $T/T_c\approx 0.91-0.41$, from
red to purple.
$\Im(\sigma^{11})$ clearly blows up as $\w\rightarrow 0$.}
\end{figure}

Figure \ref{fig:sig11}
shows the conductivity for several values of the temperature. We find that
these conductivities also show delta-function singularities at $\omega=0$.

The strength of the delta function can also be computed. It
is given by the residue of the imaginary part of the conductivity at $\omega
=0$,
\begin{align}
\lim_{\w \rightarrow 0} \w \Im(\s^{11}) \sim n_s\,.
\end{align}

\begin{figure}[htp!]
\centering
\includegraphics[width=200pt]{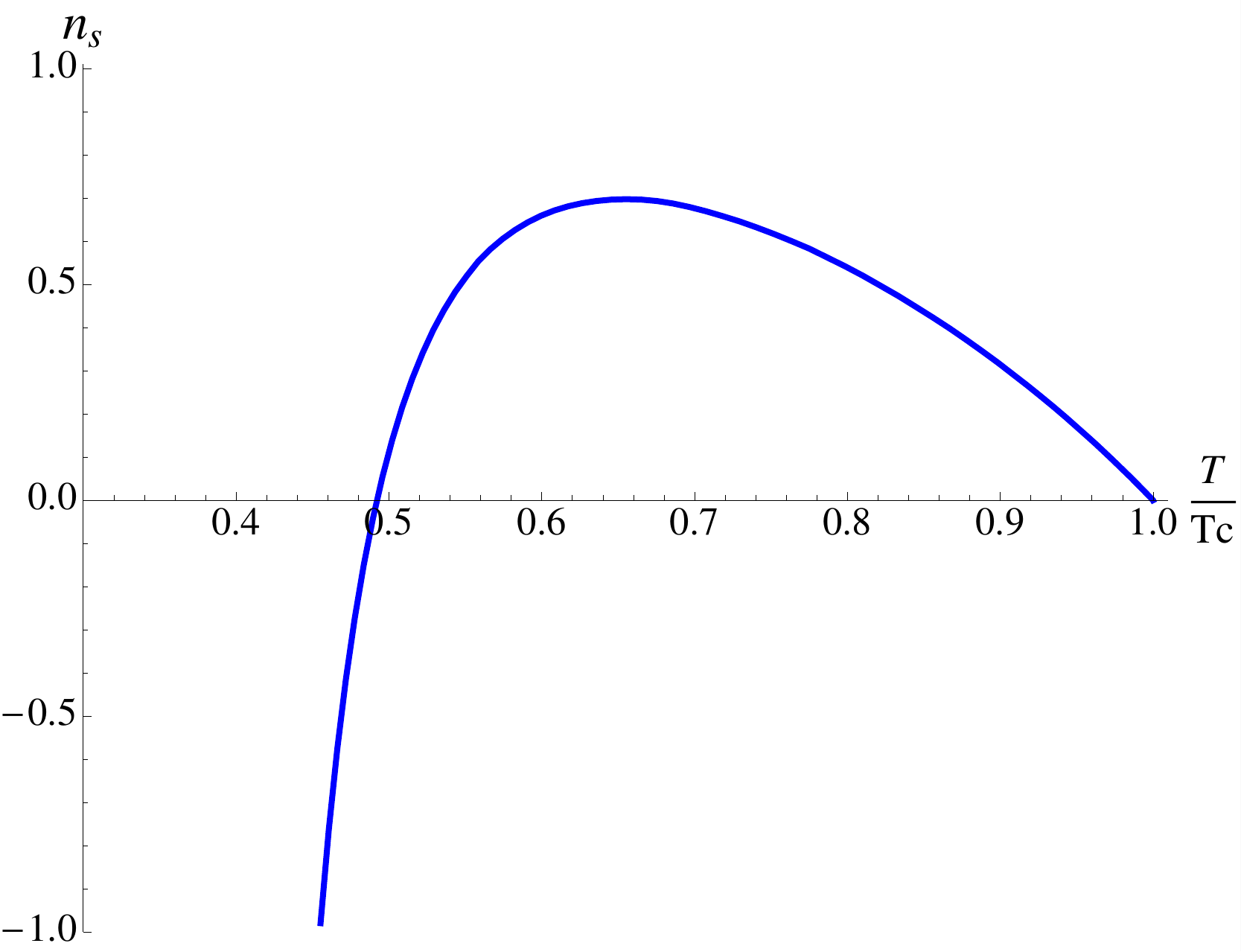}
\caption{\label{fig:supercharge}Residue at $\omega=0$ as a function
of $T/T_c$.}
\end{figure}

The residue is plotted in Figure \ref{fig:supercharge}
as a function of $T/T_c$. As expected, it starts growing from a zero value. At $T/T_c
\approx 0.65$, $n_s$ reaches a maximum and starts decreasing fast, changing sign at $T/T_c =
0.49$. To study $n_s$ down to very low temperature we would need to go beyond the proble limit.
However, as we will comment below, this behaviour of $n_s$ can be understood in light of the QNM spectra.


Let us look in detail at the behavior of the  real part of the conductivity (left plot in figure \ref{fig:sig11}). 
For high enough temperatures the optical conductivity is almost constant, $\Re(\s^{11})=1$, which is expected
since in that regime the dynamics is described essentially by (\ref{eqnax1unb}). As soon as we decrease the
temperature, the onset of the DC conductivity also decreases and only approaches the constant value asymptotically, when $\w$
becomes large enough and thus the term $\frac{\w^2}{f}$ dominates, turning equations (\ref{eqna1x}),(\ref{eqna2x})
approximately into (\ref{eqnax1unb}).  According to the Ferrell-Glover sum rule, the area missing as we lower the temperature
is proportional to $n_s$. 

Interestingly, at low temperatures the real part of $\s^{11}$ starts developing a bump at small values of
$\omega$ ($0<\omega \lesssim 2$).
The bump leaves less area for the delta function to cover, which explains why $n_s$
starts decreasing approximately at this temperature.
Moreover, the appearance of these bumps can be traced back to the fact that for a subleading gauge QNM with small
$|\Im(\omega)|$, $\Re\omega(T) >> \Im \omega(T)$ holds.
Hence, the conductivities affected by this mode start developing the reminiscence of a resonance at a particular frequency.
We have studied the spectrum of low lying QNM for the gauge sector and found that this mode corresponds in the normal phase
to the lowest excitation of $a^{(1,2)}_\mu$, $\w=-1.5 i$.
But it is at lower temperatures where one finds a remarkable fact:
at $T/T_c \approx 0.395$ the mode becomes unstable, and  indeed, as we will see, several physical quantities
modify their behavior at that temperature.

Therefore, we expect a new phase transition around $T/T_c \approx 0.395$, due
entirely to the $(1)-(2)$ sector. 
Since this phase transition seems to be triggered by an unstable mode in the
vector sector it most likely leads to the formation of a p-wave condensate.
This surely requires further investigation that is currently underway \cite{[8]}.

\subsubsection{Off-diagonal conductivities $\s^{12}\  \&\  \s^{21}$}

The off-diagonal elements of the conductivity matrix are also related
via the symmetry (\ref{eq:condsym}) and therefore obey
$\sigma^{12}=-\sigma^{21}$.
Therefore, it is enough to comment on $\s^{12}$, although the
conclusions are valid for both components.

\begin{figure}[htp!]
\centering
\includegraphics[width=230pt]{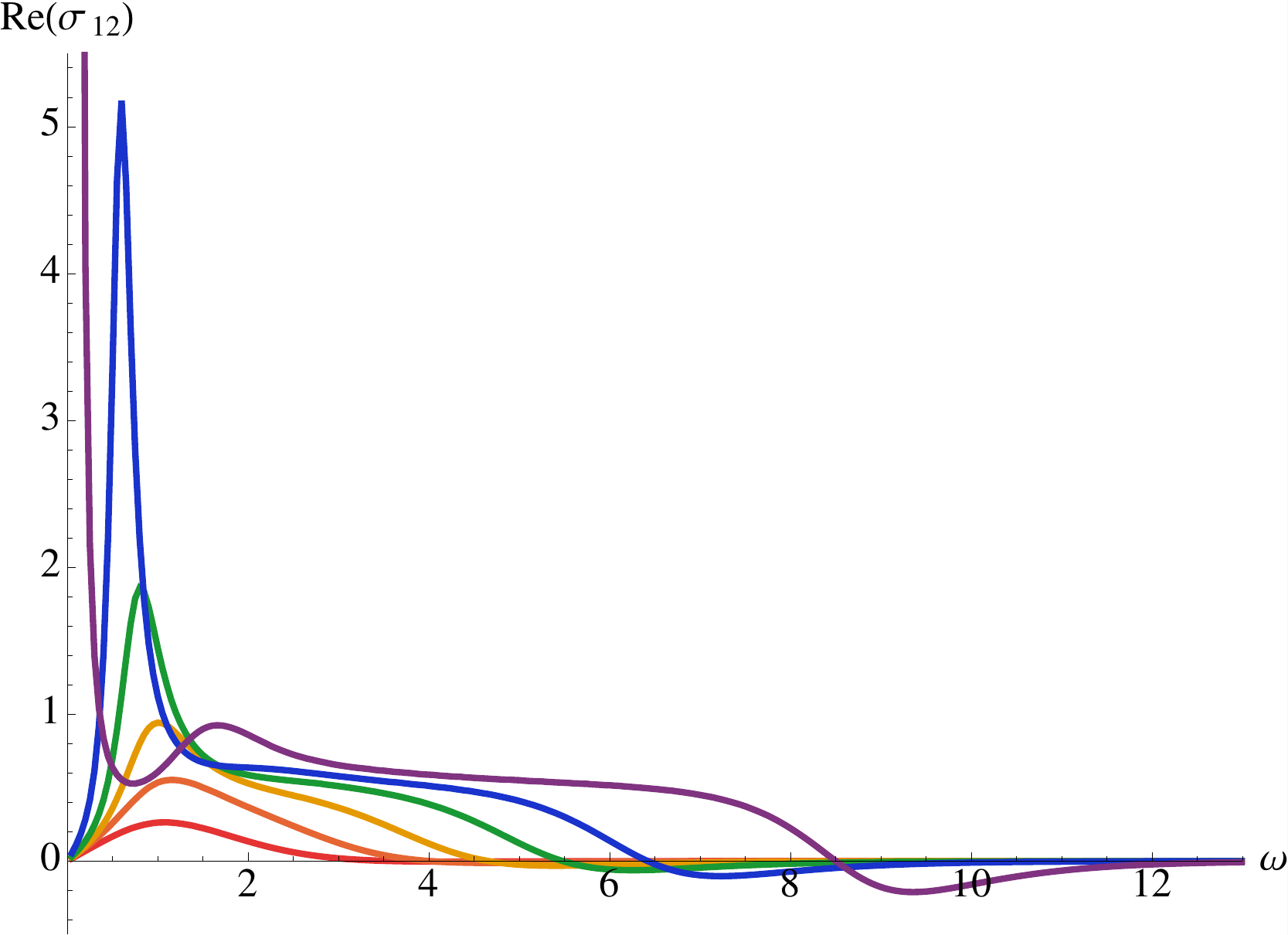}\hfill
\includegraphics[width=230pt]{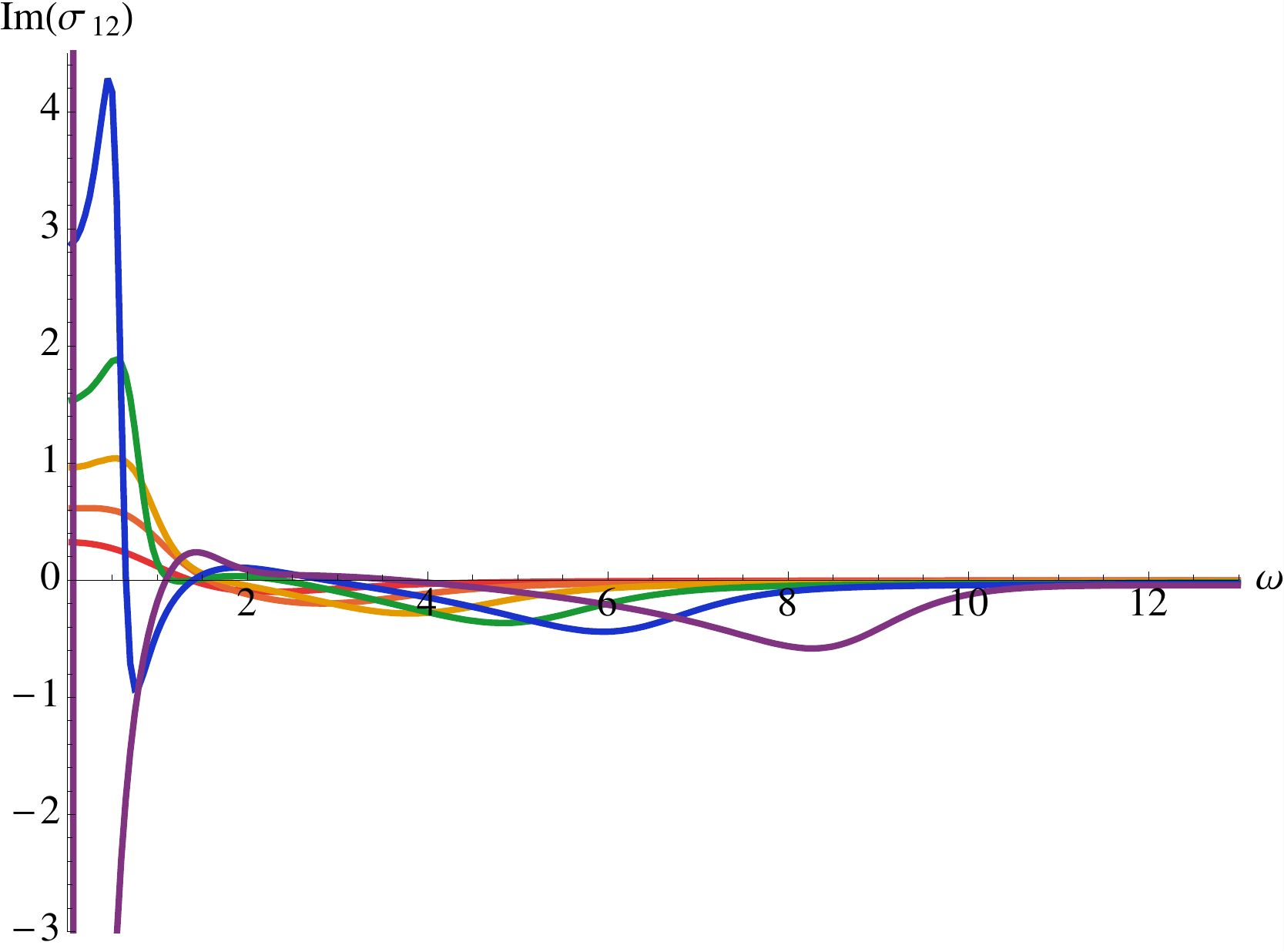}
\caption{\label{fig:sigma1221} Real (left) and imaginary (right) part of
$\s^{12}$ as a function of
$\w$ for $T/T_c \approx 0.91-0.41$, from red to purple.}
\end{figure}

The form of $\s^{12}$ is plotted in Figure \ref{fig:sigma1221} for various
different
temperatures as a function of frequency. At $T/T_c=1$ the system is practically
decoupled,
so for all temperatures the off-diagonal conductivity goes to zero as $\w$
increases.

Observe that $\s^{12}(\w)$ behaves as a normal conductivity. Its real part
vanishes as $\w \rightarrow 0$,  whereas the imaginary part tends
to a constant value.

\subsubsection{Conductivities $\sigma_{+-}$ and $\sigma_{-+}$}

It is worth to notice that the equations \erf{eqna1x}-\erf{eqna2x} decouple if
we define
a new vector field
\begin{equation}
\tilde{\varphi}= \left(\begin{array}{c}
{A_+}\\
{A_-}\\
\end{array}
\right)\ = \left(
\begin{array}{cc}
{1}&{i}\\
{1}&{-i}\\
\end{array}
\right)\, \left(\begin{array}{c}
{a^{(1)}_x}\\
{a^{(2)}_x}\\
\end{array}
\right)\, = S\,\varphi\,.
\end{equation}
In this basis, the equations of motion become
\begin{equation}
\label{eqnadec}0=fA''_\pm +f'A'_\pm+ \left(\frac{(\omega\mp\Theta)^2}{f} -
\Psi^2 \right)A_\pm \,.
\end{equation}
It is easy to check that the relation between the conductivity matrices in the
two basis is given by
\begin{equation}
 \tilde\sigma = \left(S^T\right)^{-1} \sigma S^{-1}\,,
\end{equation}
and that only the off-diagonal components of $\tilde\sigma$ are non vanishing.

\begin{figure}[htp!]
\centering
\includegraphics[width=230pt]{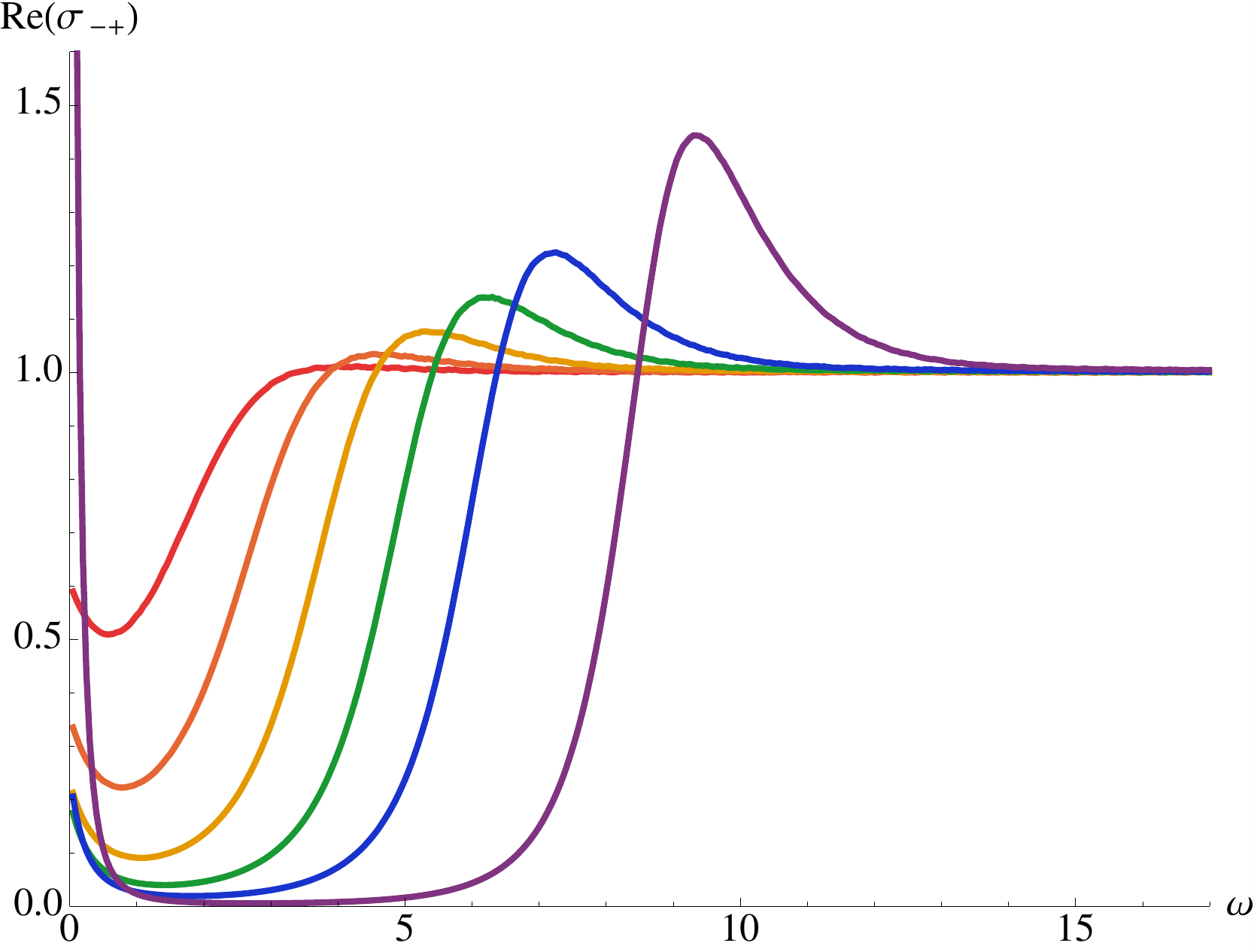}\hfill
\includegraphics[width=230pt]{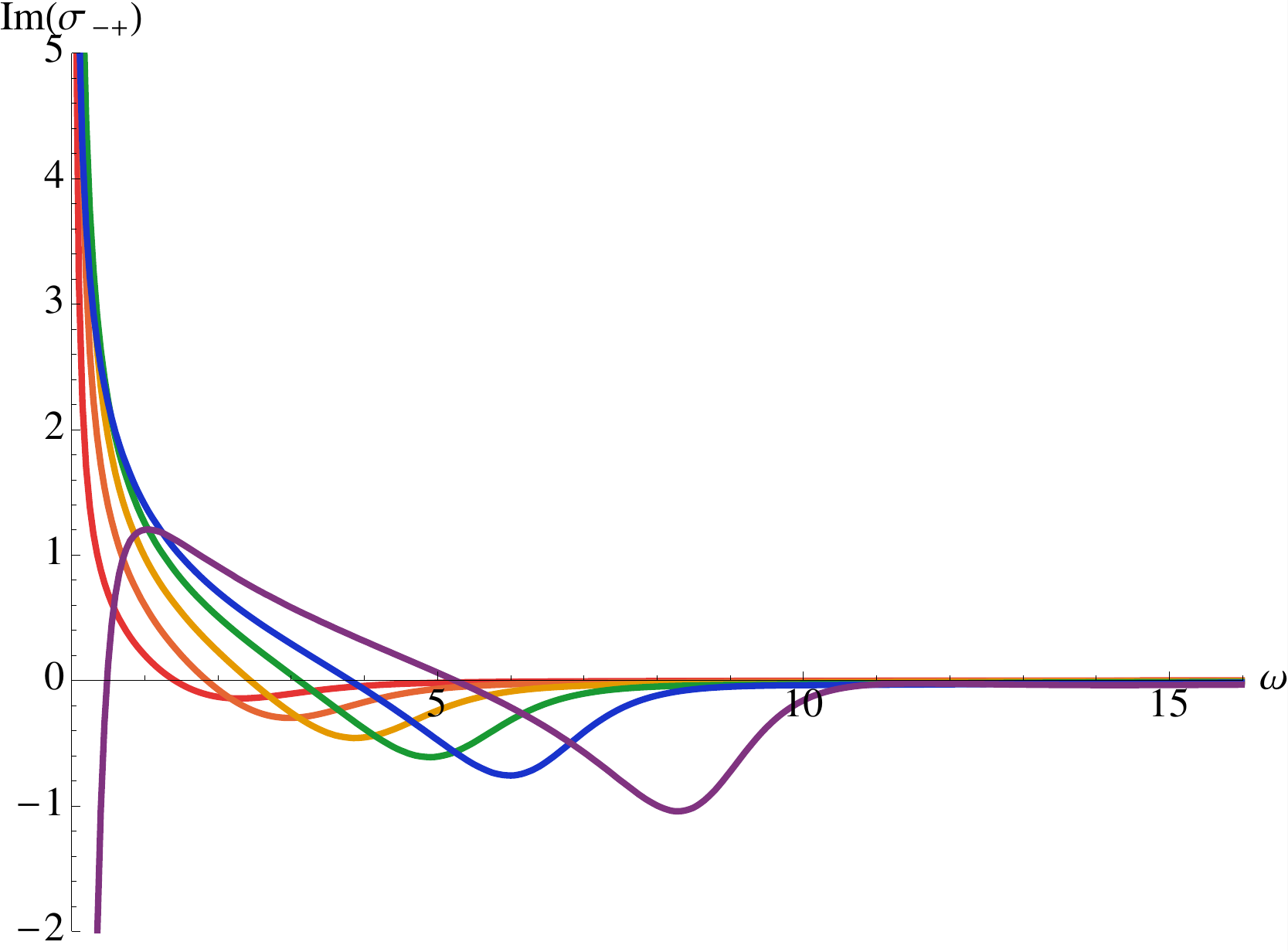}
\caption{\label{fig:smp} Real (left) and imaginary (right) part of the
conductivity
$\s_{-+}$ for temperatures in the range $T/T_c \approx 0.91-0.41$, from red to
purple.}
\end{figure}

\begin{figure}[htp!]
\centering
\includegraphics[width=230pt]{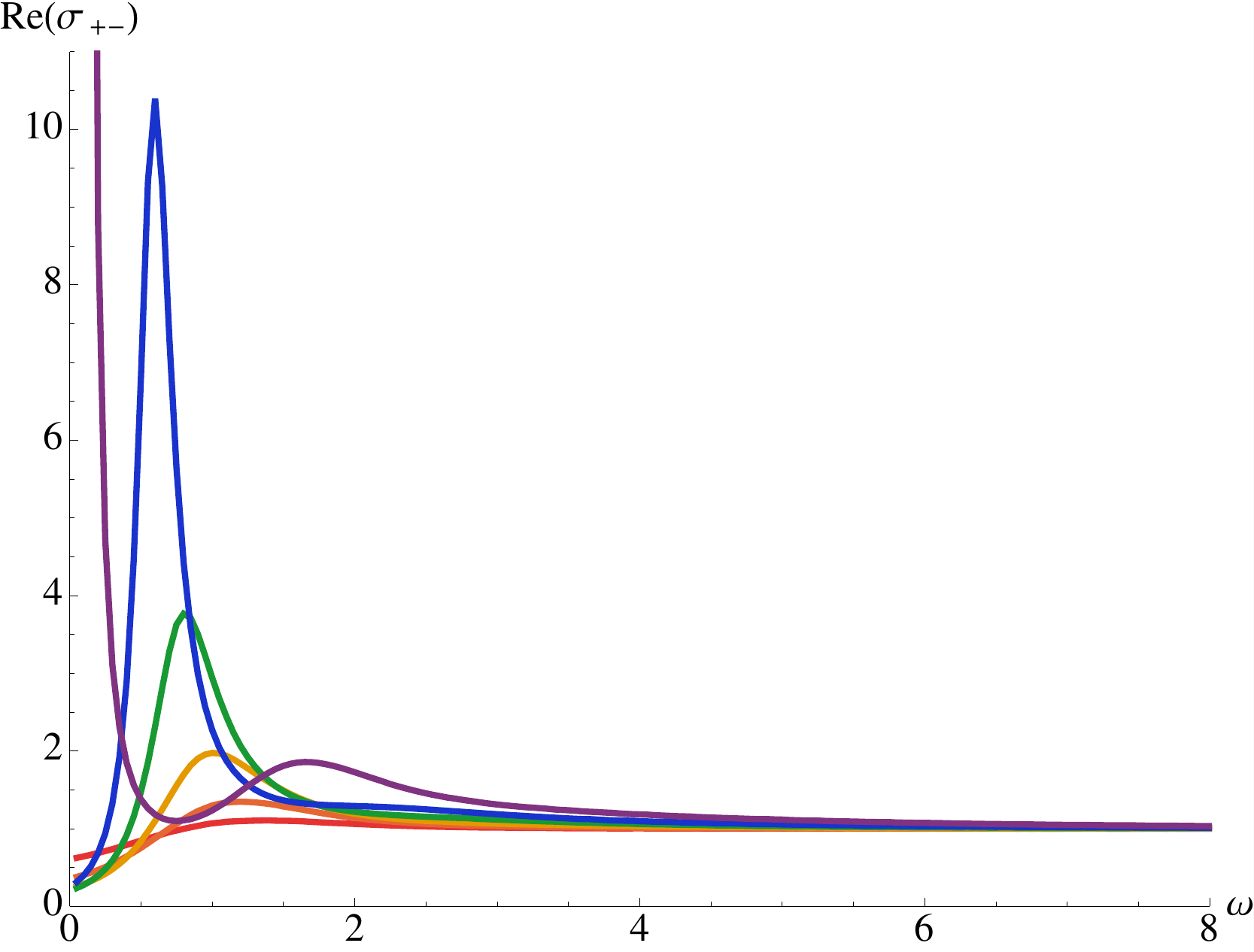}\hfill
\includegraphics[width=230pt]{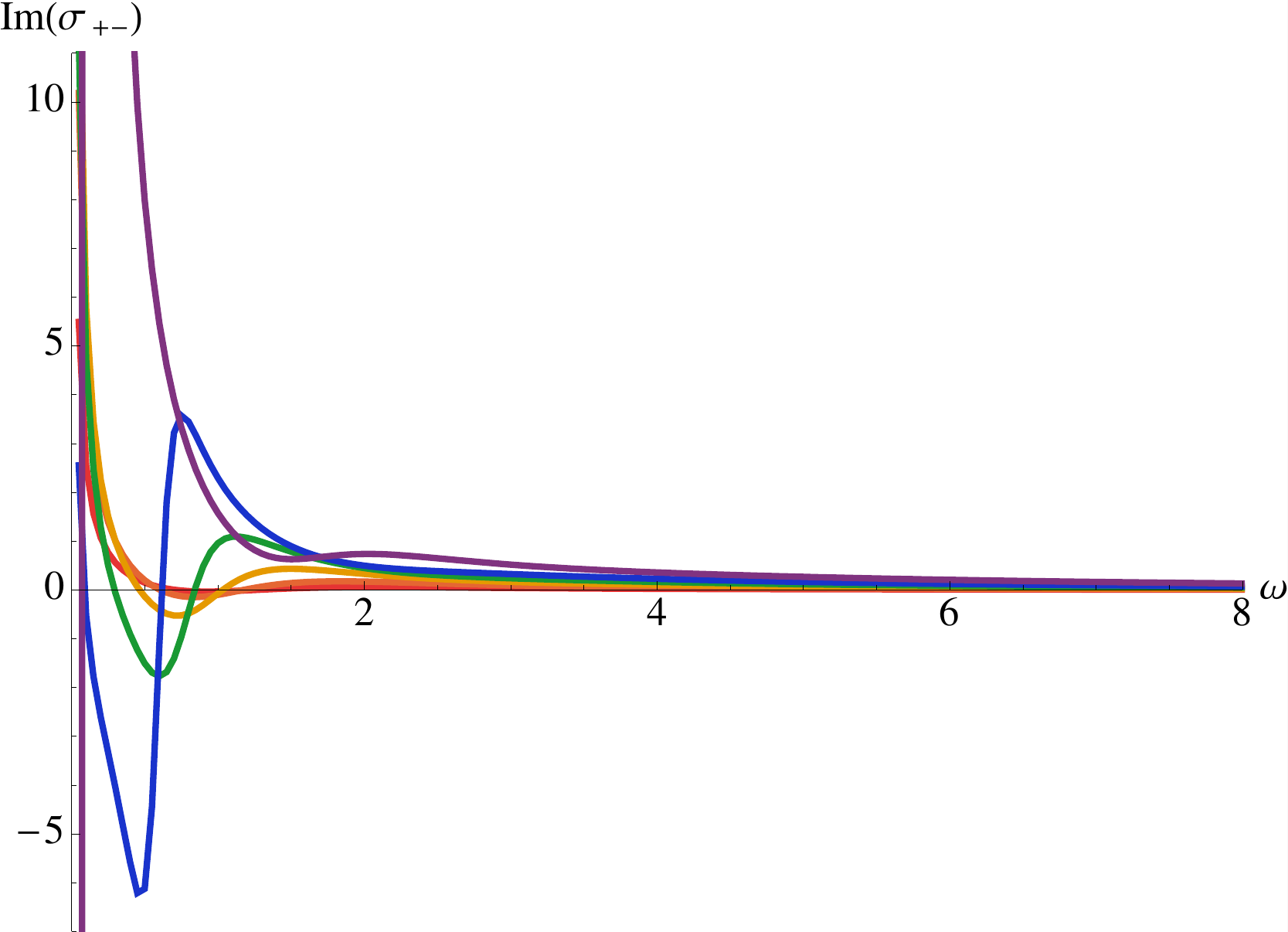}
\caption{\label{fig:spm} Real (left) and imaginary (right) part of the
conductivity
$\s_{+-}$ for temperatures in the range $T/T_c \approx 0.91-0.41$, from red to
purple.}
\end{figure}

The conductivities $\s_{-+}$
and $\s_{+-}$ are represented in Figure
\ref{fig:smp} and \ref{fig:spm}, respectively. The plot of the conductivity
$\s_{-+}$ is particularly suggestive.
Besides the superconducting delta of the DC conductivity, it resembles the
behavior observed in Graphene \cite{Graphene}.
Such a resemblance of the conductivities of holographic superconductors to the
one of graphene has been pointed
our already in \cite{Hartnoll:2009sz}. We emphasize however that the
conductivities shown in figure \ref{fig:smp} have an
even closer resemblance to \cite{Graphene}.
In particular, at small frequencies we see that a Drude-like peak develops. This
kind of behavior in metals is usually
due to the presence of impurities or lattices, whereas in our case, momentum
relaxation would be due to the non-vanishing
expectation value of the charge density operator $\vev{Q_3}=\bar n_\Theta$.
The resemblance holds for not too low temperatures.
When lowering the temperature, a gap opens up as for the $(0)-(3)$ sector. The
real part of $\s_{+-}$ shows the
same peak already observed for $\sigma_{11}$ when decreasing the temperature.
For temperatures below $T\approx 0.49 T_c$, the pole in
the imaginary part of both conductivities changes sign. Of course, it
corresponds to
the temperature at which the residue changes sign. The onset of the DC
conductivity at low temperatures grows very fast,
becoming divergent at $T/T_c\approx 0.395$. The presence of such a pole in the
conductivity is related to the
appearance of an instability in the spectrum of excitations of the gauge field
and therefore with a phase transition
to another superconducting phase, as already discussed.

\subsection{Quasinormal Modes}

Let us finally study the QNM spectrum in the $(1)-(2)$ sector. This sector
contains the
fluctuations $\eta,\,a_\mu^i$ with $i=1,2$, therefore in the unbroken phase the
spectrum will
contain two diffusive modes associated with the two gauge fields. The
fluctuations of the scalar
field in the normal phase were already discussed in section \ref{sec:ungauged}.
Analyzing the quasinormal mode spectrum in the broken phase amounts to solving
the
system of
equations \erf{qnmalpha}-\erf{cons2}. Details of the computation can be
found in appendix \ref{appB}.

\subsubsection{Type II Goldstone mode}

\begin{figure}[htp!]
\centering
\includegraphics[width=230pt]{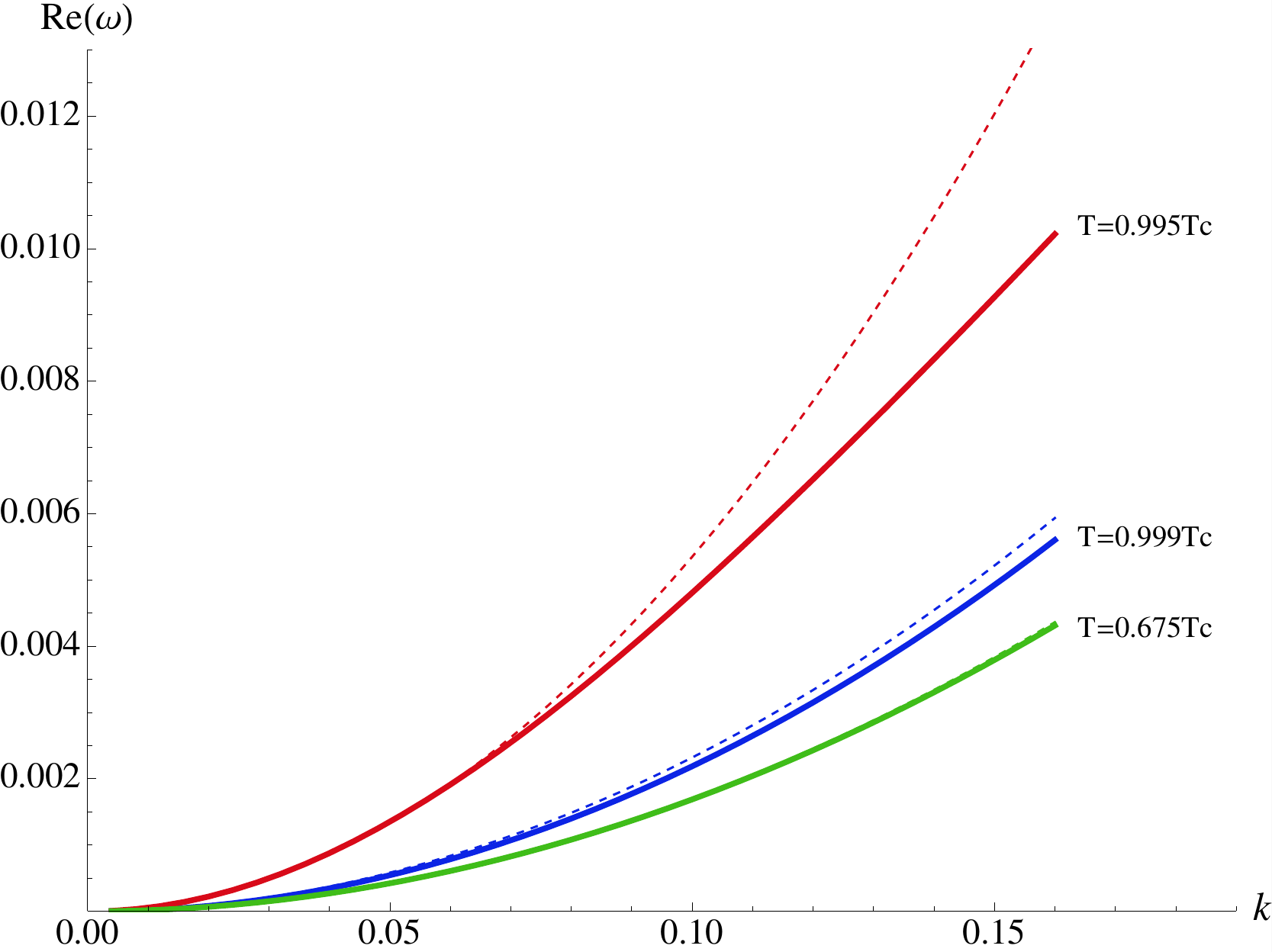}\hfill
\includegraphics[width=230pt]{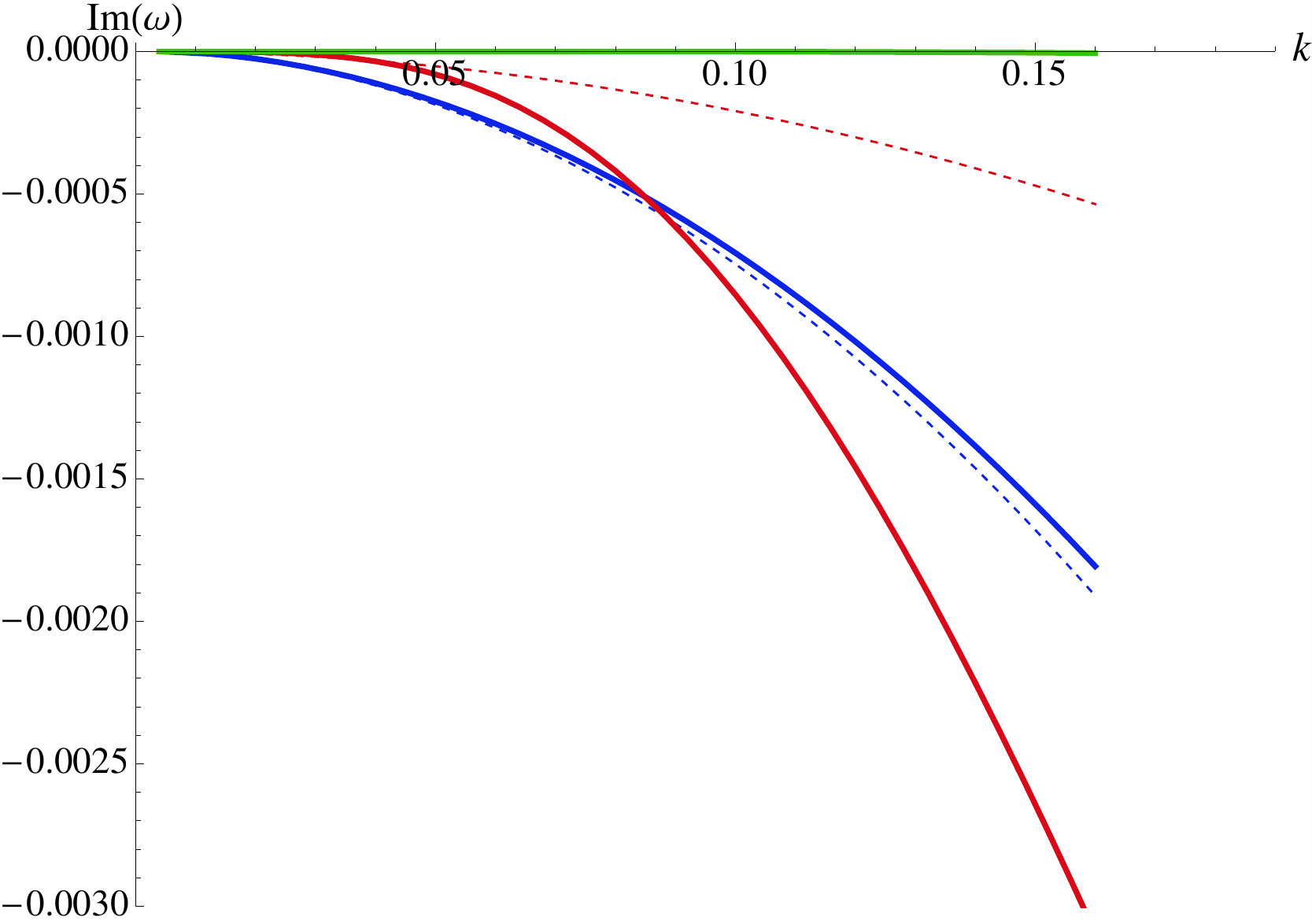}
\caption{\label{fig:bcgauged}Plots of $\Re(\w)$ (left) and $\Im(\w)$
(right) as a function of the momentum. Thick lines correspond to data and thin
lines to quadratic fit.
At $T=0.995\,T_c$  the real quadratic parameter $\cB(T)$ shows a maximum, see
Figure \ref{fig:b&cvsT}.  Relation (\ref{disr12}) is
fulfilled with high accuracy.}
\end{figure}

As expected within the $(1)-(2)$ sector we find a type II Goldstone mode.
As in the ungauged model for small enough momentum its dispersion relation can
be fitted to
\begin{eqnarray}
\label{disr12} \w =\pm \cB k^2-i \cC k^2\,.
\end{eqnarray}
Figure \ref{fig:bcgauged} shows the dispersion relation for various values of
the temperature in the hydrodynamic regime.
The quadratic behavior with momentum is apparent.

\begin{figure}[htp!]
\centering
\includegraphics[width=230pt]{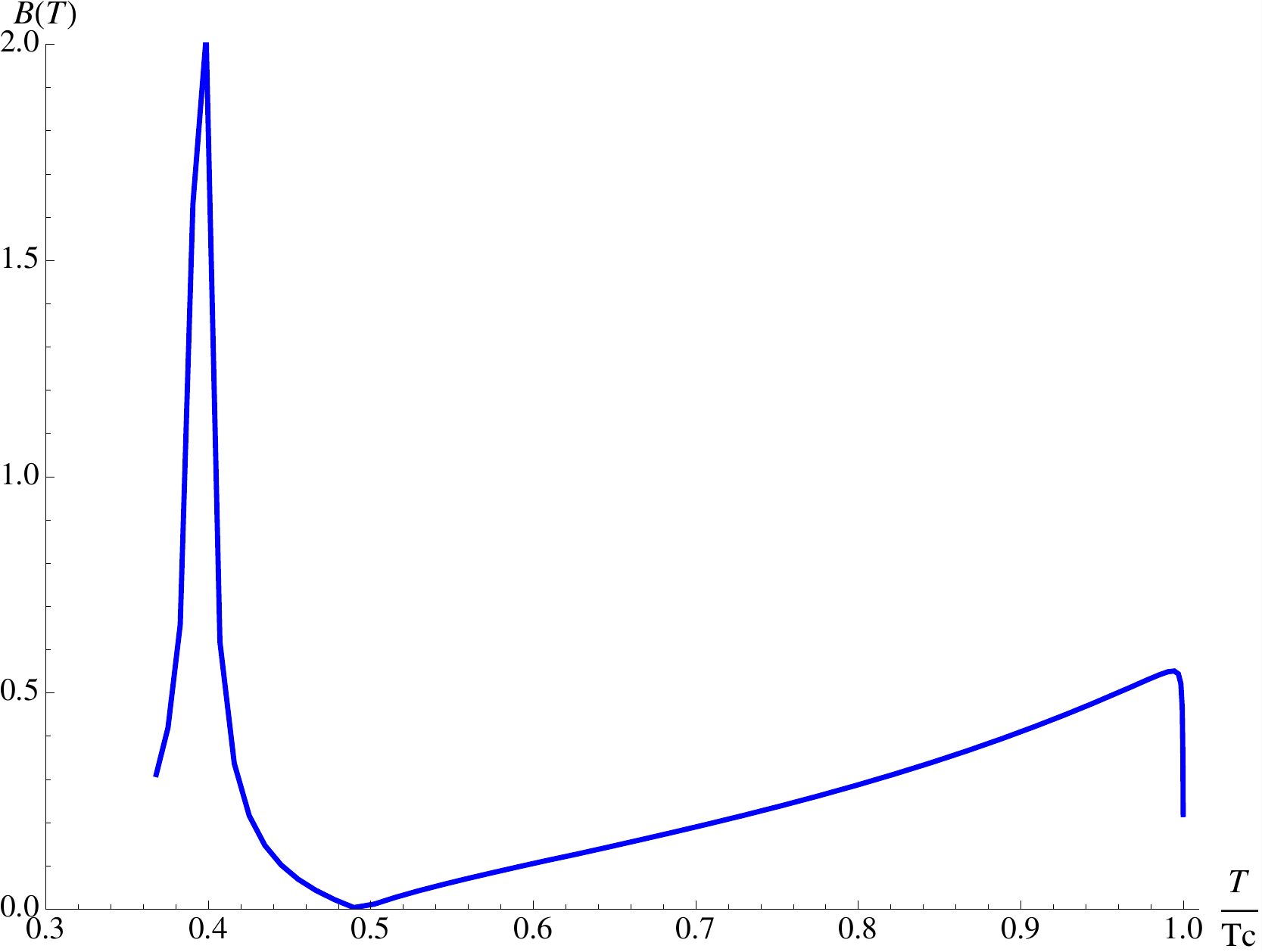}\hfill
\includegraphics[width=230pt]{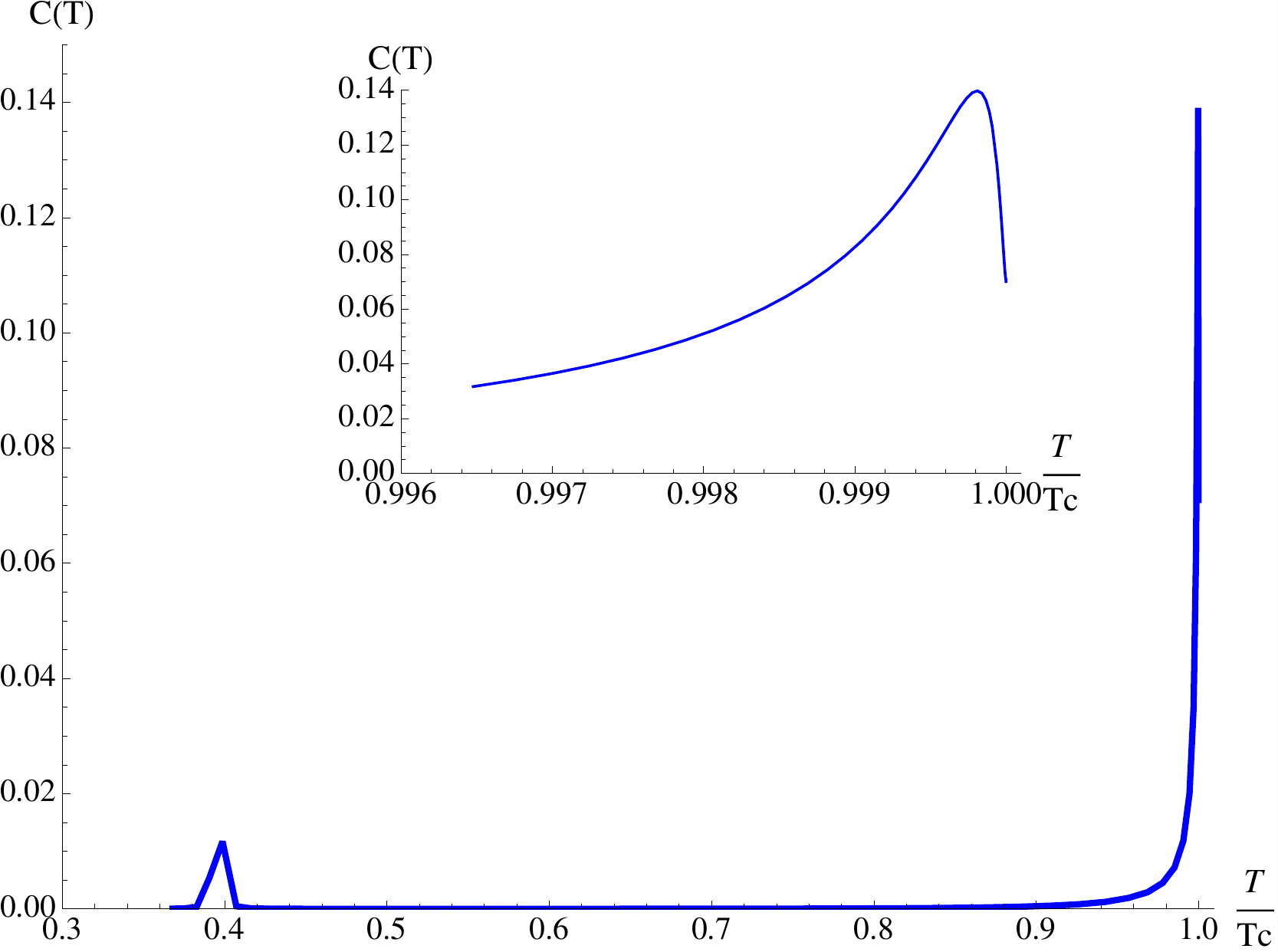}
\caption{\label{fig:b&cvsT}$\cB$ (left) and $\cC$ (right)
as a function of $T/T_c$. The zoom-in shows the peak of $\cC$ close to the
transition.
Furthermore at $T\simeq 0.4 T_c$ a sharp peak shows up in both coefficients. We
relate this
feature also to the instability arising in the vector sector.}
\end{figure}

The temperature dependence of $\cB$ and $\cC$ is plotted in Figure
\ref{fig:b&cvsT}.
The value at $T=T_c$ is given by the same value as in the ungauged model
\erf{bcIIung} and
in fact can also be cross checked by calculating the scalar mode dispersion
relation in the unbroken phase at $T=T_c$ since the QNMs must be continuous
through the phase transition. We find a rather surprising dependence of $\cB$
with the temperature.
It starts at a finite value at the transition and then it rises rather sharply
and falls off slower. It reaches a minimum
at $T\approx 0.49\, T_c$, temperature at which we found the change of sign in
the residue of current-current correlators. We also find
another peak around $T\approx0.4\, T_c$. We expect that it is again related with
the instability found in the gauge sector around that temperature.
It would also be interesting to  calculate $\cB(T)$ using an
alternative
method e.g. as the sound velocity can be calculated from thermodynamic
considerations
alone. In order to do this one would need to formulate the hydrodynamics
of type II Goldstone modes. We are however not aware of such a hydrodynamic
formulation and leave this for future research.

The attenuation $\mathcal{C}(T)$ decreases rapidly with temperature. For
temperatures $T/T_c  < 0.9$ it is negligible and the width of the type II
Goldstone scales with $k^4$ in the hydrodynamic limit.  This fast decreasing
with temperature reflects that this mode propagates almost ideally in the fluid
at low temperature.
No further ungapped modes can be found in this sector.

\subsubsection{Higher quasinormal modes}

Higher quasinormal modes correspond to gapped modes in the QNM spectrum and thus
represent subleading contributions to the low energy Green's functions.
We will focus here only on two of them: the continuation of the
two diffusive modes of the unbroken phase and the special gapped mode that
appears as the partner mode of the type II Goldstone mode in the field
theoretical model.

Analyzing the first one is interesting in order to understand if
also a qualitative change in the response pattern, such as that characterized
by $T_*$ in the $U(1)$ superconductor sector, exists in the $(1)-(2)$ sector.
Since in this sector there exist however
two diffusive modes in the unbroken phase it is also possible that the diffusive
modes do not simply develop a gap but that they pair up and move off the
imaginary axis in the broken phase. Indeed as we will see this is what happens.

The special gapped mode corresponds to a mode that is associated to the complex
conjugate of the scalar perturbation in the unbroken phase. At $k=0$ and $\mu=0$
the scalar
mode and its complex conjugate are degenerated. As we lower the
temperature they split into two different modes. When we reach $T=T_c$,
the lowest scalar mode becomes the type II Goldstone mode whereas the
mode of the complex conjugate scalar field turns into the special gapped mode.
The gap of this mode is expected to be given by the tree level result
(\ref{eq:spmode})
\cite{Nicolis:2012vf}.

\paragraph{Fate of diffusive modes:}

As already mentioned, in the $(1)-(2)$ sector we have two degenerate
diffusive modes in the unbroken phase. When going through the phase transition
these modes can therefore pair up and move off the imaginary axes such that
their
quasinormal frequencies develop real parts and lie symmetrically around the
imaginary axis.
We expect therefore that in the low energy limit the dispersion relation takes
the form
\begin{equation}
\label{eq:psedo}\omega = \Gamma(T) + \mathcal{M}(T) k^2\,,
\end{equation}
where both coefficients are complex functions and the second mode is located at
$\omega'=-\omega^*$.
Besides, we expect the QNMs to be continuous through the phase
transition, which in particular means that for $T= T_c$, our pseudo-diffusive
modes should match the unbroken phase values,
i.e. $\Gamma(T_c) =  0$ and $\mathcal{M}(T_c) = -i$.

\begin{figure}[htp!]
\centering
\includegraphics[width=250pt]{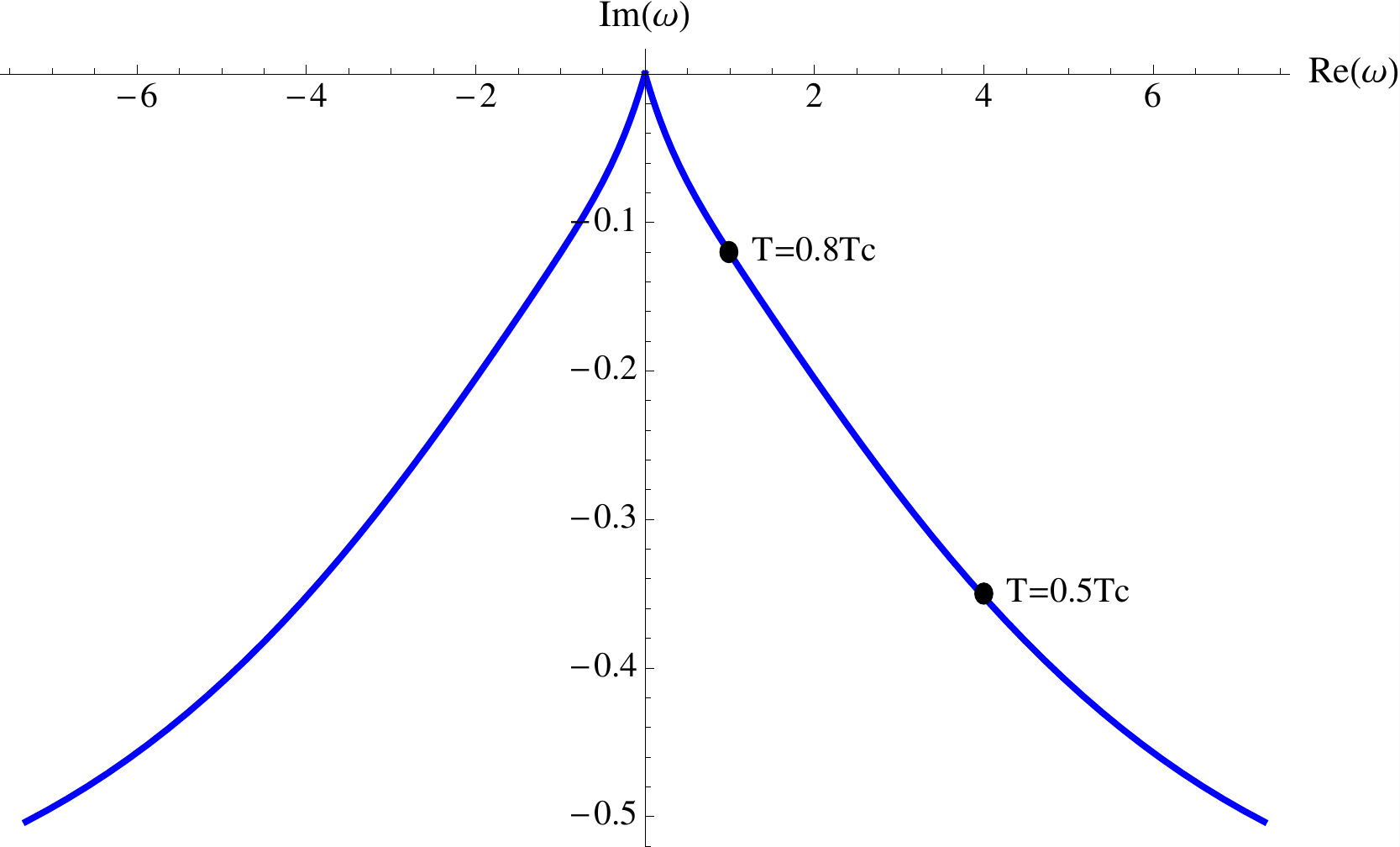}
\caption{\label{fig:ImvsRepseudo}$\Im\w $ versus $\Re\w$ at $k=0$ as a function
of
the temperature. The shape of the figure is compatible with $T$ symmetry,
since there are two pseudo-diffusive modes. Having $\Re\w (k=0)\ne 0$ is
characteristic of the non-Abelian case.}
\end{figure}

The modes at zero momentum are plotted in Figure \ref{fig:ImvsRepseudo}.
We see that indeed the gap vanishes as $T\rightarrow T_c$, whereas the modes
split and develop a real
part as we decrease the temperature. This last feature is exclusive of the
non-Abelian system
and thus does not take place in the usual $U(1)$ holographic superconductor,
where the gap is purely imaginary (see
\cite{Amado:2009ts} and comments above). Close to the phase transition, they
present a linear behavior in temperature,
\be
\Gamma(T) =\left(4.1-0.8\,i\right)\left(1-\frac{T}{T_c}\right)\quad{\rm
near}\,\,T_c\,.
\ee

\begin{figure}[htp!]
\centering
\includegraphics[width=230pt]{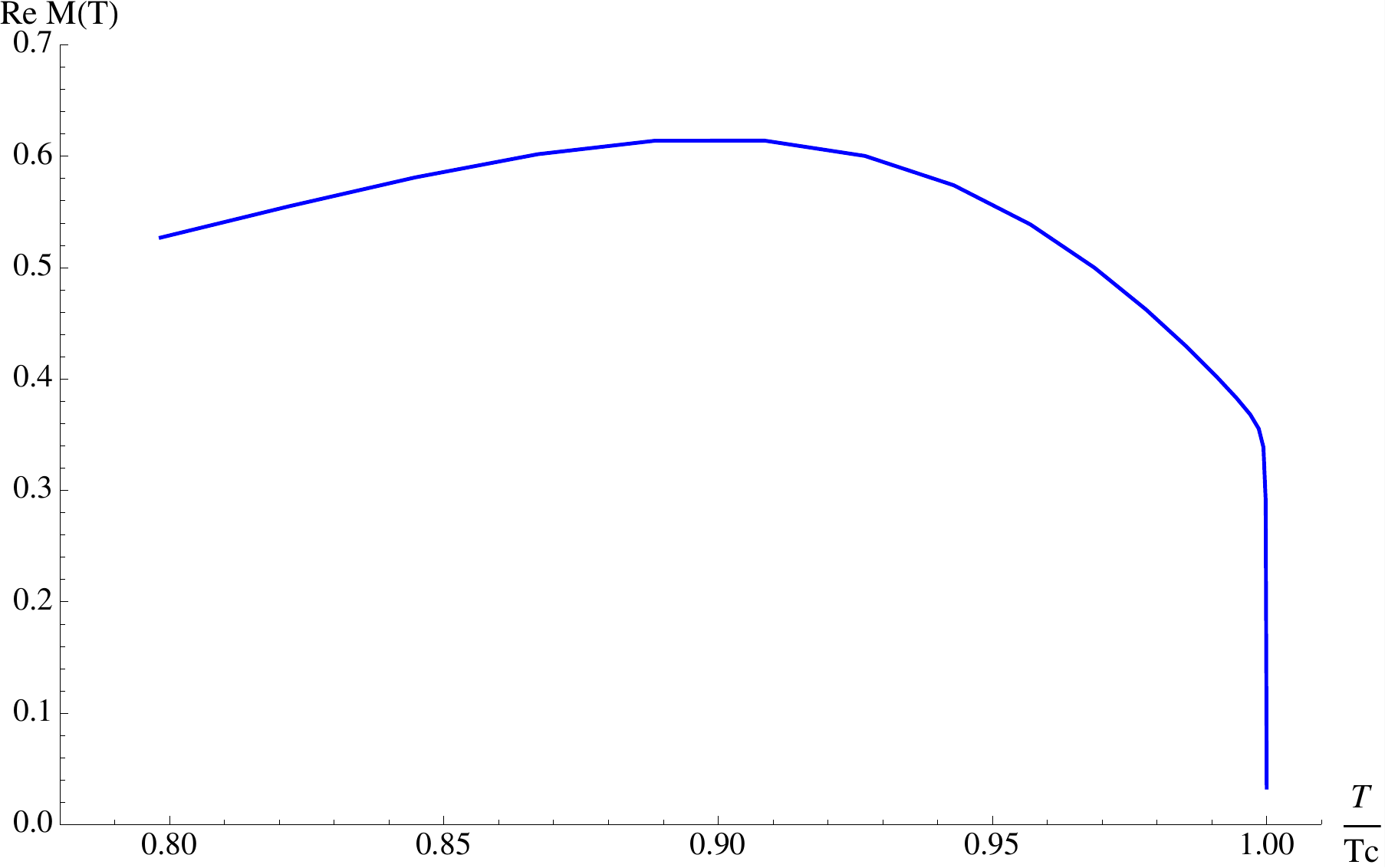}\hfill
\includegraphics[width=230pt]{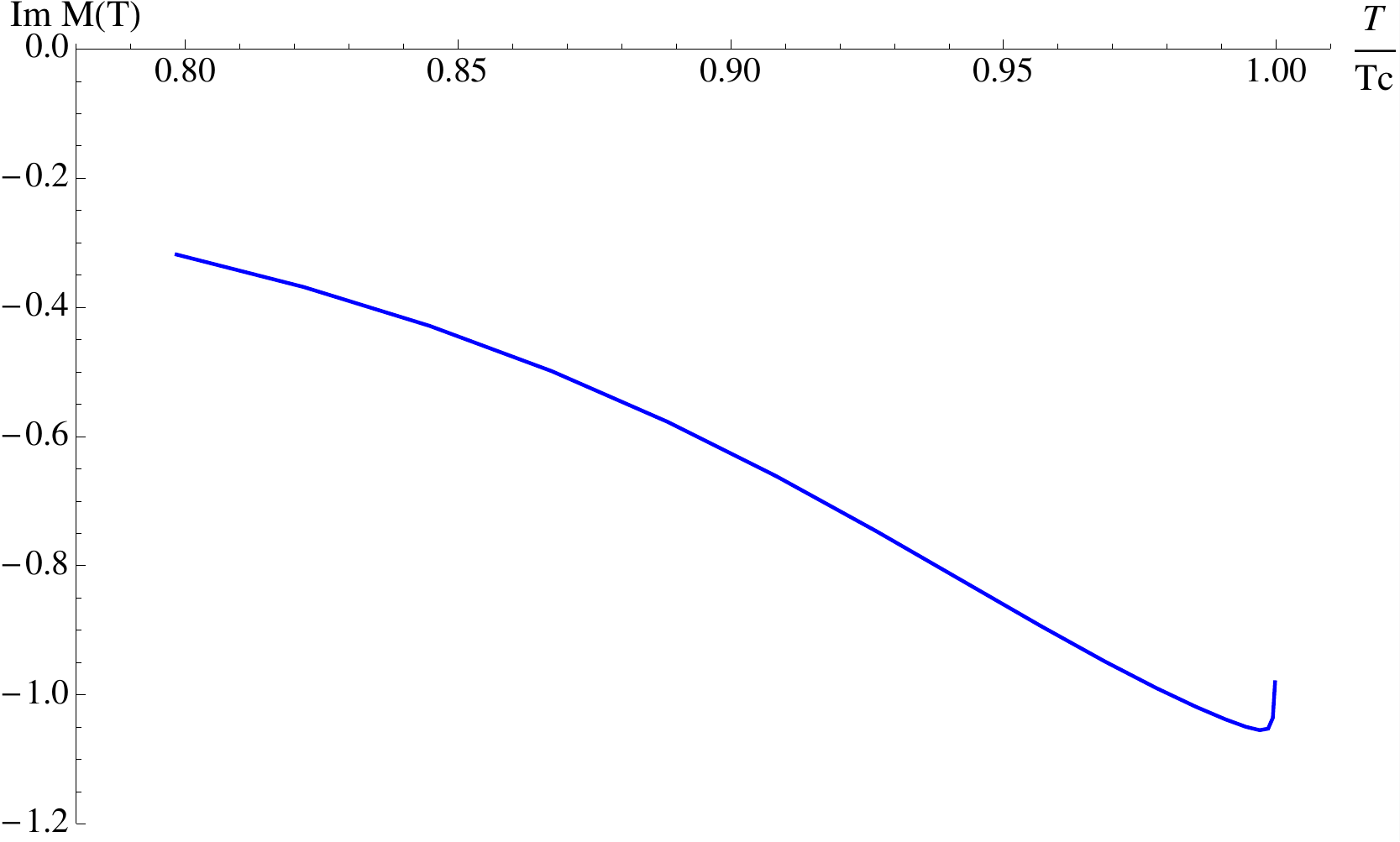}
\caption{\label{fig:M(T)} Real (left) and imaginary (right) part of $\cM(T)$
as a function of $T/T_c$. As the temperature approaches $T_c$, the value of
$\mathcal{M}(T)$ reaches the one prescribed by continuity through the phase
transition.}
\end{figure}

The temperature dependence of the coefficient of the momentum in \erf{eq:psedo},
 $\cM(T)$,
is shown in Figure \ref{fig:M(T)}. The real part rises very steeply just below
the phase transition. The imaginary part approaches the unbroken phase value at
the critical temperature, i.e. $\cM(T_c)=-i$, as is expected for the
pseudo-diffusion modes to continuously
connect to the normal diffusion modes through the phase transition. Notice  $\Im
\cM (T)$ decreases when lowering the temperature.

Another check of the fact that the pseudo diffusion modes come from the pairing
up of the diffusion modes
of the normal phase is that their dispersion relation at the phase transition
matches.
Therefore the two diffusive modes are continuous through the transition,
as expected for second order phase transitions, however instead of simply
developing an imaginary gap to
drop out of the hydrodynamic spectrum as for the usual $U(1)$ superconductor,
they pair up in two modes that on top of
this gap also develop a real part.

The fact that $\Re(\w)$ does not vanish for these modes implies that
sufficiently close to $T_c$ and in the limit $k=0$, the late-time response of
the perturbed state will present an oscillatory decay of the perturbations,
meaning that, contrary to the $U(1)$ case, there will not be a temperature at
which the late-time behavior changes \emph{qualitatively}.

\paragraph{Special Gapped mode:}

Seeking for this mode is computationally much more involved. Its behavior is
characterized by a gap that is proportional to $\mu$. In particular,
 in \cite{Nicolis:2012vf} it was argued that a type II Goldstone
mode is accompanied by a gapped mode obeying $\omega(0) = q \mu$ with $q$
being the charge of the corresponding field. In our conventions here we have
$q=1$.
So we have to look for a mode with $\w(k=0) = \mu$. Furthermore we expect that
it connects to the lowest mode of the complex conjugate scalar in the unbroken
phase.

\begin{figure}[htp!]
\centering
\includegraphics[width=230pt]{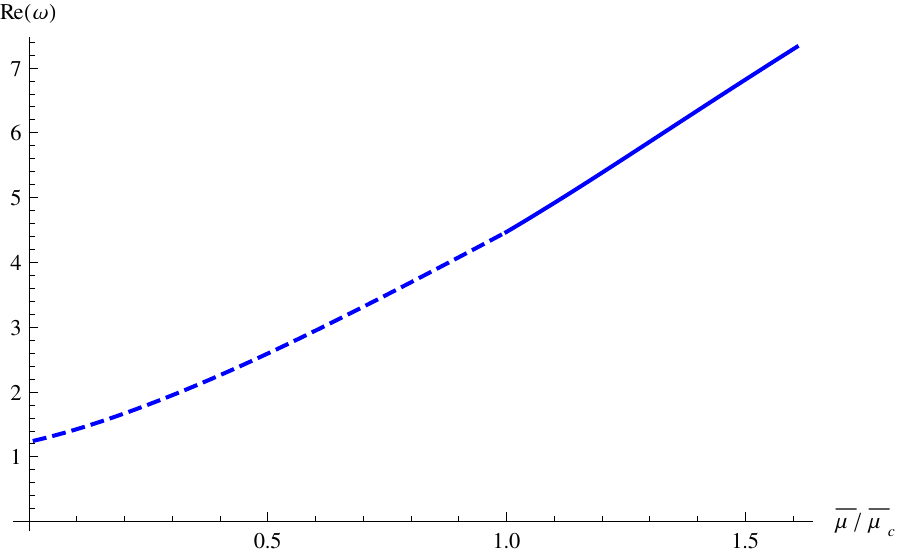}\hfill
\includegraphics[width=230pt]{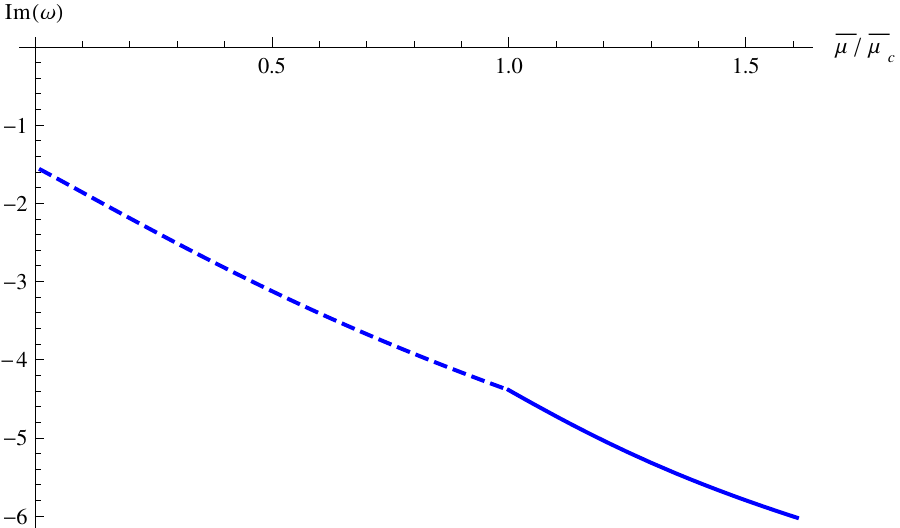}
\caption{\label{fig:Sgapped}Real (left) and imaginary (right) part of the
special gapped mode
versus the chemical
potential. We encounter the expected
linear behavior with $\mu$. The plot covers both the
unbroken (dashed line) and the broken (solid line) phases.}
\end{figure}

In Figure \ref{fig:Sgapped} we depict such mode at zero momentum with
respect to the chemical potential $\bar\mu$ in numerical units. Notice that the
mode is
continuous at the phase transition, as expected. We observe the
linear behavior with the chemical potential that is predicted theoretically, at
least near $\bar\mu_c$. It is very difficult to do the analysis when $\bar\mu >
6$ due to the high computational power demanded to carry out the computation.
The mode shows of course also a non-vanishing imaginary part which is due to the
dissipation at finite temperature. We find that the real part above the phase
transition
can be approximated by
\begin{equation}
 \Re\omega = 1.10\, \bar\mu\quad{\rm near}\,\,\bar\mu_c\,.
\end{equation}
This result shows a deviation from the conjectured behavior which could
nevertheless  be due to uncertainties
in the numerics. Let us emphasize here  that the numerics involved in tracking
this mode through the phase transition
were rather challenging.

\section{Discussion and Outlook}\label{sec:conclusions}

The main focus of this work was to establish the existence of
type II Goldstone modes in the quasinormal mode spectrum of
a holographic theory dual to a strongly coupled superfluid with
$U(2)$ symmetry.

We studied two models, one in which only the overall $U(1)$ symmetry is
gauged in the AdS bulk and another in which all the $U(2)$ symmetry is
gauged. The most important finding is that indeed there exist ungapped
excitations represented by quasinormal modes in the AdS bulk that show
the expected but somewhat unusual quadratic dispersion relation of type II
Goldstone bosons.

For the ungauged model this does constitute a surprising result.
After all, the field theory dual to this model does not contain the necessary
conserved
currents that would correspond to the generators of the global $SU(2)$ symmetry.
Standard proofs of the Goldstone theorem take the existence of such conserved
currents
for granted. On the other hand it is basically guaranteed that one can construct
an effective field theory, a simple Landau-Ginzburg type model, that captures
the
essential dynamics of the light modes, i.e. the lowest lying quasinormal modes.
Such a model would be essentially given by the field theoretical model of
section
\ref{sec:fieldth} and this guarantees the existence of the type II Goldstone
modes.
However one can expect that such an effective field theory approach can capture
only
the physics of the low lying QNMs but not the higher modes. This is indeed what
happens: the partner mode of the type II Goldstone mode in the ungauged model
does
not behave in the supposed universal way $\omega = q \mu$.
In contrast the corresponding mode in the gauged model does obey this relation
approximately and the deviation we found could very well be attributed to
numerical
difficulties and uncertainties that arise in the study of the higher QNMs.

One rather interesting perspective on the ungauged model opens up if we vary the
masses
of the scalar fields in the AdS bulk. If the masses are slightly different, then
at the
critical temperature only one of the two scalars will feature an ungapped QNM
(the one
with smaller mass). The lowest scalar mode of the second one will still be
gapped
at that temperature. As one goes through the phase transition we do not expect
this mode
to become massless at lower temperatures. Rather it should become a
pseudo-Goldstone
mode with a gap that is proportional to the mass splitting. The appearance of
the type II
Goldstone mode can then be interpreted as the effect of a symmetry enhancement
at the
point in parameter space where the masses of the scalars become degenerate.
Since this symmetry
is not represented by bulk-gauge fields we might call it an accidental symmetry.
At this point it is difficult to resist the temptation to draw a parallel to the
conjectured symmetry enhancement of high $T_c$ superconductors. In
\cite{Zhang:1997} it
was suggested that the phase diagram of high $T_c$ superconductors can be
captured
by a unified model with and enhancement of the $SO(3)\times U(1)$ symmetry of
rotations and electromagnetism to a larger $SO(5)$ symmetry. Since high $T_c$
superconductors
are d-wave rather than s-wave it remains to be seen how our symmetry enhancement
mechanism and the resulting type II Goldstone mode can be combined with
holographic
models of d-wave superfluids such as \cite{Benini:2010pr, Chen:2010mk}
\footnote{The appearance of unexpected massless modes related to symmetry 
enhancement in the context of Bose condensates was as well found in
\cite{Uchino:2010}}.

The second model we studied has bulk gauge fields for all of the $U(2)$
symmetry.
There are several important differences compared to the ungauged model. The most
eye-jumping one is that now we can also define and study the full set of
conductivities corresponding to the $U(2)$ symmetry. Nothing special occurs of
course
in the unbroken phase, there are simply four diagonal conductivities for all the
four
bulk gauge fields. In the broken phase there are however interesting new
phenomena.
In particular there are now off-diagonal conductivities that do not simply
vanish.
In addition we have found that also the diagonal conductivities in the $(1)-(2)$
sector, the
one containing the type II Goldstone mode, have delta-function poles at zero
frequency.
In this sense this sector is still superconducting. Moreover, going to a
decoupling basis for 
this sector leads to a very suggestive result: the conductivity develops a
Drude-like peak characteristic 
of metals on top of the infinite DC conductivity. On the other hand Landau's
criterion
for superfluidity does not hold in this sector. Recall that this says that
superfluidity
takes place for flow velocities $v$ that are smaller than the critical velocity
$v_c$ where
$v_c = \mathrm{min}_i \omega_i(k)/k$ for all excitation branches $i$ and over
all momenta
$k$ \cite{Khalatnikov, LanLif}. For a type II Goldstone mode the critical flow
velocity is clearly zero.

A second difference concerns the fate of the diffusive modes. In the unbroken
phase there
are simply four diffusive modes, one for each gauge field in the AdS bulk. In
the broken
phase there is one purely imaginary gapped `pseudo-diffusive' mode in the
$(0)-(3)$ sector,
i.e. in the sector isomorphic to the $U(1)$ s-wave superfluid. Since there is
still one
unbroken $U(1)$ symmetry there is also a normal diffusive mode for the preserved
$U(1)$ symmetry.
In the $(1)-(2)$ sector we have however two diffusive modes in the unbroken
phase. Going through
the phase transition these two modes can pair up and move off the imaginary
axis, becoming a pair
of usual gapped quasinormal modes with real and imaginary parts in their
frequencies. Generically
the imaginary part of this gap is smaller (i.e. it lies closer to the real axis)
then the gap
of the purely imaginary mode in the $(0)-(3)$ sector. A large, generic
perturbation
will in its late time response pattern excite both the $(0)-(3)$ and the
$(1)-(2)$ sector.
The late time response of the $U(2)$ invariant order parameter
$\sqrt{|\cO_1|^2+|\cO_2|^2}$
will therefore be dominated by these paired modes and show an oscillatory
behavior in contrast
to the response pattern of the order parameter in the $U(1)$ case
\cite{Bhaseen:2012gg}.

Another remarkable QNM is the special gapped mode, i.e. the partner mode of the
type II
Goldstone boson. At very high temperatures this
mode and the one which at $T=T_c$ leads to the sound mode are
degenerate. As we lower the temperature the gap of these modes becomes different
and, for $T<T_c$, it is expected that $\Re(\omega(k=0))$ for the Special Gapped
mode is proportional to $q \mu$ \cite{ Kapustin:2012cr, Nicolis:2012vf}. In
particular
we find $\omega \sim 1.1 \mu$ even if $q=1$ in our conventions. Unfortunately
with the numerical
methods employed in this paper we found it very difficult to study this mode and
the discrepancy
can therefore very well be a consequence of insufficient numerical accuracy. It
is probably worth
the effort to study this mode with alternative methods such as the relaxation
method developed
in \cite{Hoyos:2006gb, Kaminski:2009ce}.

There are several generalizations of the $U(2)$ model that seem interesting
and could be investigated in the future. A straightforward one would be to
analyze the $p$-wave like instability commented in section 4 and to look for a
stable background at low temperatures. The existence of a perturbation that
becomes tachyonic means that the system will suffer a second order phase
transition into a new phase, probably with the remaining $U(1)$ symmetry broken.
Such an investigation is currently underway. A similar scenario has been found
in 
\cite{Gusynin:2003yu, Gusynin:2004xr}, in which a gauged version of the field
theoretical sigma model undergoes 
a phase transition driven by an anisotropic vector condensate.

Another possible generalization would be to analyze the model when the
backreaction onto the metric is taken into account.
This introduces the energy-momentum tensor as an operator of the dual field
theory and thus we expect the usual sound and shear modes to stem from bulk
metric fluctuations. Moreover, this would allow us to obtain reliable results
even at very low temperatures and for instance compute the density of
superconducting charge densities at zero temperature, as well as
$\mathcal{B}(T=0)$.

We have constructed here a simply model with type II Goldstone bosons using
a ``bottom-up'' strategy. It is however also interesting to ask if such
models can be realized via ``top-down'' D-brane, string theory or M-theory
constructions \cite{Ammon:2008fc, Gauntlett:2009dn, Gubser:2009qm, Bobev:2010ib,
Bobev:2011rv}.

Another possible direction of research involves using the Fluid/Gravity
correspondence
\cite{Bhattacharyya:2008jc} in order to derive the Hydrodynamic expansion of the
current and upon including backreaction also the constitutive relation for the
energy-momentum tensor. This will throw light on the hydrodynamic behavior of
non-relativistic superfluids and in particular should result in the formulation
of the hydrodynamics of relativistic type II Goldstone modes. Up to our
knowledge
this is not even known to the leading, i.e. zeroth order in derivatives.

Another direction of investigation concerns the Landau criterion of
superfluidity. According to this criterion, the dispersion relation
(\ref{disr12}) prevents the system from accommodating a superflow. Therefore,
even though $\sigma_{11}$ and $\sigma_{22}$ are superconducting, as soon as a
non-vanishing supercurrent/superflow is switched on the system should in
principle be taken out of the superfluid phase. It should be noted however
that in a holographic superfluid the condensate and its flow are of leading
order in a large $N$ expansion and the excitation spectrum, the QNMs, are
subleading. Therefore it seems not clear if Landau's criterion can be applied
straightforwardly. It is known however that for the $U(1)$ superfluid there
exists a critical superflow or a critical supercurrent above which the
condensate
vanishes \cite{Herzog:2008he, Basu:2008st, Arean:2010xd}.
It would be very interesting to
analyze if this is so (along the lines of \cite{Arean:2010xd}) and also to study
how the whole mechanism takes place. Such an investigation is currently underway
\cite{[8]}.

Finally it is also interesting to ask the question if holographic models
featuring Goldstone
modes with higher order dispersion relation $\omega = c k^n$ with $n>2$ can be
constructed.

\appendix

\section{Matrix valued Kramers-Kronig relation}
\label{app.a}
The generically {\em matrix-valued} spectral function is defined as
\begin{equation}
\rho_{ij}( x) =\langle [\cO_i(x), \cO_j(0) ]\rangle\, , \label{spfu}
\end{equation}
where $\cO_i$ are Hermitian operators.
Its behavior under Hermitian conjugation is
\begin{equation}
\rho(x )^\dagger  =  \rho( -x )  = - \rho(x )^t\, .
\end{equation}
Correspondingly,  the Fourier transform $\tilde \rho(k)=\int d^4 x\,
e^{-ikx}\rho(x) $ also satisfies a set of identities
\begin{equation}
\tilde\rho(k)^\dagger = \tilde\rho(k) = -\tilde\rho(-k)^t \label{proprhok} \, .
\end{equation}
In particular this means that the diagonal components are real and antisymmetric
under $k\to -k$.
One may also be interested in the behavior under $\omega\to -\omega$. We take
now $k=(\omega, \bq )$.
For theories with rotational invariance
the spectral function can depend only on $\bq^2$. Consequently the diagonal
components will also be real and
odd in $\omega$
\begin{equation}
\rho_{ii}(\omega,\bq^2) = \rho_{ii}(\omega,\bq^2)^* =- \rho_{ii}(-\omega,\bq^2) 
\, .
\end{equation}
For the off-diagonal components however, only if one also imposes time reversal
or parity symmetry
can one prove that the off-diagonal entries must be either even or odd functions
of the frequency.
In the present case time reversal symmetry is broken by the presence of the
chemical potential.
Further constraints can however by obtained by supposing that the theory is
invariant under
$\bx \rightarrow -\bx$.
For an odd number of spatial dimensions we could use the parity operators $P$ to
take $\bx \rightarrow -\bx$.
In the two spatial dimensions we study in this paper we can take $P$ to by a
rotation by $\pi$
(for an arbitrary even  number of spatial dimensions $D=2n$
we could take the angle $\pi$ for all the rotations in the $i, i+1$-th plane for
all $i\leq n$).
This P-operator acts as $P \cO_i(t,\bx)P^{-1} = \sigma_i  \cO_i(t,-\bx)$ with
$\sigma_i = \pm1$.
In odd spatial dimensions $\sigma_i$ is the parity of the operator. In even
spatial dimension
$\sigma_i=-1$ if $\cO_i$ is the component of a spatial vector. Hence
\begin{equation}
P[ \rho_{ij}(t,  \bx )] = \sigma_i \sigma_j  \rho_{ij}( t, -  \bx ) \,.
\end{equation}
P-invariance implies $\rho_{ij}( t,  \bx )=\sigma_i \sigma_j  \rho_{ij}( t, - 
\bx )$, which for the Fourier transform implies that
\begin{equation}\label{eq:Pspec}
\tilde{\rho}_{ij}(\omega, \bq ) = - \sigma_i
\sigma_j\tilde{\rho}_{ij}(-\omega,\bq)^*\,.
\end{equation}
So the off-diagonal entries are either odd or even functions of $\omega$
depending on the signs $\sigma_i$. In the case where the fields transform in the
same way under the parity operator this means that the real (imaginary) part of
the off-diagonal components is an odd (even) function of the frequency.

From the spectral function, as defined in (\ref{spfu}) we can define two causal
propagators, namely
the retarded and advanced Green's functions
\begin{align}
G_R( x) =& -i \Theta(t) \rho( x )\,,\\
G_A( x) =& ~i \Theta(-t) \rho( x )\, ,
\end{align}
where $x = (t,\bx)$.
Using (\ref{proprhok}), one can prove the following relation among the  Fourier
transforms of these
\begin{equation}
\tilde G_R(k) = \tilde G_{R}(-k)^* = \tilde G_{A}(k)^\dagger \, . \label{reldag}
\end{equation}
From here, we see that the real (imaginary) part, $\Re(G_R)$ ($\Im(G_R)$), is
even (odd) under $k\to -k$.
We can compute the Fourier transform of the retarded Green's function, which is
given by the convolution of the Fourier transform of the Heaviside step function
$\tilde\Theta(\omega)$ with the Fourier transform of the spectral function
$\tilde\rho(k)$,
\begin{equation}
\tilde G_R(\omega,\bq) = -i \int_{-\infty}^\infty \tilde \Theta (\omega - \mu)
\tilde \rho (\mu,\bq) \frac{d\mu}{2\pi}\, .
\end{equation}
Using the Fourier transform of the step function
\begin{equation}
\tilde\Theta(\omega) = \frac{i}{\omega + i \epsilon}\,,\nonumber
\end{equation}
and the Sokhatsky-Weierstrass theorem we get
\begin{equation}
\tilde{G}_R(\omega,\bq) =  \cP\!\! \int_{-\infty}^\infty \frac{
\tilde{\rho}(\omega',\bq) }{\omega - \omega'}\frac{d\omega'}{2\pi}
- \frac{i}{2} \tilde{\rho}(\omega,\bq)\,, \label{grsplit}
\end{equation}
where $\cP$ denotes the principle value.
From the Hermiticity of $\tilde \rho(k)$ we see that  we can regard
(\ref{grsplit}) as a split  of $\tilde G^R(k)$ into its Hermitian and
anti-Hermitian parts, and find that the spectral function can be computed
from the anti-Hermitian part of the Fourier transform of the retarded Green's
function
\begin{equation} \label{specfuncdef}
\tilde\rho(k)= i [\tilde{G}_R(k) - \tilde{G}_R(k)^\dagger] \equiv  2 i
\tilde{G}_R^{(A)}(k)\,,
\end{equation}
where the $(A)$ stands for anti-Hermitian\footnote{Using (\ref{reldag}) we can
always work with retarded Green's functions $G_R$.}. Plugging this back into
(\ref{grsplit}) and taking the Hermitian part $(H)$ on both sides we arrive at
\begin{equation}\label{KKrel}
\tilde G_R^{(H)}(\omega) =  \frac{i}{\pi} \cP \!\! \int_{-\infty}^\infty \frac{
G_R^{(A)}(\omega') }{\omega - \omega'} d\omega'\, ,
\end{equation}
which is nothing but the Kramers-Kr\"onig relation for the matrix Green's
function.
It is complemented by the conjugate relation interchanging the Hermitian and
anti-Hermitian parts.
Imposing P-invariance and using (\ref{eq:Pspec}) and (\ref{grsplit}) if follows
that the Green's function satisfies
\begin{equation}\label{eq:Gparity}
\tilde G^R_{ij} (\omega, \bq) = \sigma_i \sigma_j \tilde G^R_{ij}(-\omega,\bq)^*
\, .
\end{equation}
This constrains the QNM spectrum. Taking for example a diagonal Green's function
with $i=j$ and writing
it as a sum over quasinormal frequencies \cite{Amado:2008ji,Stephanov:2011wf}
one seems that the quasinormal frequencies have to
come either in pairs obeying $\omega_n$ and $\tilde\omega_n = - \omega_n^*$ or
are confined to lie on the imaginary
axis. The residues of the pairs are related by complex conjugation and the
purely imaginary
ones have to have also purely imaginary residue.

\section{Solving the fluctuation equations}\label{appB}
The $(1)-(2)$ sector of the gauged model in the broken phase consists of a
system of coupled equations \erf{qnmalpha}-\erf{cons2}. In order to extract the
spectrum of quasinormal
modes we made use of the techniques detailed in
\cite{Amado:2009ts,Kaminski:2009dh}, where a method to
compute the poles of the Green functions in terms of non-gauge invariant fields
was developed. The quasinormal
frequencies are given by the zeroes of the determinant of the field matrix
spanned by a maximal set of linearly
independent solutions satisfying infalling boundary conditions on the horizon
evaluated at the boundary.

Imposing infalling boundary conditions, the near horizon behavior of the fields
solving the
mentioned equations reads
\bea
\alpha&=& (\rho-1)^\kappa \left(\alpha_{(0)} +\alpha_{(1)} (\rho-1)  + \dots
\right) \,,\\
\beta&=& (\rho-1)^\kappa \left(\beta_{(0)} +\beta_{(1)} (\rho-1)  + \dots
\right) \,,\\
a_t^{(i)}&=& (\rho-1)^{\kappa+1} \left(a^{(i)}_{t\,(0)} +a^{(i)}_{t\,(1)}
(\rho-1) + \dots \right) \,,\\
a_x^{(i)}&=& (\rho-1)^\kappa \left(a^{(i)}_{x\,(0)} +a^{(i)}_{x\,(1)} (\rho-1) +
\dots \right) \,,
\eea
where $\kappa=-i\omega/3$ and $i=1,2$. Since the system is subject to two
constraints, we can only choose
four of the six parameters at the horizon. Without loss of generality, solutions
can be parametrized
by $\{\alpha_{(0)},\beta_{(0)},a^{(i)}_{x\,(0)}\}$. In this way it is possible
to construct four
independent solutions to the field equations. We can label them as $I,\,
II,\,III,\,IV$.

Two additional solutions, $V,\,VI$, can be obtained by performing gauge
transformations of the trivial solution,
\begin{eqnarray}
\label{gsol1} &&\a \rightarrow 0,\,  \b \rightarrow i\frac{\l_1 \Psi}{2},\,
a^{(1)}_x \rightarrow -k \l_1, \, a^{(2)}_x \rightarrow 0,\,  a^{(1)}_t
\rightarrow \w \l_1,\, a^{(2)}_t \rightarrow i \Theta \l_1 \,,\\
\label{gsol2} &&\a \rightarrow i\frac{\l_2 \Psi}{2},\, \b \rightarrow 0,\,
a^{(1)}_x \rightarrow 0, \, a^{(2)}_x \rightarrow -k \l_2,\,  a^{(1)}_t
\rightarrow - i \Theta \l_2,\, a^{(2)}_t \rightarrow \w \l_2 \,,
\end{eqnarray}
where $\l_i$ are arbitrary constants. Notice that these pure gauge solutions are
not algebraic
since they have a nontrivial dependence on the bulk coordinate $\rho$.

The most general solution for each field $\varphi_i =
\{\tilde\alpha,\,\tilde\beta,\,a_t^{(i)},\,a_x^{(i)}\}$
is given by a linear combination of the above solutions, including the pure
gauge modes,
\be
\varphi_i = c_{I} \varphi_i^{I} + c_{II} \varphi_i^{II} + c_{III}
\varphi_i^{III} +c_{IV} \varphi_i^{IV}+c_{V} \varphi_i^{V}+c_{VI}
\varphi_i^{VI}\,,
\ee
where we have defined $\{\tilde\alpha(\rho),\,\tilde\beta(\rho)\}
=\{\rho\alpha(\rho),\,\rho\beta(\rho)\}$. This convenient
choice allows us to identify the asymptotic boundary values $\varphi_i$ with the
sources of the
gauge invariant operators of the dual field theory.

As shown in \cite{Amado:2009ts}, the poles of the retarded Green functions will
be given by the values of the
frequency for which the determinant of the matrix spanned by $\varphi_i^N$
vanishes asymptotically. Expanding the determinant
and evaluating it at a cutoff $\rho=\Lambda$, it reads
\begin{eqnarray} \label{eq:detBroken}
0 &=& \frac{1}{\lambda_1\lambda_2}\det \left(
\begin{array}{cccccc}
{\varphi_\alpha}^I&{\varphi_\alpha}^{II}&{\varphi_\alpha}^{III}&{\varphi_\alpha}
^{IV}&{\varphi_\alpha}^{V}&{\varphi_\alpha}^{VI}\\
{\varphi_\beta}^I&{\varphi_\beta}^{II}&{\varphi_\beta}^{III}&{\varphi_\beta}^{IV
}&{\varphi_\beta}^{V}&{\varphi_\beta}^{VI}\\
{\varphi_{t(1)}}^I&{\varphi_{t(1)}}^{II}&{\varphi_{t(1)}}^{III}&{\varphi_{t(1)}}
^{IV}&{\varphi_{t(1)}}^{V}&{\varphi_{t(1)}}^{VI}\\
{\varphi_{t(2)}}^I&{\varphi_{t(2)}}^{II}&{\varphi_{t(2)}}^{III}&{\varphi_{t(2)}}
^{IV}&{\varphi_{t(2)}}^{V}&{\varphi_{t(2)}}^{VI}\\
{\varphi_{x(1)}}^I&{\varphi_{x(1)}}^{II}&{\varphi_{x(1)}}^{III}&{\varphi_{x(1)}}
^{IV}&{\varphi_{x(1)}}^{V}&{\varphi_{x(1)}}^{VI}\\
{\varphi_{x(2)}}^I&{\varphi_{x(2)}}^{II}&{\varphi_{x(2)}}^{III}&{\varphi_{x(2)}}
^{IV}&{\varphi_{x(2)}}^{V}&{\varphi_{x(2)}}^{VI}\\
\end{array}
\right) \, \\
&=& \omega^2 \det\left(
\begin{array}{cccc}
{\varphi_\alpha^I}&{\varphi^{II}_\alpha}&{\varphi_\alpha^{III}}&{\varphi_\alpha^
{IV}}\\
{\varphi_\beta^I}&{\varphi_\beta^{II}}&{\varphi_\beta^{III}}&{\varphi_\beta^{IV}
}\\
{\varphi_{x(1)}^I}&{\varphi_{x(1)}^{II}}&{\varphi_{x(1)}^{III}}&{\varphi_{x(1)}^
{IV}}\\
{\varphi_{x(2)}^I}&{\varphi_{x(2)}^{II}}&{\varphi_{x(2)}^{III}}&{\varphi_{x(2)}^
{IV}}\\
\end{array}
\right)
+\omega k \det\left(
\begin{array}{cccc}
{\varphi_\alpha^I}&{\varphi^{II}_\alpha}&{\varphi_\alpha^{III}}&{\varphi_\alpha^
{IV}}\\
{\varphi_\beta^I}&{\varphi_\beta^{II}}&{\varphi_\beta^{III}}&{\varphi_\beta^{IV}
}\\
{\varphi_{t(1)}^I}&{\varphi_{t(1)}^{II}}&{\varphi_{t(1)}^{III}}&{\varphi_{t(1)}^
{IV}}\\
{\varphi_{x(2)}^I}&{\varphi_{x(2)}^{II}}&{\varphi_{x(2)}^{III}}&{\varphi_{x(2)}^
{IV}}\\
\end{array}
\right)\,\nonumber\\
&&-\omega k \det\left(
\begin{array}{cccc}
{\varphi_\alpha^I}&{\varphi^{II}_\alpha}&{\varphi_\alpha^{III}}&{\varphi_\alpha^
{IV}}\\
{\varphi_\beta^I}&{\varphi_\beta^{II}}&{\varphi_\beta^{III}}&{\varphi_\beta^{IV}
}\\
{\varphi_{t(2)}^I}&{\varphi_{t(2)}^{II}}&{\varphi_{t(2)}^{III}}&{\varphi_{t(2)}^
{IV}}\\
{\varphi_{x(1)}^I}&{\varphi_{x(1)}^{II}}&{\varphi_{x(1)}^{III}}&{\varphi_{x(1)}^
{IV}}\\
\end{array}
\right)
+ k^2  \det\left(
\begin{array}{cccc}
{\varphi_\alpha^I}&{\varphi^{II}_\alpha}&{\varphi_\alpha^{III}}&{\varphi_\alpha^
{IV}}\\
{\varphi_\beta^I}&{\varphi_\beta^{II}}&{\varphi_\beta^{III}}&{\varphi_\beta^{IV}
}\\
{\varphi_{t(1)}^I}&{\varphi_{t(1)}^{II}}&{\varphi_{t(1)}^{III}}&{\varphi_{t(1)}^
{IV}}\\
{\varphi_{t(2)}^I}&{\varphi_{t(2)}^{II}}&{\varphi_{t(2)}^{III}}&{\varphi_{t(2)}^
{IV}}\\
\end{array}
\right)\, , \nonumber
\end{eqnarray}
where the background boundary conditions $\Th(\Lambda)=0$ and $\Lambda \Psi =0$
have been already imposed. This absence of background sources for the
corresponding operators makes the point
$(\w,k)=(0,0)$ a trivial
solution to the vanishing determinant condition, which ensures the existence of
a hydrodynamic
mode. Notice also that the point $(\w,k)=(0,0)$ is a double solution to the
previous determinant equation.

Solutions to the equations of motion and to the determinant condition
\erf{eq:detBroken}
have been computed numerically. It has been checked that the election of
solution basis,
i.e. of initial values of the
free parameters, does not affect the result.

\section*{Acknowledgments}
We would like to thank T. Brauner, J. Gauntlett, A. G. Grushin, I. Shovkovy, A.
Schmitt and M.A. H. Vozmediano, T. Wiseman for enjoyable
and useful discussions.
I. A. is supported by the Israel Science Foundation under grants no. 392/09 and
495/11.
A. J., L. M. and K. L. are supported by Plan Nacional de Altas Energ\'\i as FPA
2009-07890, Consolider Ingenio 2010 CPAN CSD200-00042 and
HEP-HACOS S2009/ESP-2473. L.M. has been supported by FPI-fellowship
BES-2010-041571. A. J. has been supported by FPU fellowship AP2010-5686.
D. A. thanks the FRont Of pro-Galician Scientists for unconditional support.
D. A. and I. S. would like to thank the HIPM for reminding them how fun working
in
physics can be.



\begin{thebibliography}{99}


\bibitem{Gubser:2008px}
  S.~S.~Gubser,
  ``Breaking an Abelian gauge symmetry near a black hole horizon,''
  Phys.\ Rev.\ D {\bf 78} (2008) 065034
  [arXiv:0801.2977 [hep-th]].

\bibitem{Hartnoll:2008vx}
  S.~A.~Hartnoll, C.~P.~Herzog and G.~T.~Horowitz,
  ``Building a Holographic Superconductor,''
  Phys.\ Rev.\ Lett.\  {\bf 101} (2008) 031601
  [arXiv:0803.3295 [hep-th]].

\bibitem{Hartnoll:2008kx}
  S.~A.~Hartnoll, C.~P.~Herzog and G.~T.~Horowitz,
  ``Holographic Superconductors,''
  JHEP {\bf 0812} (2008) 015
  [arXiv:0810.1563 [hep-th]].

\bibitem{Gubser:2008wv}
  S.~S.~Gubser and S.~S.~Pufu,
  ``The Gravity dual of a p-wave superconductor,''
  JHEP {\bf 0811} (2008) 033
  [arXiv:0805.2960 [hep-th]].

\bibitem{Benini:2010pr}
  F.~Benini, C.~P.~Herzog, R.~Rahman and A.~Yarom,
  ``Gauge gravity duality for d-wave superconductors: prospects and
challenges,''
  JHEP {\bf 1011} (2010) 137
  [arXiv:1007.1981 [hep-th]].

\bibitem{Chen:2010mk}
  J.~-W.~Chen, Y.~-J.~Kao, D.~Maity, W.~-Y.~Wen and C.~-P.~Yeh,
  ``Towards A Holographic Model of D-Wave Superconductors,''
  Phys.\ Rev.\ D {\bf 81} (2010) 106008
  [arXiv:1003.2991 [hep-th]].

\bibitem{Horowitz:2010gk}
  G.~T.~Horowitz,
  ``Introduction to Holographic Superconductors,''
  arXiv:1002.1722 [hep-th].

\bibitem{Kaminski:2010zu}
  M.~Kaminski,
 ``Flavor Superconductivity \& Superfluidity,''
  Lect.\ Notes Phys.\  {\bf 828} (2011) 349
  [arXiv:1002.4886 [hep-th]].

\bibitem{Son:2002sd}
  D.~T.~Son and A.~O.~Starinets,
  ``Minkowski space correlators in AdS / CFT correspondence: Recipe and
applications,''
  JHEP {\bf 0209} (2002) 042
  [hep-th/0205051].

\bibitem{Herzog:2002pc}
  C.~P.~Herzog and D.~T.~Son,
  ``Schwinger-Keldysh propagators from AdS/CFT correspondence,''
  JHEP {\bf 0303} (2003) 046
  [hep-th/0212072].

\bibitem{Horowitz:1999jd}
  G.~T.~Horowitz and V.~E.~Hubeny,
  ``Quasinormal modes of AdS black holes and the approach to thermal
equilibrium,''
  Phys.\ Rev.\ D {\bf 62} (2000) 024027
  [hep-th/9909056].

\bibitem{Birmingham:2001pj}
  D.~Birmingham, I.~Sachs and S.~N.~Solodukhin,
  ``Conformal field theory interpretation of black hole quasinormal modes,''
  Phys.\ Rev.\ Lett.\  {\bf 88} (2002) 151301
  [hep-th/0112055].

\bibitem{Berti:2009kk}
  E.~Berti, V.~Cardoso and A.~O.~Starinets,
  ``Quasinormal modes of black holes and black branes,''
  Class.\ Quant.\ Grav.\  {\bf 26} (2009) 163001
  [arXiv:0905.2975 [gr-qc]].

\bibitem{Landsteiner:2012gn}
  K.~Landsteiner,
 ``The Sound of Strongly Coupled Field Theories: Quasinormal Modes In AdS,''
  AIP Conf.\ Proc.\  {\bf 1458} (2011) 174
  [arXiv:1202.3550 [gr-qc]].

\bibitem{Amado:2009ts}
  I.~Amado, M.~Kaminski and K.~Landsteiner,
  ``Hydrodynamics of Holographic Superconductors,''
  JHEP {\bf 0905} (2009) 021
  [arXiv:0903.2209 [hep-th]].

\bibitem{Yarom:2009uq}
  A.~Yarom,
  ``Fourth sound of holographic superfluids,''
  JHEP {\bf 0907} (2009) 070
  [arXiv:0903.1353 [hep-th]].

\bibitem{Bhaseen:2012gg}
  M.~J.~Bhaseen, J.~P.~Gauntlett, B.~D.~Simons, J.~Sonner and T.~Wiseman,
  ``Holographic Superfluids and the Dynamics of Symmetry Breaking,''
  arXiv:1207.4194 [hep-th].


\bibitem{Halperin} B.I. Halperin, ``Dynamic properties of the multicomponent
Bose fluid," \emph{Phys. Rev. B} \textbf{11}, 178–190 (1975).

\bibitem{Schafer:2001bq}
  T.~Schafer, D.~T.~Son, M.~A.~Stephanov, D.~Toublan and J.~J.~M.~Verbaarschot,
  ``Kaon condensation and Goldstone's theorem,''
  Phys.\ Lett.\ B {\bf 522} (2001) 67
  [hep-ph/0108210].

\bibitem{Miransky:2001tw}
  V.~A.~Miransky and I.~A.~Shovkovy,
  ``Spontaneous symmetry breaking with abnormal number of Nambu-Goldstone bosons
and kaon condensate,''
  Phys.\ Rev.\ Lett.\  {\bf 88} (2002) 111601
  [hep-ph/0108178].

\bibitem{Filev:2009xp}
  V.~G.~Filev, C.~V.~Johnson and J.~P.~Shock,
  ``Universal Holographic Chiral Dynamics in an External Magnetic Field,''
  JHEP {\bf 0908} (2009) 013
  [arXiv:0903.5345 [hep-th]].

\bibitem{Brauner:2010wm}
  T.~Brauner,
  ``Spontaneous Symmetry Breaking and Nambu-Goldstone Bosons in Quantum
Many-Body Systems,''
  Symmetry {\bf 2} (2010) 609
  [arXiv:1001.5212 [hep-th]].

\bibitem{Nielsen:1975hm}
  H.~B.~Nielsen and S.~Chadha,
  ``On How to Count Goldstone Bosons,''
  Nucl.\ Phys.\ B {\bf 105} (1976) 445.

\bibitem{Watanabe:2011ec}
  H.~Watanabe and T.~Brauner,
  ``On the number of Nambu-Goldstone bosons and its relation to charge
densities,''
  Phys.\ Rev.\ D {\bf 84} (2011) 125013
  [arXiv:1109.6327 [hep-ph]].

\bibitem{Watanabe:2012hr}
  H.~Watanabe and H.~Murayama,
  ``Unified Description of Nambu-Goldstone Bosons without Lorentz Invariance,''
  Phys.\ Rev.\ Lett.\  {\bf 108} (2012) 251602
  [arXiv:1203.0609 [hep-th]].

\bibitem{Watanabe:2013iia}
  H.~Watanabe and H.~Murayama,
  ``Redundancies in Nambu-Goldstone Bosons,''
  arXiv:1302.4800 [cond-mat.other].

\bibitem{Hidaka:2012ym}
  Y.~Hidaka,
  ``Counting rule for Nambu-Goldstone modes in nonrelativistic systems,''
  Phys.\ Rev.\ Lett.\  {\bf 110} (2013)  091601
  arXiv:1203.1494 [hep-th].


\bibitem{Kapustin:2012cr}
   A.~Kapustin,
   ``Remarks on nonrelativistic Goldstone bosons,''
   arXiv:1207.0457 [hep-ph].

\bibitem{Graphene}
Z. Q. Li, E. A. Henriksen, Z. Jiang, Z. Hao, M. C. Martin, P. Kim, H. L. Stormer
and D. N. Basov,
``Dirac charge dynamics in graphene by infrared spectroscopy,''
Nature Physics 4, 532 - 535 (2008). 

\bibitem{Brauner:2006xm}
  T.~Brauner,
  ``Spontaneous symmetry breaking in the linear sigma model at finite chemical
potential: One-loop corrections,''
  Phys.\ Rev.\ D {\bf 74} (2006) 085010
  [hep-ph/0607102].

\bibitem{Nicolis:2012vf}
  A.~Nicolis and F.~Piazza,
  ``A relativistic non-relativistic Goldstone theorem: gapped Goldstones at
finite charge density,''
  Phys.\ Rev.\ Lett.\  {\bf 110} (2013) 011602
  [arXiv:1204.1570 [hep-th]].

\bibitem{Klebanov:1999tb}
  I.~R.~Klebanov and E.~Witten,
  ``AdS / CFT correspondence and symmetry breaking,''
  Nucl.\ Phys.\ B {\bf 556} (1999) 89
  [hep-th/9905104].
in which $\psi_1$ is


\bibitem{Amado:2008ji}
  I.~Amado, C.~Hoyos-Badajoz, K.~Landsteiner and S.~Montero,
  ``Hydrodynamics and beyond in the strongly coupled N=4 plasma,''
  JHEP {\bf 0807} (2008) 133
  [arXiv:0805.2570 [hep-th]].

\bibitem{Kaminski:2009dh}
  M.~Kaminski, K.~Landsteiner, J.~Mas, J.~P.~Shock and J.~Tarrio,
  ``Holographic Operator Mixing and Quasinormal Modes on the Brane,''
  JHEP {\bf 1002} (2010) 021
  [arXiv:0911.3610 [hep-th]].

\bibitem{Davison:2011ek}
  R.~A.~Davison and A.~O.~Starinets,
  ``Holographic zero sound at finite temperature,''
  Phys.\ Rev.\ D {\bf 85} (2012) 026004
  [arXiv:1109.6343 [hep-th]].
  
\bibitem{francescos}
F. Bigazzi, A. L. Cotrone, D. Musso, N. P. Fokeeva and D. Seminara,
``Unbalanced Holographic Superconductors and Spintronics,''
JHEP {\bf 1202} (2012) 078
  [arXiv:1111.6601 [hep-th]].
  
\bibitem{Hartnoll:2009sz}
  S.~A.~Hartnoll,
  ``Lectures on holographic methods for condensed matter physics,''
  Class.\ Quant.\ Grav.\  {\bf 26} (2009) 224002
  [arXiv:0903.3246 [hep-th]].
  
\bibitem{Zhang:1997}
  S.~C. `Zhang. 1997.
 `` A Unified Theory Based on SO(5) Symmetry of Superconductivity and
Antiferromagnetism,'' Science,275,1089
  
\bibitem{Uchino:2010}
  S.~Uchino, M.~Kobayashi and M.~Ueda,
  ``Bogoliubov Theory and Lee-Huang-Yang Corrections in Spin-1 and Spin-2
Bose-Einstein Condensates in the Presence of the Quadratic Zeeman Effect,''
  Phys.\ Rev.\ A {\bf 81} (2010)  063632.
  
\bibitem{Khalatnikov}
 I.~M.~ Khalatnikov,
 ``An Introduction to the Theory of Superfluidity,''
 Advanced Book Classics, Westview Press.

\bibitem{LanLif}
 L.~.D.~ Landau and E.~M.~ Lifshitz,
 ``Course of Theoretical Physcis,``
 Vol. 9, Statistical Physics, Part 2, Chap. III,
 Pergamon Press.

\bibitem{Hoyos:2006gb}
  C.~Hoyos-Badajoz, K.~Landsteiner and S.~Montero,
  ``Holographic meson melting,''
  JHEP {\bf 0704} (2007) 031
  [hep-th/0612169].

\bibitem{Kaminski:2009ce}
  M.~Kaminski, K.~Landsteiner, F.~Pena-Benitez, J.~Erdmenger, C.~Greubel and
P.~Kerner,
  ``Quasinormal modes of massive charged flavor branes,''
  JHEP {\bf 1003} (2010) 117
  [arXiv:0911.3544 [hep-th]].

\bibitem{Gusynin:2003yu} 
  V.~P.~Gusynin, V.~A.~Miransky and I.~A.~Shovkovy,
 ``Spontaneous rotational symmetry breaking and roton - like excitations in
gauged sigma model at finite density,''
  Phys.\ Lett.\ B {\bf 581}, 82 (2004)
  [hep-ph/0311025].
 
\bibitem{Gusynin:2004xr}
  V.~P.~Gusynin, V.~A.~Miransky and I.~A.~Shovkovy,
  ``Surprises in nonperturbative dynamics in sigma-model at finite density,''
  Mod.\ Phys.\ Lett.\ A {\bf 19} (2004) 1341
  [hep-ph/0406219].

\bibitem{Ammon:2008fc}
  M.~Ammon, J.~Erdmenger, M.~Kaminski and P.~Kerner,
  ``Superconductivity from gauge/gravity duality with flavor,''
  Phys.\ Lett.\ B {\bf 680} (2009) 516
  [arXiv:0810.2316 [hep-th]].

\bibitem{Gauntlett:2009dn}
  J.~P.~Gauntlett, J.~Sonner and T.~Wiseman,
  ``Holographic superconductivity in M-Theory,''
  Phys.\ Rev.\ Lett.\  {\bf 103} (2009) 151601
  [arXiv:0907.3796 [hep-th]].

\bibitem{Gubser:2009qm}
  S.~S.~Gubser, C.~P.~Herzog, S.~S.~Pufu and T.~Tesileanu,
  ``Superconductors from Superstrings,''
  Phys.\ Rev.\ Lett.\  {\bf 103} (2009) 141601
  [arXiv:0907.3510 [hep-th]].

\bibitem{Bobev:2010ib}
  N.~Bobev, N.~Halmagyi, K.~Pilch and N.~P.~Warner,
  ``Supergravity Instabilities of Non-Supersymmetric Quantum Critical Points,''
  Class.\ Quant.\ Grav.\  {\bf 27} (2010) 235013
  [arXiv:1006.2546 [hep-th]].
  
\bibitem{Bobev:2011rv}
  N.~Bobev, A.~Kundu, K.~Pilch and N.~P.~Warner,
  ``Minimal Holographic Superconductors from Maximal Supergravity,''
  JHEP {\bf 1203} (2012) 064
  [arXiv:1110.3454 [hep-th]].
  

\bibitem{Bhattacharyya:2008jc}
  S.~Bhattacharyya, V.~EHubeny, S.~Minwalla and M.~Rangamani,
  ``Nonlinear Fluid Dynamics from Gravity,''
  JHEP {\bf 0802} (2008) 045
  [arXiv:0712.2456 [hep-th]].

\bibitem{Herzog:2008he}
  C.~P.~Herzog, P.~K.~Kovtun and D.~T.~Son,
  ``Holographic model of superfluidity,''
  Phys.\ Rev.\ D {\bf 79} (2009) 066002
  [arXiv:0809.4870 [hep-th]].

\bibitem{Basu:2008st}
  P.~Basu, A.~Mukherjee and H.~-H.~Shieh,
  ``Supercurrent: Vector Hair for an AdS Black Hole,''
  Phys.\ Rev.\ D {\bf 79} (2009) 045010
  [arXiv:0809.4494 [hep-th]].

\bibitem{Arean:2010xd}
  D.~Arean, M.~Bertolini, J.~Evslin and T.~Prochazka,
  ``On Holographic Superconductors with DC Current,''
  JHEP {\bf 1007} (2010) 060
  [arXiv:1003.5661 [hep-th]].

\bibitem{[8]} \emph{In preparation}.


\bibitem{Stephanov:2011wf}
  M.~A.~Stephanov and Y.~Yin,
  ``Conductivity and quasinormal modes in holographic theories,''
  JHEP {\bf 1202} (2012) 017
  [arXiv:1111.5303 [hep-ph]].


\end{thebibliography}
\end{document}